\documentclass[a4paper,10pt]{article}
\usepackage{jheppub}
\usepackage[utf8]{inputenc}
\usepackage{amsmath,amssymb,bm,slashed,braket}
\usepackage{dsfont}
\usepackage{amsfonts}
\usepackage{bm,bbm}
\usepackage{graphicx}
\usepackage{slashed} 
\usepackage[dvipsnames,table]{xcolor}
\usepackage[normalem]{ulem}
\usepackage{hyperref}
\hypersetup{colorlinks,citecolor= blue,linkcolor= blue, urlcolor=blue}
\usepackage{array}
\usepackage{verbatim}
\usepackage{epsfig}
\usepackage{multirow, booktabs}
\usepackage{makecell}
\usepackage{hhline}
\usepackage{bbold}
\usepackage[version=4]{mhchem}
\usepackage[capitalise]{cleveref}
\usepackage{orcidlink}
\usepackage{subcaption}
\usepackage{enumitem}

\allowdisplaybreaks[4]
\usepackage{tikz} 
\usepackage{tkz-euclide}
\usetikzlibrary{backgrounds} 
\usetikzlibrary{decorations.pathmorphing}
\usetikzlibrary{arrows.meta}
\tikzset{
mystyle/.style={line width=1, baseline, scale=0.6, every node/.style={scale=1}},
v/.style={decorate, draw, decoration={snake, segment length=2.mm, amplitude=0.5mm}},
f/.style={draw, decoration={markings,mark=at position #1 with {\arrow[]{Latex[length=1.5mm,width=1.5mm]}}},
    postaction={decorate},node contents=#1},
f/.default=.6,
fb/.style={draw,decoration={markings,mark=at position #1 with {\arrowreversed[]{Latex[length=1.5mm,width=1.5mm]}}},
    postaction={decorate},node contents=#1},
fb/.default=.6,
s/.style={dashed,draw, decoration={markings,mark=at position #1 with {\arrow[]{Latex[length=1.5mm,width=1.5mm]}}},
    postaction={decorate},node contents=#1},
s/.default=.6,    
sb/.style={dashed,draw,decoration={markings,mark=at position #1 with {\arrowreversed[]{Latex[length=1.5mm,width=1.5mm]}}},
    postaction={decorate},node contents=#1},
sb/.default=.4,
snar/.style={dashed,draw,line width =1.25pt},
cross/.style={cross out, draw=black, minimum size=2*(#1-\pgflinewidth), inner sep=0pt, outer sep=0pt}, 
         }
         
\newcommand{\calO}{\mathcal{O}}
\newcommand{\C}{ {\tt C} }
\newcommand{\tL}{ {\tt L} }
\newcommand{\tR}{ {\tt R} }
\newcommand{\hc}{\text{H.c.}}
\newcommand{\GeV}{\rm GeV}
\newcommand{\MeV}{\rm MeV}

\title{Comprehensive investigation on baryon number violating nucleon decays involving an axion-like particle}

\author[a]{Wei-Qi Fan\,\orcidlink{0009-0001-5778-2571},}
\emailAdd{fanweiqi@mail.nankai.edu.cn}
\affiliation[a]{School of Physics, Nankai University, Tianjin 300071, China}
\author[a,b,c]{Yi Liao\,\orcidlink{0000-0002-1009-5483},}
\emailAdd{liaoy@m.scnu.edu.cn}
\affiliation[b]{State Key Laboratory of Nuclear Physics and
Technology, Institute of Quantum Matter, South China Normal
University, Guangzhou 510006, China}
\affiliation[c]{Guangdong Basic Research Center of Excellence for
Structure and Fundamental Interactions of Matter, Guangdong
Provincial Key Laboratory of Nuclear Science, Guangzhou
510006, China}
\author[b,c]{Xiao-Dong Ma\,\orcidlink{0000-0001-7207-7793},}
\emailAdd{maxid@scnu.edu.cn}
\author[b,c]{and Hao-Lin Wang\,\orcidlink{0000-0002-2803-5657}}
\emailAdd{whaolin@m.scnu.edu.cn}

\abstract{
In this study, we systematically investigate baryon number violating (BNV) nucleon decays into an axion-like particle (ALP), within a low energy effective field theory extended with an ALP, which is referred to as aLEFT.
Unlike previous studies in the literature, we consider contributions to nucleon decays from a complete set of  
dimension-eight BNV aLEFT operators involving light $u,\,d,$ and $s$ quarks. 
We perform the chiral irreducible representation (irrep) decomposition of these interactions under the QCD chiral group $\rm SU(3)_{\tt L}\times SU(3)_{\tt R}$, and match them onto the recently developed chiral framework to obtain nucleon-level effective interactions among the ALP, octet baryons, and octet pseudoscalar mesons. 
Within this framework, we derive general expressions for the decay widths of nucleon two- and three-body decays involving an ALP. 
Subsequently, we analyze momentum distributions for three-body modes and find that operators belonging to the newly identified chiral irreps $\pmb{6}_{\tt L(R)}\times \pmb{3}_{\tt R(L)}$ exhibit markedly different behavior compared to that in the usual irreps $\pmb{8}_{\tt L(R)}\times \pmb{1}_{\tt R(L)}$ and $\pmb{3}_{\tt L(R)}\times \bar{\pmb{3}}_{\tt R(L)}$.
In addition, we reanalyze experimental data collected by Super-Kamiokande and establish bounds on the inverse decay widths of these new modes by properly accounting for experimental efficiencies and Cherenkov threshold effects because of the lack of direct constraints on those exotic decay modes. Our recasting constraints are several orders of magnitude more stringent than inclusive bounds used in the literature. Based on these improved bounds, we set conservative limits on associated effective scales across a broad range of ALP mass and predict stringent bounds on certain neutron and hyperon decays involving an ALP.
}


\makeatletter
\gdef\@fpheader{}
\makeatother
\begin{document} 

\maketitle
\setcounter{page}{2}
\section{Introduction}

Baryon number violating (BNV) interactions have played a crucial role in explaining the matter-antimatter asymmetry of the Universe \cite{Sakharov:1967dj}, and have been predicted in various scenarios beyond the standard model (SM). The primary method to test BNV interactions is through searches of nucleon decays caused by their unique experimental signature. 
Over the past several decades, many large-fiducial-mass experiments including IMB~\cite{Irvine-Michigan-Brookhaven:1983iap}, SNO+~\cite{SNO:2018ydj}, KamLAND~\cite{KamLAND:2015pvi}, Kamiokande~\cite{Hirata:1988ad}, and Super-Kamiokande (Super-K)~\cite{Takhistov:2016eqm} have conducted extensive searches for nucleon decays involving only SM particles, which placed very stringent limits on their occurrence.    
In recent years, with the advent of next-generation neutrino experiments such as Hyper-Kamiokande \cite{Hyper-Kamiokande:2018ofw}, DUNE \cite{DUNE:2020ypp}, JUNO \cite{JUNO:2015zny}, and THEIA \cite{Theia:2019non}, BNV nucleon decays have attracted considerable attention, including in particular exotic modes involving new light invisible particles in the final state \cite{Davoudiasl:2014gfa,Helo:2018bgb,Heeck:2020nbq,Fajfer:2020tqf,Liang:2023yta,Fridell:2023tpb,Domingo:2024qoj,Li:2024liy,Li:2025slp,Heeck:2025uwh}.

One such well-motivated light particle is the axion, or more generally, axion-like particles (ALP) 
\footnote{An ALP is a CP-odd pseudo-Nambu-Goldstone boson that arises from spontaneous breaking of a global shift symmetry~\cite{Weinberg:1977ma,Wilczek:1977pj}. A generic pseudoscalar is not connected to such a symmetry.}.
The axion was originally proposed as a solution to the strong CP problem via the Peccei-Quinn mechanism \cite{Peccei:1977hh,Peccei:1977ur,Weinberg:1977ma,Wilczek:1977pj}. By relaxing the coupling-mass relationship specific to the QCD axion, a general class of ALPs has emerged \cite{Bagger:1994hh,Branco:2011iw,Witten:1984dg,Ringwald:2012cu}. 
These particles are not necessarily tied to the strong CP problem, and thus exhibit more flexible masses and couplings. 
Both axions and ALPs have been widely incorporated into scenarios beyond the standard model, where they can make connections to neutrino mass generation \cite{Chikashige:1980ui,Mohapatra:1982tc} and serve as viable dark matter candidates \cite{Preskill:1982cy,Dine:1982ah,Abbott:1982af,Hui:2016ltb,Niemeyer:2019aqm}. Especially, ALP models involving BNV interactions have been pursued recently \cite{Quevillon:2020hmx,Arias-Aragon:2022byr,Brugeat:2024rxe}, offering potential explanations for the long-standing neutron lifetime anomaly \cite{Fornal:2018eol,Liang:2023yta,Bastero-Gil:2024kjo}.
Therefore, combining BNV interactions with an ALP presents a very interesting direction to explore. 

Within the framework of the low energy effective field theory (LEFT) extended by an ALP referred to as aLEFT, Ref.\,\cite{Li:2024liy} studied BNV nucleon and hyperon decays involving an ALP in the final state. 
In the aLEFT with a shift symmetry in the ALP, relevant BNV operators first appear at dimension 8 (dim 8) \cite{Grojean:2023tsd} with a total of 20 operators without counting flavors. Under the QCD chiral symmetry $\rm SU(3)_\tL\otimes SU(3)_\tR$ of the light $u,~d,~s$ quarks, these dim-8 operators can be classified into three irreducible representations (irreps) and their chiral partners: $\pmb{8}_{\tL(\tR)} \otimes \pmb{1}_{\tR(\tL)}$, $\pmb{3}_{\tL(\tR)} \otimes \bar{\pmb{3}}_{\tR(\tL)}$, and $\pmb{3}_{\tL(\tR)} \otimes \pmb{6}_{\tR(\tL)}$.
Only 8 operators in the first two usual irreps are considered in \cite{Li:2024liy}, while the remaining 12 operators associated with the new chiral irreps $\pmb{3}_{\tL(\tR)} \otimes \pmb{6}_{\tR(\tL)}$ are discarded as sub-leading contributions.
Recently, we found that operators in the new chiral irreps can contribute to nucleon decays at the same leading chiral order as the other two \cite{Liao:2025vlj}.
Further, processes that change isospin by $3/2$ units, such as $n\to \pi^+ e^- a$ and $n\to \pi^+ \mu^- a$, can only be mediated by operators belonging to the irreps $\pmb{3}_{\tL(\tR)} \otimes \pmb{6}_{\tR(\tL)}$ because of their unique flavor and Lorentz structures.
Motivated by these considerations, in this work, we systematically investigate nucleon decays involving an ALP based on the complete set of all 20 BNV aLEFT operators.

We start by collecting the relevant dim-8 BNV aLEFT operators involving an ALP, a SM lepton, and three light quark fields, which are invariant under the QCD and QED symmetries $\rm SU(3)_c\otimes U(1)_{\rm em}$. Hadronic matrix elements are calculated using a systematic approach that employs the chiral perturbation theory (ChPT) framework \cite{Weinberg:1978kz,Gasser:1983yg,Gasser:1984gg} by matching the quark-level operators onto their hadronic counterparts.
The leading-order chiral matching for all relevant nucleon decay operators in the LEFT has been systematically established in \cite{Claudson:1981gh,Liao:2025vlj,Liao:2025sqt} by ensuring that the hadron-level operators share the same chiral and Lorentz transformation properties as their quark-level counterparts.
Following these works, we decompose all aLEFT operators into irreps under the chiral group and determine the relevant spurion fields stemmed from aLEFT interactions.
We obtain the desired hadron-level BNV operators by substituting these spurion fields into chiral results in \cite{Liao:2025vlj} and expanding to the appropriate order in pseudoscalar meson fields. 
Together with the standard baryon ChPT interactions \cite{Jenkins:1990jv,Bijnens:1985kj}, we calculate the amplitudes and decay widths for both octet baryon two-body decays and nucleon three-body decays involving an ALP.
To extract more meaningful information from experimental data, we analyze the momentum distributions of the final-state charged leptons and mesons in nucleon three-body decays.

After establishing the theoretical framework for nucleon decays involving an ALP, we examine the experimental constraints on the relevant aLEFT interactions. Given the limited experimental search for these exotic decay modes, we constrain these channels by reanalyzing existing data from proton decay searches conducted by the Super-K experiment \cite{Super-Kamiokande:2015pys,Super-Kamiokande:2016exg,Super-Kamiokande:2024qbv,Super-Kamiokande:2013rwg,Super-Kamiokande:2022egr}. We simulate proton decay processes using analytical decay distributions by accounting for the resolution and efficiency of the detector, and we compare the results with Super-K data to extract lower bounds on the partial lifetime
($\Gamma^{-1}$) for certain decay modes. Subsequently, these lower limits are translated into constraints on the effective scale $\Lambda_{\tt eff}$ associated with the relevant Wilson coefficients (WCs) of aLEFT operators. Finally, we predict new bounds on the occurrence of BNV neutron and hyperon decay modes involving an ALP.
These predictions provide valuable guidance for future experimental searches.

The remainder of this paper is organized as follows. We first introduce the dim-8 BNV aLEFT operators and derive their hadron-level counterparts within the ChPT framework in \cref{sec:EFTframework}. 
In \cref{sec:dewidth}, we formulate the general expressions for the decay widths of nucleon decays involving an ALP and study the momentum distributions of the charged leptons and/or pseudoscalar mesons for the three-body decays.
\cref{sec:constraints} reinterprets existing proton decay data to establish bounds on corresponding modes involving an additional ALP. 
Subsequently, we combine these results with the available inclusive limits to set conservative constraints on the relevant WCs and further explore their implications for other decay modes.
Finally, we summarize our results in \cref{sec:Conclusion}. 
In addition, \cref{app:ChiL} collects the relevant BNV vertices involving a spurion field and a baryon field, while
\cref{app:DW-aLEFT} summarizes complete expressions for decay widths expressed in terms of the aLEFT WCs.

\section{BNV ALP interactions in the EFTs}
\label{sec:EFTframework}

In this section, we present a detailed effective field
theory (EFT) description of the BNV interactions involving an ALP field, from quark-level interactions to matching them onto hadron-level interactions based on the QCD chiral symmetry. We start by collecting the relevant dim-8 local operators within the LEFT framework extended by an ALP (aLEFT). To perform chiral matching, we decompose these operators into irreducible representations of the chiral group
and identify the corresponding spurion fields involved in the matching. 
We incorporate these spurion fields into the formalism established in Ref.\,\cite{Liao:2025vlj} and obtain hadronic counterparts of the BNV aLEFT operators. 

\subsection{BNV ALP interactions in the aLEFT}

To investigate low energy processes, the LEFT is a good starting point. When the LEFT is extended by a light axion-like particle, we refer to it as aLEFT. 
In this work, we focus on two- and three-body BNV nucleon decays, whose final states contain both an ALP and a lepton, plus an additional pseudoscalar meson for three-body modes. 
For such $\Delta B = |\Delta L| = 1$ processes, the required lowest dimensional aLEFT operators must contain three quark fields and one lepton field. 
We assume that the interactions preserve the shift symmetry in the axion field $a$, which results in the ALP field appearing only through its derivative, $\partial_\mu a$.
\footnote{ Usually, a general pseudoscalar is coupled to the SM fermions through a non-derivative form. 
However, when focusing on the leading-order dim-5 terms, 
non-derivative and derivative forms are physically equivalent if and only if the Wilson coefficients of the former obey specific constraints derived from field redefinitions \cite{Bonnefoy:2022rik}. Otherwise, the non-derivative form cannot be transformed into the derivative form and represent genuine symmetry-breaking effects. 
}
Consequently, the relevant leading-order operators appear at dim 8 and take the form, $lqqq\partial a$.

A complete set of these dim-8 operators was recently constructed  in \cite{Grojean:2023tsd}. 
However, to facilitate chiral matching in the subsequent parts, we constructed a slightly modified operator basis. 
The relevant dim-8 aLEFT operators adopted in our analysis are listed in the third column of \cref{tab:BNVaLEFToperator},
where $e_{\tL,\tR}$ represents the SM charged lepton chiral fields, $\nu_\tL$ represent neutrinos, and $u_{\tL,\tR}$ and $d_{\tL,\tR}$ represent up- and down-type quarks, respectively.
In the rightmost column, we indicate the correspondence with the operators given in \cite{Grojean:2023tsd,Li:2024liy}, 
where those 8 operators highlighted by underlines are considered in Ref.\,\cite{Li:2024liy} to study BNV nucleon decays while the remaining 12 operators are neglected. 
When the quark flavor is considered, there are a
total of $68$ operators associated with the $u,d,s$ quarks, 
among which only $22$ are considered in that paper.
Our operator basis exhibits a clear chirality-flip symmetry, where one half of the operators are chirality partners of the other half under the interchange $\tL \leftrightarrow \tR$,
with $\nu_\tL \leftrightarrow \nu_\tL^\C$ applied in the neutrino case. This structure simplifies the chiral matching and serves as a useful cross-check for the final results.
Operators in the $\Delta(B-L)=0$ sector involve a lepton field, whereas those in the $\Delta(B+L)=0$ sector involve a conjugate lepton filed. This difference in the global lepton number implies that they originate from distinct UV-complete scenarios, wherein the generation of operators in one sector can simultaneously forbid those in the other.

\begin{table}[t]
\center
\resizebox{\linewidth}{!}{
\renewcommand{\arraystretch}{1.3}
\begin{tabular}{|c|c|c|c|c|c|c|}
\hline   
& Notation & Operator & Chiral Irrep. & 
\# of operators & Comparison with \cite{Li:2024liy} 
\\\hline
\multirow{12}*{\rotatebox[origin=c]{90}{
\pmb{$\Delta(B-L)=0$} }}
&$ \calO_{\partial a e uud}^{\tt VL,SL} $  
& $\partial_\mu a (\overline{e_\tR^\C}\gamma^\mu u_\tL^\alpha) (\overline{u_\tL^{\beta\C}}d_\tL^\gamma) \epsilon_{\alpha\beta\gamma}$ 
& ${\color{purple}\pmb{8}_\tL \otimes  \pmb{1}_\tR}$ 
& $n_e n_u^2 n_d \,[2n_e]$
& $-$ $\calO^{\tt VL,SL}_{\underline{\partial a eudu}}$
\\
& $ \calO_{\partial a e uud}^{\tt SL,VR} $
& $\partial_\mu a  (\overline{e_\tL^\C} u_\tL^\alpha) (\overline{u_\tL^{\beta \C}}\gamma^\mu  d_\tR^\gamma) \epsilon_{\alpha\beta\gamma}$
& ${\color{Green}\pmb{6}_\tL \otimes \pmb{3}_\tR}$
& $n_e n_u^2 n_d \,[2n_e]$
& $- [\calO^{{\tt VL,SR}}_{\partial a udue}]^\dagger$
\\
& $\calO_{\partial a e udu}^{\tt SL,VR}$
&$\partial_\mu a  (\overline{e_\tL^\C} u_\tL^\alpha) (\overline{d_\tL^{\beta \C}}\gamma^\mu  u_\tR^\gamma) \epsilon_{\alpha\beta\gamma}$
& ${\color{Green}\pmb{6}_\tL \otimes \pmb{3}_\tR} \oplus 
{\color{Purple}\bar{\pmb{3}}_\tL \otimes \pmb{3}_\tR}$
&  $n_e n_u^2 n_d \,[2n_e]$
& $- [\calO^{{\tt VL,SR}}_{\partial a duue}]^\dagger$
\\
& $\calO_{\partial a e duu}^{\tt SL,VR} $
&$ \partial_\mu a (\overline{e_\tL^\C} d_\tL^\alpha) (\overline{u_\tL^{\beta \C}}\gamma^\mu  u_\tR^\gamma) \epsilon_{\alpha\beta\gamma}$
& ${\color{Green}\pmb{6}_\tL \otimes \pmb{3}_\tR} \oplus 
{\color{Purple}\bar{\pmb{3}}_\tL \otimes \pmb{3}_\tR}$
& $n_e n_u^2 n_d \,[2n_e]$
& $- [\calO^{{\tt VL,SR}}_{\partial a uude}]^\dagger$
\\
& $\calO_{\partial a e uud}^{\tt VR,SR}$
&$ \partial_\mu a (\overline{e_\tL^\C} \gamma^\mu u_\tR^\alpha) (\overline{u_\tR^{\beta \C}} d_\tR^\gamma) \epsilon_{\alpha\beta\gamma}$
& ${\color{purple}\pmb{1}_\tL \otimes  \pmb{8}_\tR}$ 
& $n_e n_u^2 n_d \,[2n_e]$
& $\calO^{\tt VR,SR}_{\underline{\partial a euud}}$
\\
& $\calO_{\partial a e uud}^{\tt SR,VL}$
& $\partial_\mu a  (\overline{e_\tR^\C} u_\tR^\alpha) (\overline{u_\tR^{\beta \C}} \gamma^\mu  d_\tL^\gamma) \epsilon_{\alpha\beta\gamma}$
& ${\color{Green}\pmb{3}_\tL \otimes \pmb{6}_\tR}$
& $n_e n_u^2 n_d \,[2n_e]$
& $\calO^{\tt VR,SR}_{\partial a duue}$ 
\\
& $ \calO_{\partial a e udu}^{\tt SR,VL}$ 
&$ \partial_\mu a (\overline{e_\tR^\C} u_\tR^\alpha) (\overline{d_\tR^{\beta \C}} \gamma^\mu  u_\tL^\gamma) \epsilon_{\alpha\beta\gamma}$
& ${\color{Green}\pmb{3}_\tL \otimes \pmb{6}_\tR} \oplus 
{\color{Purple} \pmb{3}_\tL \otimes \bar{\pmb{3}}_\tR}$
& $n_e n_u^2 n_d \,[2n_e]$
& $\calO^{\tt VL,SR}_{\partial a dueu}$
\\
& $ \calO_{\partial a e duu}^{\tt SR,VL} $
& $ \partial_\mu a  (\overline{e_\tR^\C} d_\tR^\alpha) (\overline{u_\tR^{\beta \C}} \gamma^\mu  u_\tL^\gamma) \epsilon_{\alpha\beta\gamma}$
& ${\color{Green}\pmb{3}_\tL \otimes \pmb{6}_\tR} \oplus 
{\color{Purple} \pmb{3}_\tL \otimes \bar{\pmb{3}}_\tR}$
& $n_e n_u^2 n_d \,[2n_e]$
&$ \calO^{\tt VL,SR}_{\partial a dueu} - \calO^{\tt VL,SR}_{\underline{\partial a eudu}}$
\\\cline{2-6}
&$ \calO_{\partial a \nu ddu}^{\tt VR,SR}$
&$\partial_\mu a  (\overline{\nu_\tL^\C} \gamma^\mu d_\tR^\alpha) (\overline{d_\tR^{\beta \C}} u_\tR^\gamma) \epsilon_{\alpha\beta\gamma}$
& ${\color{purple}\pmb{1}_\tL \otimes  \pmb{8}_\tR}$ 
& $n_\nu n_u n_d^2 \,[4n_\nu]$
&$-$ $\calO^{\tt VL,SR}_{\underline{\partial a d\nu du}}$
\\
& $\calO_{\partial a \nu ddu}^{\tt SL,VR} $
&$ \partial_\mu a  (\overline{\nu_\tL^\C} d_\tL^\alpha) (\overline{d_\tL^{\beta \C}} \gamma^\mu  u_\tR^\gamma) \epsilon_{\alpha\beta\gamma}$
& ${\color{Green}\pmb{6}_\tL \otimes \pmb{3}_\tR} \oplus 
{\color{Purple}\bar{\pmb{3}}_\tL \otimes \pmb{3}_\tR}$
& $n_\nu n_u n_d^2 \,[4n_\nu]$
& $- [\calO^{{\tt VL,SR}}_{\partial a dud\nu}]^\dagger$
\\
& $\calO_{\partial a \nu dud}^{\tt SL,VR} $
& $\partial_\mu a  (\overline{\nu_\tL^\C} d_\tL^\alpha) (\overline{u_\tL^{\beta \C}} \gamma^\mu  d_\tR^\gamma) \epsilon_{\alpha\beta\gamma}$
& ${\color{Green}\pmb{6}_\tL \otimes \pmb{3}_\tR} \oplus 
{\color{Purple}\bar{\pmb{3}}_\tL \otimes \pmb{3}_\tR}$
& $n_\nu n_u n_d^2 \,[4n_\nu]$
& $-$ ( $ \calO^{\tt VL,SL}_{\underline{\partial a d\nu ud}} $ $ +[\calO^{{\tt VL,SR}}_{\partial a dd\nu u}]^\dagger$ )
\\
& $ \calO_{\partial a \nu udd}^{\tt SL,VR} $
&$ \partial_\mu a  (\overline{\nu_\tL^\C} u_\tL^\alpha) (\overline{d_\tL^{\beta \C}} \gamma^\mu  d_\tR^\gamma) \epsilon_{\alpha\beta\gamma}$
& ${\color{Green}\pmb{6}_\tL \otimes \pmb{3}_\tR} \oplus 
{\color{Purple}\bar{\pmb{3}}_\tL \otimes \pmb{3}_\tR}$
& $n_\nu n_u n_d^2 \,[4n_\nu]$
& $- [\calO^{{\tt VL,SR}}_{\partial a dd\nu u}]^\dagger$
\\\hline
\cellcolor{gray!15} & $\calO_{\partial a \nu ddu}^{\tt VL,SL} $ 
& $ \partial_\mu a  (\overline{\nu_\tL} \gamma^\mu d_\tL^\alpha) (\overline{d_\tL^{\beta \C}} u_\tL^\gamma) \epsilon_{\alpha\beta\gamma}$
& ${\color{purple}\pmb{8}_\tL \otimes  \pmb{1}_\tR}$
& $n_\nu n_u n_d^2 \,[4n_\nu]$
&$\calO^{\tt VL,SL}_{\underline{\partial a \nu ddu}}$
\\
\cellcolor{gray!15}& $\calO_{\partial a \nu ddu}^{\tt SR,VL} $
 &$\partial_\mu a  (\overline{\nu_\tL} d_\tR^\alpha) (\overline{d_\tR^{\beta \C}} \gamma^\mu  u_\tL^\gamma) \epsilon_{\alpha\beta\gamma}$
& ${\color{Green}\pmb{3}_\tL \otimes \pmb{6}_\tR} \oplus 
{\color{Purple} \pmb{3}_\tL \otimes \bar{\pmb{3}}_\tR}$
& $n_\nu n_u n_d^2 \,[4n_\nu]$
& $\calO^{\tt VR,SR}_{\partial a ud\nu d} $ 
\\
\cellcolor{gray!15} &$\calO_{\partial a \nu dud}^{\tt SR,VL} $
&$\partial_\mu a  (\overline{\nu_\tL} d_\tR^\alpha) (\overline{u_\tR^{\beta \C}} \gamma^\mu  d_\tL^\gamma) \epsilon_{\alpha\beta\gamma}$
& ${\color{Green}\pmb{3}_\tL \otimes \pmb{6}_\tR} \oplus 
{\color{Purple} \pmb{3}_\tL \otimes \bar{\pmb{3}}_\tR}$
& $n_\nu n_u n_d^2 \,[4n_\nu]$
& $\calO^{\tt VR,SR}_{\partial a du\nu d}$ 
\\
\cellcolor{gray!15} &$\calO_{\partial a \nu udd}^{\tt SR,VL} $
& $\partial_\mu a  (\overline{\nu_\tL} u_\tR^\alpha) (\overline{d_\tR^{\beta \C}} \gamma^\mu  d_\tL^\gamma) \epsilon_{\alpha\beta\gamma}$
& ${\color{Green}\pmb{3}_\tL \otimes \pmb{6}_\tR} \oplus 
{\color{Purple} \pmb{3}_\tL \otimes \bar{\pmb{3}}_\tR}$
& $n_\nu n_u n_d^2 \,[4n_\nu]$
& $\calO^{\tt VR,SR}_{\partial a dd\nu u}$
\\\cline{2-6}
\cellcolor{gray!15} & $\calO_{\partial a e ddd}^{\tt VR,SR} $
& $\partial_\mu a (\overline{e_\tR} \gamma^\mu d_\tR^\alpha) (\overline{d_\tR^{\beta \C}} d_\tR^\gamma) \epsilon_{\alpha\beta\gamma}$
& ${\color{purple}\pmb{1}_\tL \otimes  \pmb{8}_\tR}$ 
& ${1\over 3} n_e n_d(n_d^2-1) \,[2n_e]$
&$\calO^{\tt VR,SR}_{\underline{\partial a eddd}}$
\\
\cellcolor{gray!15} &$ \calO_{\partial a e ddd}^{\tt SR,VL} $
&$ \partial_\mu a (\overline{e_\tL} d_\tR^\alpha) (\overline{d_\tR^{\beta \C}} \gamma^\mu  d_\tL^\gamma) \epsilon_{\alpha\beta\gamma}$
& ${\color{Green}\pmb{3}_\tL \otimes \pmb{6}_\tR} \oplus 
{\color{Purple} \pmb{3}_\tL \otimes \bar{\pmb{3}}_\tR}$
& $n_e n_d^3 \,[8n_e]$
& $\tilde{\calO}^{\tt VR,SR}_{\partial a eddd}$
\\
\cellcolor{gray!15} & $\calO_{\partial a e ddd}^{\tt VL,SL}$
&$\partial_\mu a (\overline{e_\tL} \gamma^\mu d_\tL^\alpha) (\overline{d_\tL^{\beta \C}} d_\tL^\gamma) \epsilon_{\alpha\beta\gamma}$
& ${\color{purple}\pmb{8}_\tL \otimes  \pmb{1}_\tR}$ 
&  ${1\over 3} n_e n_d(n_d^2-1)\,[2n_e]$
&$\calO^{\tt VL,SL}_{\underline{\partial a eddd}}$ 
\\
\multirow{-8}*{\rotatebox[origin=c]{90}{
\cellcolor{gray!15}\pmb{$\Delta(B+L)=0$} } } & $ \calO_{\partial a e ddd}^{\tt SL,VR} $ 
&$\partial_\mu a (\overline{e_\tR} d_\tL^\alpha) (\overline{d_\tL^{\beta \C}} \gamma^\mu  d_\tR^\gamma) \epsilon_{\alpha\beta\gamma}$
& ${\color{Green}\pmb{6}_\tL \otimes \pmb{3}_\tR} \oplus 
{\color{Purple}\bar{\pmb{3}}_\tL \otimes \pmb{3}_\tR}$
&  $n_e n_d^3 \,[8n_e]$
& $- [\calO^{{\tt VL,SR}}_{\partial a ddde}]^\dagger$
\\\hline
\end{tabular}
}
\caption{Dim-8 aLEFT BNV operators involving an ALP. $\alpha,\beta,\gamma$ are color indices while flavor indices are omitted for simplicity.  
The chiral irrep. column lists irreducible representations under the chiral group $\rm SU(3)_\tL\otimes SU(3)_\tR$ for the three light quarks $u,d,s$.
The \# column counts the number of independent operators for general $n_{e}$ charged leptons, $n_\nu$ neutrinos,
and $n_u$ up-type and $n_d$ down-type quarks; the number in the square bracket represents the case with only $u,d,s$ quarks ($n_u=1$ and $n_d=2$). 
There are $68 n_{e,\nu}$ operators associated with the $u,d,s$ quarks, 
among which only $22 n_{e,\nu}$ were considered in \cite{Li:2024liy} which are underlined in the last column.
}
\label{tab:BNVaLEFToperator}
\end{table}

To match the aLEFT operators onto those in ChPT,
one should first decompose them into irreps of the QCD chiral group $\rm SU(3)_\tL\otimes SU(3)_\tR$ for the $u,d,s$ quarks in the massless limit. 
As indicated in \cref{tab:BNVaLEFToperator}, these operators are divided into the following two structures and their chirality partners
\begin{align}
\partial_\mu a (\overline{\psi_x} \gamma^\mu q_{\tL, y}^\alpha) (\overline{ q_{\tL, z}^{\beta \C} }  q_{\tL, w}^\gamma )\epsilon_{\alpha \beta \gamma}, 
\qquad
\partial_\mu a (\overline{\psi_x} q_{\tL,y}^\alpha) (\overline{ q_{\tL, z}^{\beta \C} } \gamma^\mu q_{\tR, w}^\gamma ) \epsilon_{\alpha \beta \gamma},
\label{eq:opeclass}
\end{align}
where $\psi( = e, e^\C, \nu, \nu^\C)$ represents a lepton field, with $x$ denoting its flavor, and $y,z,w = 1,2,3 $ indicating the light quark flavor indices with $q_{1,2,3} = u, d, s$. 
The triple-quark sector in the first structure, ${\cal N}_{yzw}^{\tL\tL}\equiv q_{\tL, y}^\alpha (\overline{ q_{\tL, z}^{\beta \C} }  q_{\tL, w}^\gamma )\epsilon_{\alpha \beta \gamma}$, already belongs to the $\pmb{8}_{\tL} \otimes \pmb{1}_{\tR}$ irrep  of the chiral group. However, the triple-quark component in the second structure generally does not form an irrep. 
As recognized in Ref.\,\cite{Liao:2025vlj}, the flavor symmetric and anti-symmetric combinations of the two like-chirality quark fields form irreps $\pmb{6}_{\tL} \otimes \pmb{3}_{\tR}$ and $\bar{\pmb{3}}_{\tL} \otimes \pmb{3}_{\tR}$, respectively. 
The antisymmetric combination can be Fierz-transformed into a similar form to ${\cal N}_{yzw}^{\tL\tL}$
with $q_{\tL,y}$ being replaced by $q_{\tR,y}$.
Following Ref.\,\cite{Liao:2025vlj}, we parametrize them as
\begin{align}
{\cal N}_{yzw}^{\tR\tL} \equiv q_{\tR,y}^\alpha (\overline{ q_{\tL, z}^{\beta \C} } q_{\tL, w}^\gamma ) \epsilon_{\alpha \beta \gamma}
\in \bar{\pmb{3}}_{\tL} \otimes \pmb{3}_{\tR}, \quad 
{\cal N}_{yzw}^{\tL\tR, \mu}   \equiv q_{\tL,\{y}^\alpha (\overline{ q_{\tL, z\}}^{\beta \C} } \gamma^\mu q_{\tR, w}^\gamma ) \epsilon_{\alpha \beta \gamma}
 \in \pmb{6}_{\tL} \otimes \pmb{3}_{\tR}.
\end{align}
Then, the second structure in \cref{eq:opeclass} can be converted into these irreps using the Fierz identity,
\begin{align}
 (\overline{\psi_x} q_{\tL,[y}^\alpha) (\overline{ q_{\tL, z]}^{\beta \C} } \gamma^\mu q_{\tR, w}^\gamma ) \epsilon_{\alpha \beta \gamma}  
 = {1 \over 2}(\overline{\psi_x}\gamma^\mu q_{\tR,w}^\alpha) (\overline{ q_{\tL, z}^{\beta \C} } q_{\tL, y}^\gamma ) \epsilon_{\alpha \beta \gamma},
\end{align}
which leads to 
\begin{align}
(\overline{\psi_x} q_{\tL,y}^\alpha) (\overline{ q_{\tL, z}^{\beta \C} } \gamma^\mu q_{\tR, w}^\gamma ) \epsilon_{\alpha \beta \gamma} 
=
\overline{\psi_x} \, {\cal N}_{yzw}^{\tL\tR, \mu}
 + \frac{1}{2} \overline{\psi_x}\gamma^\mu 
{\cal N}_{wzy}^{\tR\tL}.
\end{align}
The curly and square brackets denote  symmetrization and antisymmetrization over two flavor indices, $A_{\{y}B_{z\}} \equiv (1/2)(A_{y} B_{z} + A_{z}B_{y})$ and $A_{[y}B_{z]} \equiv (1/2)(A_{y} B_{z} - A_{z}B_{y})$, respectively.  
Similarly, chirality-flipped operators with $\tL\leftrightarrow\tR$ can be decomposed into $\pmb{3}_{\tL} \otimes \pmb{6}_{\tR}$ and $\pmb{3}_{\tL} \otimes \bar{\pmb{3}}_{\tR}$.
For example, 
\begin{subequations}
\begin{align}
[\calO_{\partial a euud}^{\tt SL,VR}]_{x11y}
= & \partial_\mu a \, \overline{e_{\tL,x}^\C} 
{\cal N}^{\tL\tR,\mu}_{11y},
\\
[\calO_{\partial a eudu}^{\tt SL,VR}]_{x1y1}
= & \partial_\mu a \, \overline{e_{\tL,x}^\C} 
{\cal N}^{\tL\tR,\mu}_{1y1}
+\frac{1}{2} \partial_\mu a \, \overline{e_{\tL,x}^\C} \gamma^\mu
{\cal N}^{\tR\tL}_{1y1},
\\
[\calO_{\partial a \nu ddu}^{\tt SL,VR}]_{xyz1}
= & \partial_\mu a \, \overline{\nu_{\tL,x}^\C} 
{\cal N}^{\tL\tR,\mu}_{yz1}
+\frac{1}{2} \partial_\mu a \, \overline{\nu_{\tL,x}^\C} \gamma^\mu
{\cal N}^{\tR\tL}_{1zy},
\label{eq:decom}
\\
[\calO_{\partial a eddd}^{\tt SL,VR}]_{xyzw}
= & \partial_\mu a \, \overline{e_{\tR,x}} 
{\cal N}^{\tL\tR,\mu}_{yzw}
+\frac{1}{2} \partial_\mu a \, \overline{e_{\tR,x}} \gamma^\mu
{\cal N}^{\tR\tL}_{wzy}.
\end{align}
\end{subequations} 
Based on the arguments outlined above, we indicate the chiral irreps of each operator under consideration in the fourth column of \cref{tab:BNVaLEFToperator}. 
Such a chiral decomposition was neglected in \cite{Li:2024liy}, which results in many operators belonging to $\pmb{3}_{\tL(\tR)} \otimes \bar{\pmb{3}}_{\tR(\tL)}$ being also omitted from their analysis.

\begin{table}[htbp]
\center
\resizebox{\linewidth}{!}{
\renewcommand{\arraystretch}{1.08}
\begin{tabular}{|c|c|c|c|c|}
\hline
~Irrep.~ & ~~~Spurion~~~ &  Expression   & ~~~Spurion~~~ &  Expression  
\\\hline
\multirow{8}*{\rotatebox[origin=c]{90}{
\color{purple}$\pmb{8}_{\tL(\tR)}\otimes \pmb{1}_{\tR(\tL)}$}}
& ${\cal P}_{dds}^{\tL\tL}$ 
& \cellcolor{gray!15}$ [C^{\tt VL,SL}_{\partial a eddd}]_{x223} 
(\partial_\mu a) \overline{e_{\tL,x}} \gamma^\mu$
& ${\cal P}_{dds}^{\tR\tR}$ 
& \cellcolor{gray!15}$ [C^{\tt VR,SR}_{\partial a eddd}]_{x223} 
(\partial_\mu a) \overline{e_{\tR,x}} \gamma^\mu$
\\\hhline{~----}
& ${\cal P}_{sds}^{\tL\tL}$ 
& \cellcolor{gray!15}$ [C^{\tt VL,SL}_{\partial a eddd}]_{x323} 
(\partial_\mu a) \overline{e_{\tL,x}} \gamma^\mu$
& ${\cal P}_{sds}^{\tR\tR}$ 
& \cellcolor{gray!15}$ [C^{\tt VR,SR}_{\partial a eddd}]_{x323} 
(\partial_\mu a) \overline{e_{\tR,x}} \gamma^\mu$
\\\hhline{~----}
& ${\cal P}_{usu}^{\tL\tL}$ 
& ~~~$ - [C^{\tt VL,SL}_{\partial a euud}]_{x113} 
(\partial_\mu a) \overline{e_{\tR,x}^\C} \gamma^\mu$~~~
& ${\cal P}_{usu}^{\tR\tR}$ 
&~~~ $ - [C^{\tt VR,SR}_{\partial a  euud}]_{x113} 
(\partial_\mu a) \overline{e_{\tL,x}^\C} \gamma^\mu$~~~
\\\hhline{~----}
& ${\cal P}_{dsu}^{\tL\tL}$ 
& \cellcolor{gray!15}$ [C^{\tt VL,SL}_{\partial a \nu ddu}]_{x231} 
(\partial_\mu a) \overline{\nu_{\tL,x}} \gamma^\mu$
& ${\cal P}_{dsu}^{\tR\tR}$ 
& $ [C^{\tt VR,SR}_{\partial a  \nu ddu}]_{x231} 
(\partial_\mu a) \overline{\nu_{\tL,x}^\C} \gamma^\mu$
\\\hhline{~----}
& ${\cal P}_{ssu}^{\tL\tL}$ 
& \cellcolor{gray!15}$ [C^{\tt VL,SL}_{\partial a \nu ddu}]_{x331} 
(\partial_\mu a) \overline{\nu_{\tL,x}} \gamma^\mu$
& ${\cal P}_{ssu}^{\tR\tR}$ 
& $ [C^{\tt VR,SR}_{\partial a  \nu ddu}]_{x331} 
(\partial_\mu a) \overline{\nu_{\tL,x}^\C} \gamma^\mu$
\\\hhline{~----}
& ${\cal P}_{uud}^{\tL\tL}$ 
&  $ [C^{\tt VL,SL}_{\partial a euud}]_{x112} 
(\partial_\mu a) \overline{e_{\tR,x}^\C} \gamma^\mu$
& ${\cal P}_{uud}^{\tR\tR}$ 
& $ [C^{\tt VR,SR}_{\partial a  euud}]_{x112} 
(\partial_\mu a) \overline{e_{\tL,x}^\C} \gamma^\mu$
\\\hhline{~----}
& ${\cal P}_{dud}^{\tL\tL}$ 
& \cellcolor{gray!15}$ - [C^{\tt VL,SL}_{\partial a \nu ddu}]_{x221} 
(\partial_\mu a) \overline{\nu_{\tL,x}} \gamma^\mu$
& ${\cal P}_{dud}^{\tR\tR}$ 
& $ - [C^{\tt VR,SR}_{\partial a  \nu ddu}]_{x221} 
(\partial_\mu a) \overline{\nu_{\tL,x}^\C} \gamma^\mu$
\\\hhline{~----}
& ${\cal P}_{sud}^{\tL\tL}$ 
& \cellcolor{gray!15}$ - [C^{\tt VL,SL}_{\partial a \nu ddu}]_{x321} 
(\partial_\mu a) \overline{\nu_{\tL,x}} \gamma^\mu$
& ${\cal P}_{sud}^{\tR\tR}$ 
& $ - [C^{\tt VR,SR}_{\partial a  \nu ddu}]_{x321} 
(\partial_\mu a) \overline{\nu_{\tL,x}^\C} \gamma^\mu$
\\\hline
\multirow{9}*{\rotatebox[origin=c]{90}{
\color{Purple}$\pmb{3}_{\tL(\tR)}\otimes \bar{\pmb{3}}_{\tR(\tL)}$}} 
& ${\cal P}_{uds}^{\tR\tL}$ 
& $-[C^{\tt SL,VR}_{\partial a  \nu ddu}]_{x231}^- 
(\partial_\mu a) \overline{\nu_{\tL,x}^\C} \gamma^\mu$
& ${\cal P}_{uds}^{\tL\tR}$ 
& \cellcolor{gray!15}$ - [C^{\tt SR,VL}_{\partial a \nu ddu}]_{x231}^- 
(\partial_\mu a) \overline{\nu_{\tL,x}} \gamma^\mu$
\\\hhline{~----}
& ${\cal P}_{dds}^{\tR\tL}$ 
& \cellcolor{gray!15}$- [C^{\tt SL,VR}_{\partial a eddd}]_{x232}^-
(\partial_\mu a) \overline{e_{\tR,x}} \gamma^\mu$
& ${\cal P}_{dds}^{\tL\tR}$ 
& \cellcolor{gray!15}$ - [C^{\tt SR,VL}_{\partial a e ddd}]_{x232}^- 
(\partial_\mu a) \overline{e_{\tL,x}} \gamma^\mu$
\\\hhline{~----}
& ${\cal P}_{sds}^{\tR\tL}$ 
& \cellcolor{gray!15}$ - [C^{\tt SL,VR}_{\partial a eddd}]_{x233}^- 
(\partial_\mu a) \overline{e_{\tR,x}} \gamma^\mu$
& ${\cal P}_{sds}^{\tL\tR}$ 
& \cellcolor{gray!15}$- [C^{\tt SR,VL}_{\partial a e ddd}]_{x233}^-
(\partial_\mu a) \overline{e_{\tL,x}} \gamma^\mu$
\\\hhline{~----}
& ${\cal P}_{usu}^{\tR\tL}$ 
& $[C^{\tt SL,VR}_{\partial a  eudu}]_{x131}^- 
(\partial_\mu a) \overline{e_{\tL,x}^\C} \gamma^\mu$
& ${\cal P}_{usu}^{\tL\tR}$ 
& $[C^{\tt SR,VL}_{\partial a  eudu}]_{x131}^- 
(\partial_\mu a) \overline{e_{\tR,x}^\C} \gamma^\mu$
\\\hhline{~----}
& ${\cal P}_{dsu}^{\tR\tL}$ 
& $[C^{\tt SL,VR}_{\partial a  \nu udd}]_{x132}^-
(\partial_\mu a) \overline{\nu_{\tL,x}^\C} \gamma^\mu$
& ${\cal P}_{dsu}^{\tL\tR}$ 
& \cellcolor{gray!15}$[C^{\tt SR,VL}_{\partial a \nu udd}]_{x132}^-
(\partial_\mu a) \overline{\nu_{\tL,x}} \gamma^\mu$
\\\hhline{~----}
& ${\cal P}_{ssu}^{\tR\tL}$ 
& $[C^{\tt SL,VR}_{\partial a  \nu udd}]_{x133}^- 
(\partial_\mu a) \overline{\nu_{\tL,x}^\C} \gamma^\mu$
& ${\cal P}_{ssu}^{\tL\tR}$ 
& \cellcolor{gray!15}$[C^{\tt SR,VL}_{\partial a \nu udd}]_{x133}^-
(\partial_\mu a) \overline{\nu_{\tL,x}} \gamma^\mu$
\\\hhline{~----}
& ${\cal P}_{uud}^{\tR\tL}$ 
& $- [C^{\tt SL,VR}_{\partial a  eudu}]_{x121}^-
(\partial_\mu a) \overline{e_{\tL,x}^\C} \gamma^\mu$
& ${\cal P}_{uud}^{\tL\tR}$ 
& $- [C^{\tt SR,VL}_{\partial a  eudu}]_{x121}^-
(\partial_\mu a) \overline{e_{\tR,x}^\C} \gamma^\mu$
\\\hhline{~----}
& ${\cal P}_{dud}^{\tR\tL}$ 
& $- [C^{\tt SL,VR}_{\partial a  \nu udd}]_{x122}^- 
(\partial_\mu a) \overline{\nu_{\tL,x}^\C} \gamma^\mu$
& ${\cal P}_{dud}^{\tL\tR}$ 
& \cellcolor{gray!15}$- [C^{\tt SR,VL}_{\partial a\nu udd}]_{x122}^-
(\partial_\mu a) \overline{\nu_{\tL,x}} \gamma^\mu$
\\\hhline{~----}
& ${\cal P}_{sud}^{\tR\tL}$ 
& $- [C^{\tt SL,VR}_{\partial a  \nu udd}]_{x123}^-
(\partial_\mu a) \overline{\nu_{\tL,x}^\C} \gamma^\mu$
& ${\cal P}_{sud}^{\tL\tR}$ 
& \cellcolor{gray!15}$ - [C^{\tt SR,VL}_{\partial a\nu udd}]_{x123}^-
(\partial_\mu a) \overline{\nu_{\tL,x}} \gamma^\mu$
\\\hline
\multirow{18}*{\rotatebox[origin=c]{90}{
\color{Green}$\pmb{6}_{\tL(\tR)}\otimes \pmb{3}_{\tR(\tL)}$}} 
& ${\cal P}_{uuu}^{\tL\tR,\mu}$ 
&  ---
& ${\cal P}_{uuu}^{\tR\tL,\mu}$ 
&  ---
\\\hhline{~----}
& ${\cal P}_{uud}^{\tL\tR,\mu}$ 
& $ [C^{\tt SL,VR}_{\partial a  euud}]_{x112} 
(\partial^\mu a) \overline{e_{\tL,x}^\C}$
& ${\cal P}_{uud}^{\tR\tL,\mu}$ 
& $ [C^{\tt SR,VL}_{\partial a  euud}]_{x112} 
(\partial^\mu a) \overline{e_{\tR,x}^\C}$ 
\\\hhline{~----}
& ${\cal P}_{uus}^{\tL\tR,\mu}$ 
& $ [C^{\tt SL,VR}_{\partial a  euud}]_{x113} 
(\partial^\mu a) \overline{e_{\tL,x}^\C}$
& ${\cal P}_{uus}^{\tR\tL,\mu}$ 
& $ [C^{\tt SR,VL}_{\partial a  euud}]_{x113} 
(\partial^\mu a) \overline{e_{\tR,x}^\C}$
\\\hhline{~----}%
& ${\cal P}_{udu}^{\tL\tR,\mu}$ 
& $ [C^{\tt SL,VR}_{\partial a  eudu}]_{x121}^+
(\partial^\mu a) \overline{e_{\tL,x}^\C}$
& ${\cal P}_{udu}^{\tR\tL,\mu}$ 
& $[C^{\tt SR,VL}_{\partial a  eudu}]_{x121}^+
(\partial^\mu a) \overline{e_{\tR,x}^\C}$
\\\hhline{~----}
& ${\cal P}_{udd}^{\tL\tR,\mu}$ 
& $[C^{\tt SL,VR}_{\partial a  \nu udd}]_{x122}^+
(\partial^\mu a) \overline{\nu_{\tL,x}^\C}$
& ${\cal P}_{udd}^{\tR\tL,\mu}$ 
& \cellcolor{gray!15}$[C^{\tt SR,VL}_{\partial a \nu udd}]_{x122}^+
(\partial^\mu a) \overline{\nu_{\tL,x}}$
\\\hhline{~----}
& ${\cal P}_{uds}^{\tL\tR,\mu}$ 
& $[C^{\tt SL,VR}_{\partial a  \nu udd}]_{x123}^+
(\partial^\mu a) \overline{\nu_{\tL,x}^\C}$
& ${\cal P}_{uds}^{\tR\tL,\mu}$ 
& \cellcolor{gray!15}$[C^{\tt SR,VL}_{\partial a \nu udd}]_{x123}^+
(\partial^\mu a) \overline{\nu_{\tL,x}}$
\\\hhline{~----}%
& ${\cal P}_{usu}^{\tL\tR,\mu}$ 
& $[C^{\tt SL,VR}_{\partial a  eudu}]_{x131}^+ 
(\partial^\mu a) \overline{e_{\tL,x}^\C}$
& ${\cal P}_{usu}^{\tR\tL,\mu}$ 
& $[C^{\tt SR,VL}_{\partial a  eudu}]_{x131}^+
(\partial^\mu a) \overline{e_{\tR,x}^\C}$
\\\hhline{~----}
& ${\cal P}_{usd}^{\tL\tR,\mu}$ 
& $[C^{\tt SL,VR}_{\partial a  \nu udd}]_{x132}^+
(\partial^\mu a) \overline{\nu_{\tL,x}^\C}$
& ${\cal P}_{usd}^{\tR\tL,\mu}$ 
& \cellcolor{gray!15}$[C^{\tt SR,VL}_{\partial a \nu udd}]_{x132}^+
(\partial^\mu a) \overline{\nu_{\tL,x}}$
\\\hhline{~----}
& ${\cal P}_{uss}^{\tL\tR,\mu}$ 
& $[C^{\tt SL,VR}_{\partial a  \nu udd}]_{x133}^+ 
(\partial^\mu a) \overline{\nu_{\tL,x}^\C}$
& ${\cal P}_{uss}^{\tR\tL,\mu}$ 
&\cellcolor{gray!15}$[C^{\tt SR,VL}_{\partial a \nu udd}]_{x133}^+ 
(\partial^\mu a) \overline{\nu_{\tL,x}}$
\\\hhline{~----}%
& ${\cal P}_{ddu}^{\tL\tR,\mu}$ 
& $ [C^{\tt SL,VR}_{\partial a \nu ddu}]_{x221} 
(\partial^\mu a) \overline{\nu_{\tL,x}^\C}$
& ${\cal P}_{ddu}^{\tR\tL,\mu}$ 
& \cellcolor{gray!15}$[C^{\tt SR,VL}_{\partial a \nu ddu}]_{x221} 
(\partial^\mu a) \overline{\nu_{\tL,x}}$
\\\hhline{~----}
& ${\cal P}_{ddd}^{\tL\tR,\mu}$ 
& \cellcolor{gray!15}$ [C^{\tt SL,VR}_{\partial a eddd}]_{x222} 
(\partial^\mu a) \overline{e_{\tR,x}}$
& ${\cal P}_{ddd}^{\tR\tL,\mu}$ 
& \cellcolor{gray!15}$ [C^{\tt SR,VL}_{\partial a e ddd}]_{x222} 
(\partial^\mu a) \overline{e_{\tL,x}}$
\\\hhline{~----}
& ${\cal P}_{dds}^{\tL\tR,\mu}$ 
& \cellcolor{gray!15}$ [C^{\tt SL,VR}_{\partial a eddd}]_{x223} 
(\partial^\mu a) \overline{e_{\tR,x}}$
& ${\cal P}_{dds}^{\tR\tL,\mu}$ 
& \cellcolor{gray!15}$ [C^{\tt SR,VL}_{\partial a e ddd}]_{x223} 
(\partial^\mu a) \overline{e_{\tL,x}}$
\\\hhline{~----}%
& ${\cal P}_{dsu}^{\tL\tR,\mu}$ 
& $[C^{\tt SL,VR}_{\partial a  \nu ddu}]_{x231}^+
(\partial^\mu a) \overline{\nu_{\tL,x}^\C}$
& ${\cal P}_{dsu}^{\tR\tL,\mu}$ 
& \cellcolor{gray!15}$[C^{\tt SR,VL}_{\partial a\nu ddu}]_{x231}^+ 
(\partial^\mu a) \overline{\nu_{\tL,x}}$
\\\hhline{~----}
& ${\cal P}_{dsd}^{\tL\tR,\mu}$ 
& \cellcolor{gray!15}$[C^{\tt SL,VR}_{\partial a eddd}]_{x232}^+
(\partial^\mu a) \overline{e_{\tR,x}}$
& ${\cal P}_{dsd}^{\tR\tL,\mu}$ 
& \cellcolor{gray!15}$[C^{\tt SR,VL}_{\partial a e ddd}]_{x232}^+
(\partial^\mu a) \overline{e_{\tL,x}}$
\\\hhline{~----}
& ${\cal P}_{dss}^{\tL\tR,\mu}$ 
& \cellcolor{gray!15}$[C^{\tt SL,VR}_{\partial a eddd}]_{x233}^+
(\partial^\mu a) \overline{e_{\tR,x}}$
& ${\cal P}_{dss}^{\tR\tL,\mu}$ 
& \cellcolor{gray!15}$[C^{\tt SR,VL}_{\partial a e ddd}]_{x233}^+
(\partial^\mu a) \overline{e_{\tL,x}}$
\\\hhline{~----}%
& ${\cal P}_{ssu}^{\tL\tR,\mu}$ 
& $ [C^{\tt SL,VR}_{\partial a  \nu ddu}]_{x331} 
(\partial^\mu a) \overline{\nu_{\tL,x}^\C}$
& ${\cal P}_{ssu}^{\tR\tL,\mu}$ 
& \cellcolor{gray!15}$ [C^{\tt SR,VL}_{\partial a\nu ddu}]_{x331}
(\partial^\mu a) \overline{\nu_{\tL,x}}$
\\\hhline{~----}
& ${\cal P}_{ssd}^{\tL\tR,\mu}$ 
& \cellcolor{gray!15}$ [C^{\tt SL,VR}_{\partial a eddd}]_{x332} 
(\partial^\mu a) \overline{e_{\tR,x}}$
& ${\cal P}_{ssd}^{\tR\tL,\mu}$ 
& \cellcolor{gray!15}$ [C^{\tt SR,VL}_{\partial a e ddd}]_{x332} 
(\partial^\mu a) \overline{e_{\tL,x}}$
\\\hhline{~----}
& ${\cal P}_{sss}^{\tL\tR,\mu}$ 
& \cellcolor{gray!15}$ [C^{\tt SL,VR}_{\partial a eddd}]_{x333} 
(\partial^\mu a) \overline{e_{\tR,x}}$
& ${\cal P}_{sss}^{\tR\tL,\mu}$ 
& \cellcolor{gray!15}$ [C^{\tt SR,VL}_{\partial a e ddd}]_{x333} 
(\partial^\mu a) \overline{e_{\tL,x}}$
\\\hline%
\end{tabular}
}
\caption{
Expressions of spurion fields from dim-8 aLEFT interactions.
The white and gray cells correspond to the $\Delta(B-L)=0$ and $\Delta(B+L)=0$ interactions, respectively. 
The subscripts 1, 2, and 3 stand for $u$, $d$, and $s$, and $x$ serves as a lepton flavor index.
For the WCs related to representations $\pmb{3}_{\tL(\tR)}\otimes \bar{\pmb{3}}_{\tR(\tL)}$ and $\pmb{6}_{\tL(\tR)}\otimes \pmb{3}_{\tR(\tL)}$, the following abbreviations are used:
$[C^{\tt SL,VR}_{\partial a e ddq}]_{xyzw}^\pm \equiv 
(1/2)([C^{\tt SL,VR}_{\partial a e ddq}]_{xyzw} 
\pm [C^{\tt SL,VR}_{\partial a e ddq}]_{xzyw} )$ 
and $[C^{\tt SL,VR}_{\partial a e udq}]_{x1zw}^\pm \equiv 
(1/2)([C^{\tt SL,VR}_{\partial a e udq}]_{x1zw)} 
\pm [C^{\tt SL,VR}_{\partial a e duq}]_{xz1w})$, where $q=d,u$.
Similar abbreviations are used for those of parity partner operators with superscript `{\tt SR,VL}' and in the neutrino case.} 
\label{tab:spurion}
\end{table}
 
To perform the chiral realization of operators listed in \cref{tab:BNVaLEFToperator}, it is convenient to use the spurion field approach by treating the combination of the nonquark component of each operator and its corresponding WC as a spurion field $\cal P$.
Following our previous works in~\cite{Fan:2024gzc,Liao:2025vlj}, 
we organize the flavor components of the $\pmb{8}_{\tL} \otimes  \pmb{1}_{\tR}$ and $\bar{\pmb{3}}_{\tL} \otimes \pmb{3}_{\tR}$ irreps in the matrix form: 
\begin{align}
{\cal N}_{{\bf 8}_\tL\otimes {\bf 1}_\tR}
 =
\begin{pmatrix}
{\cal N}^{\tL\tL}_{uds}  &  {\cal N}^{\tL\tL}_{usu}  & {\cal N}^{\tL\tL}_{uud}  
\\[2pt]
{\cal N}^{\tL\tL}_{dds}  & {\cal N}^{\tL\tL}_{dsu} & {\cal N}^{\tL\tL}_{dud}  
\\[2pt]
{\cal N}^{\tL\tL}_{sds} & {\cal N}^{\tL\tL}_{ssu} & {\cal N}^{\tL\tL}_{sud}
\end{pmatrix}, \quad 
{\cal N}_{\bar{\pmb{3}}_\tL \otimes \pmb{3}_\tR }   = 
 \begin{pmatrix}
{\cal N}_{uds}^{\tR\tL}  
& {\cal N}_{usu}^{\tR\tL} 
& {\cal N}_{uud}^{\tR\tL} 
\\[2pt]
{\cal N}_{dds}^{\tR\tL}  
& {\cal N}_{dsu}^{\tR\tL} 
& {\cal N}_{dud}^{\tR\tL} 
\\[2pt] 
{\cal N}_{sds}^{\tR\tL}  
& {\cal N}_{ssu}^{\tR\tL} 
& {\cal N}_{sud}^{\tR\tL}
 \end{pmatrix},
\label{eq:3qpart}
\end{align}
and similar ones for their chirality partners with $\tL\leftrightarrow\tR$. 
Accordingly, we represent the corresponding spurion matrices as
\begin{align}
 {\cal P}_{\pmb{8}_\tL \otimes \pmb{1}_\tR} 
 = 
 \begin{pmatrix}
0   & {\cal P}_{dds}^{\tL\tL}  
& {\cal P}_{sds}^{ \tL\tL} 
\\
{\cal P}_{usu}^{\tL\tL}  
& {\cal P}_{dsu}^{\tL\tL} 
& {\cal P}_{ssu}^{\tL\tL} 
\\
{\cal P}_{uud}^{\tL\tL}  
& {\cal P}_{dud}^{\tL\tL} 
& {\cal P}_{sud}^{\tL\tL}
 \end{pmatrix}, 
\,\,\,
 {\cal P}_{\pmb{3}_\tL \otimes \bar{\pmb{3}}_\tR }
 = 
 \begin{pmatrix}
{\cal P}_{uds}^{\tR\tL}  
&  {\cal P}_{dds}^{\tR\tL} 
& {\cal P}_{sds}^{\tR\tL} 
\\
{\cal P}_{usu}^{\tR\tL}  
&  {\cal P}_{dsu}^{\tR\tL} 
& {\cal P}_{ssu}^{\tR\tL} 
\\
{\cal P}_{uud}^{\tR\tL}  
&  {\cal P}_{dud}^{\tR\tL} 
& {\cal P}_{sud}^{\tR\tL}
\end{pmatrix},  
\end{align}
and the chirality partners with $\tL \leftrightarrow \tR$.
The condition $\textrm{Tr}{\cal N}_{{\bf 8}_\tL\otimes {\bf 1}_\tR}=0$ has been used to attribute the $(1,1)$ entry of ${\cal P}_{{\bf 8}_\tL\otimes {\bf 1}_\tR}$ to its $(2,2)$ and $(3,3)$ entries by treating ${\cal N}^{\tL\tL}_{uds}$ as redundant~\cite{Liao:2025sqt}.
The accompanying spurion fields of the $\pmb{6}_{\tL(\tR)} \otimes \pmb{3}_{\tR(\tL)}$ irreps ${\cal N}_{yzw}^{\tL\tR(\tR\tL), \mu}$ are denoted by ${\cal P}_{yzw}^{\tL\tR(\tR\tL),\mu}$.
With the above conventions, all aLEFT interactions in \cref{tab:BNVaLEFToperator} can be compactly written as 
\begin{align}
{\cal L}^{\slashed{B}} = & 
{\rm Tr} \big[  
  {\cal P}_{\pmb{8}_\tL \otimes \pmb{1}_\tR }
  {\cal N}_{\pmb{8}_\tL \otimes \pmb{1}_\tR } 
+ {\cal P}_{ \pmb{1}_\tL \otimes \pmb{8}_\tR }  
  {\cal N}_{  \pmb{1}_\tL \otimes \pmb{8}_\tR }
  \big]  
+ {\rm Tr} \big[ 
  {\cal P}_{\pmb{3}_\tL \otimes \bar{\pmb{3}}_\tR }
  {\cal N}_{\bar{\pmb{3}}_\tL \otimes \pmb{3}_\tR } 
+ {\cal P}_{\bar{\pmb{3}}_\tL \otimes \pmb{3}_\tR }
  {\cal N}_{\pmb{3}_\tL \otimes \bar{\pmb{3}}_\tR }
 \big] 
\nonumber 
\\
& +  \big[
{\cal P}_{yzw}^{\tL\tR,\mu}
{\cal N}_{yzw,\mu}^{\tL\tR}
+ {\cal P}_{yzw}^{\tR\tL,\mu}
{\cal N}_{yzw,\mu}^{\tR\tL}
\big]
 +\hc,
\label{eq:q3LEFT1}
\end{align}
where `Tr' represents the trace over the flavor space as indicated in the matrix notation and the repeated indices $y,z,w$ are summed over the three flavors $u,d,s$. 

After decomposing all operators in \cref{tab:BNVaLEFToperator} into irreps (${\cal N}_{yzw}^{\tL\tL (\tR\tR)}$, ${\cal N}_{yzw}^{\tR\tL(\tL\tR)}$, and ${\cal N}_{yzw}^{\tL\tR(\tR\tL), \mu}$) of the chiral group, 
one can immediately identify the corresponding spurion fields $\cal P$ through the above conventions. 
Note that \cref{eq:q3LEFT1} sums both $y$ and $z$ over all three quark flavors, and to avoid double counting, 
a factor of $1/2$ is included for the resulting spurion fields ${\cal P}_{yzw}^{\tL\tR(\tR\tL), \mu}={\cal P}_{zyw}^{\tL\tR(\tR\tL), \mu}$ when $y\neq z$. 
For example, for operators in \cref{eq:decom} with $yz=32$ and $yz=23$, we have 
\begin{align}
\sum_{yz=23,32}[C_{\partial a \nu ddu}^{\tt SL,VR}]_{xyz1} [\calO_{\partial a \nu ddu}^{\tt SL,VR} ]_{xyz1}  
= {\cal P}_{uds}^{\tR\tL} {\cal N}_{uds}^{\tR\tL}
+ {\cal P}_{dsu}^{\tL\tR,\mu} {\cal N}_{dsu,\mu}^{\tL\tR}
+ {\cal P}_{sdu}^{\tL\tR,\mu} {\cal N}_{sdu,\mu}^{\tL\tR},
\end{align}
which leads to the spurion fields, 
${\cal P}_{uds}^{\tR\tL} =
(1/2)( [C_{\partial a \nu ddu}^{\tt SL,VR}]_{x321} 
- [C_{\partial a \nu ddu}^{\tt SL,VR}]_{x231} ) 
\partial_\mu a \, \overline{\nu_{\tL,x}^\C} \gamma^\mu $ and  
${\cal P}_{dsu}^{\tL\tR,\mu} 
={\cal P}_{sdu}^{\tL\tR,\mu} =
(1/2) ( [C_{\partial a \nu ddu}^{\tt SL,VR}]_{x321} 
+ [C_{\partial a \nu ddu}^{\tt SL,VR}]_{x231})  
\partial^\mu a \, \overline{\nu_{\tL,x}^\C} $. 
The full expressions of the resulting spurion fields for all aLEFT operators are summarized in \cref{tab:spurion}. 
The absence of ${\cal P}^{\tL\tR(\tR\tL),\mu}_{uuu}$ is a consequence of electric charge conservation.
It is interesting to note that more than half of aLEFT operators contain $\pmb{6}_{\tL(\tR)} \otimes \pmb{3}_{\tR(\tL)}$ components, generating non-zero ${\cal P}_{yzw}^{\tL\tR(\tR\tL), \mu}$ terms. 
The operators 
$ [\calO_{\partial a e uud}^{\tt SL,VR}]_{x11y}$, $[\calO_{\partial a\nu ddu}^{\tt SL,VR}]_{xyy1}$, and
$ [\calO_{\partial a e ddd}^{\tt SL,VR}]_{xyyz} $, 
and their chirality partners that contain two identical chiral quark fields belong to $\pmb{6}_{\tL(\tR)} \otimes \pmb{3}_{\tR(\tL)}$; therefore, they contribute exclusively to ${\cal P}_{yzw}^{\tL\tR(\tR\tL), \mu}$.
In the following, we will incorporate these spurion fields into the recently developed chiral framework \cite{Liao:2025vlj} for exploring their phenomenological implications, 
with a specific emphasis on interactions associated with $\pmb{6}_{\tL(\tR)} \otimes \pmb{3}_{\tR(\tL)}$ irreps overlooked thus far \cite{Li:2024liy}.

\subsection{BNV ALP interactions in the ChPT}

All aLEFT operators under consideration contain three light quarks without being acted upon by a derivative. For this class of operators, their chiral matching has been investigated systematically in~\cite{Liao:2025vlj} for the pure pseudoscalar meson case and in~\cite{Liao:2025sqt} for those also involving a vector meson. In this study, we focus on the pseudoscalar meson sector and utilize chiral results in \cite{Liao:2025vlj} for the subsequent analysis. 
We define the octet pseudoscalar field by 
$\Sigma(x) = \xi^2(x) = \exp[i\sqrt{2}\Pi(x)/F_0]$
and baryon field by $B(x)$ with
\begin{align}
\Pi(x) & =   
\begin{pmatrix}
\frac{\pi^0}{\sqrt{2}}+\frac{\eta}{\sqrt{6}} & \pi^+ & K^+
\\
\pi^- & -\frac{\pi^0}{\sqrt{2}}+\frac{\eta}{\sqrt{6}} & K^0
\\
K^- & \bar{K}^0 & -\sqrt{\frac{2}{3}}\eta
\end{pmatrix}, \quad
B(x) =
\begin{pmatrix}
{\Sigma^{0}\over \sqrt{2}}+{\Lambda^0 \over \sqrt{6}}  & \Sigma^+ & p \\
\Sigma^- & -{\Sigma^{0} \over \sqrt{2}}+{\Lambda^0 \over \sqrt{6}} &  n \\ 
\Xi^- & \Xi^0 & - \sqrt{2\over 3}\Lambda^0
\end{pmatrix},
\end{align}
where $F_0=f_{\pi}/\sqrt{2}$ represents the pion decay constant in the chiral limit with $f_{\pi}=130.41(20)~\rm MeV$~\cite{ParticleDataGroup:2024cfk}.

For the three spurion fields associated with the chiral irreps 
$\bar{\pmb{3}}_{\tL(\tR)} \otimes \pmb{3}_{\tR(\tL)}$, 
$\pmb{8}_{\tL(\tR)} \otimes \pmb{1}_{\tR(\tL)}$, and
$\pmb{6}_{\tL(\tR)} \otimes \pmb{3}_{\tR(\tL)}$, the leading-order chiral Lagrangian involving the octet baryon and pseudoscalar meson takes the form~\cite{Liao:2025vlj}
\begin{align}
{\cal L}_{B}^{\slashed{B}} & =
c_1 {\rm Tr}\big[ 
{\cal P}_{  \bar{\pmb{3}}_\tL \otimes \pmb{3}_\tR} \xi B_\tL \xi -
{\cal P}_{\pmb{3}_\tL \otimes \bar{\pmb{3}}_\tR} \xi^\dagger B_\tR \xi^\dagger 
 \big]
 + c_2 {\rm Tr}\big[ 
{\cal P}_{\pmb{8}_\tL \otimes \pmb{1}_\tR}\xi B_\tL \xi^\dagger
- {\cal P}_{ \pmb{1}_\tL \otimes  \pmb{8}_\tR} \xi^\dagger B_\tR \xi
\big] 
\notag
\\
& + {c_3 \over \Lambda_\chi} \big[ 
{\cal P}_{yzi}^{\tL\tR,\mu}
{\Gamma}_{\mu\nu}^{\tt L} 
(\xi i D^\nu B_\tL \xi)_{yj}
\Sigma_{zk} \epsilon_{ijk}
 - {\cal P}_{yzi}^{\tR\tL,\mu}
{\Gamma}_{\mu\nu}^{\tt R}
(\xi^\dagger i D^\nu B_\tR  \xi^\dagger)_{yj}
\Sigma^*_{kz} \epsilon_{ijk} \big]
+ \hc.
\label{eq:chiB}
\end{align} 
where all indices $y,z$ and $i,j,k$ are summed over the three flavors $u,d,s$, and $B_{\tL(\tR)}\equiv P_{\tL(\tR)}B$ represent the chiral baryon fields. 
$\Gamma_{\mu\nu}^{\tL,\tR}\equiv (g_{\mu\nu}-\gamma_\mu\gamma_\nu/4)P_{\tL,\tR}$ are the vector-spinor projectors with $P_{\tL,\tR}\equiv (1\mp\gamma_5)/2$.\footnote{We do not consider the chiral symmetry breaking effects due to quark masses, as they enter into the chiral matching at a higher chiral order and are therefore expected to be suppressed at least by a factor of $m_s/\Lambda_\chi\sim0.1$ at the amplitude level.}
The low energy constants (LECs) $c_{1,2}$ are the $\alpha$ and $\beta$ parameters used in the literature, with recent lattice QCD calculations yielding $c_1=\alpha=-0.01257(111)\,{\rm GeV}^3$ and $c_2=\beta=0.01269(107)\,{\rm GeV}^3$ \cite{Yoo:2021gql}, respectively.
However, there is no available LQCD computation for $c_3$, 
so we use the naive dimension analysis estimation $c_3 \sim 0.011\,{\rm GeV}^3$ \cite{Liao:2025vlj} for the  numerical analysis.

Expanding the Lagrangian in \cref{eq:chiB} to the appropriate order in the pseudoscalar meson fields, we obtain interaction vertices involving spurion fields, octet baryons, and meson fields. To this end, relevant vertices include ${\cal P}{\rm -}B$ terms without mesons and ${\cal P}{\rm-}{\texttt N}{\rm-}M$ terms containing one pseudoscalar meson field, which are summarized in \cref{app:ChiL} for reference. By substituting the spurion fields $\cal P$ in these terms with the corresponding results from \cref{tab:spurion},
we obtain the desired BNV vertices containing a single ALP field, which are displayed in \cref{tab:CB2la,tab:CN2lMa}.

\section{Nucleon decays involving an ALP}
\label{sec:dewidth}
\begin{figure}[h]
\centering
\begin{subfigure}[b]{0.3\textwidth}
\centering
\begin{tikzpicture}[mystyle,scale=0.8]
\begin{scope}[shift={(1,1.5)}] 
\draw[f] (0, 0)node[left]{$B$} -- (1.5,0);
\draw[f] (1.5, 0) -- (3,0) node[right]{$\ell/\nu$};
\draw[snar, purple] (1.5,0) -- (2.5,1.2) node[right,yshift = 2 pt]{$a$};
\filldraw [cyan] (1.5,0) circle (3pt);
\end{scope}
\end{tikzpicture}
\caption{}
\label{fig:2body}
\end{subfigure}
\hspace{-1 cm} 
\begin{subfigure}[b]{0.65\textwidth}
\centering
\begin{tikzpicture}[mystyle,scale=0.8]
\begin{scope}[shift={(1,1.2)}] 
\draw[f] (0, 0)node[left]{$\texttt{N}$} -- (1.5,0);
\draw[f] (1.5, 0) -- (3,0) node[midway,yshift = 8 pt]{$B$};
\draw[snar, black] (1.5,0) -- (2.5,-1.2) node[right,yshift = 2 pt]{$M$};
\draw[f] (3.0, 0) -- (4.5,0) node[right]{$\ell/\nu$};
\draw[snar, purple] (3,0) -- (4,1.2) node[right,yshift = 2 pt]{$a$};
\filldraw [black] (1.38,-0.12) rectangle (1.62,0.12);
\filldraw [cyan] (3,0) circle (3pt);
\end{scope}
\end{tikzpicture}
\hspace{0.5cm} 
\begin{tikzpicture}[mystyle,scale=0.8]
\begin{scope}[shift={(1,1.2)}] 
\draw[f] (0, 0)node[left]{$\texttt{N}$} -- (1.5,0);
\draw[f] (1.5, 0) -- (3,0) node[right]{$\ell/\nu$};
\draw[snar, purple] (1.5,0) -- (2.5,1.2) node[right,yshift = 2 pt]{$a$};
\draw[snar, black] (1.5,0) -- (2.5,-1.2) node[right,yshift = 2 pt]{$M$};
\filldraw [cyan] (1.5,0) circle (3pt);
\end{scope}
\end{tikzpicture}
\caption{}
\label{fig:3body}
\end{subfigure}
\caption{Diagrams for BNV octet baryon two-body (a) and nucleon three-body (b) decays involving an axion. The cyan blob (black square) represents the insertion of a BNV (usual) chiral vertex.}
\label{fig:Feyndiagram}
\end{figure}
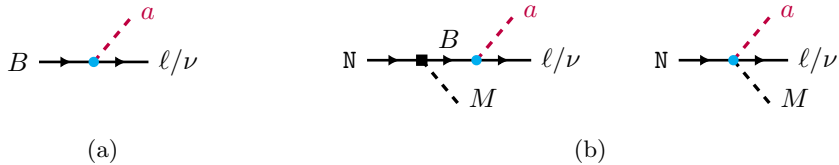

We consider BNV two-body decays of octet baryons and three-body decays of nucleons involving an ALP, with the leading-order Feynman diagrams shown in \cref{fig:Feyndiagram}.
We first derive general expressions for the amplitude squared and resulting decay width by exploiting BNV hadronic interactions obtained in the preceding section along with the usual baryon ChPT interactions. 
These results provide the necessary input for the subsequent phenomenological analysis. Further, we study differential distribution with respect to the momenta of final-state particles, which is crucial for disentangling operator structures and deriving constraints on interactions from experimental data. 

\subsection{Octet baryon two-body decays $B\to l+a$}

\begin{table}[t]
\center
\resizebox{\linewidth}{!}{
\renewcommand{\arraystretch}{1.3}
\begin{tabular}{|c|c|c|}
\hline
$B\to l+a$
&  \multicolumn{1}{c|}{ $C^{1\tL}_{B \to l}$ (upper cell) and $C^{1\tR}_{B \to l}$ (lower cell)}
& \multicolumn{1}{c|}{ $C^{3\tL}_{B \to l}$ (upper cell) and $C^{3\tR}_{B \to l}$ (lower cell)}
\\\hline
\multirow{2}*{$p \to \ell_x^+ $}
& $- [C^{\tt SR,VL}_{\partial a  eudu}]_{x121}^- 
+ \kappa_2 [C^{\tt VL,SL}_{\partial a euud}]_{x112}$ 
& $- \kappa_3 \big( [C^{\tt SL,VR}_{\partial a  eudu}]_{x121}^+ 
-[C^{\tt SL,VR}_{\partial a  euud}]_{x112} \big) $
\\\hhline{~--}
& $ [C^{\tt SL,VR}_{\partial aeudu}]_{x121}^-
- \kappa_2 [C^{\tt VR,SR}_{\partial a euud}]_{x112}$ 
& $  \kappa_3 \big( [C^{\tt SR,VL}_{\partial aeudu}]_{x121}^+ 
-[C^{\tt SR,VL}_{\partial aeuud}]_{x112} \big) $
\\\hline
\multirow{2}*{$\Sigma^+ \to \ell_x^+ $}
& $  [C^{\tt SR,VL}_{\partial a  eudu}]_{x131}^- 
- \kappa_2 [C^{\tt VL,SL}_{\partial a euud}]_{x113}$
& $ \kappa_3 \big( [C^{\tt SL,VR}_{\partial a  eudu}]_{x131}^+
- [C^{\tt SL,VR}_{\partial a  euud}]_{x113} \big) $
\\\hhline{~--}
& $ 
- [C^{\tt SL,VR}_{\partial aeudu}]_{x131}^-
+ \kappa_2 [C^{\tt VR,SR}_{\partial a euud}]_{x113}$
& $ - \kappa_3 \big( [C^{\tt SR,VL}_{\partial a  eudu}]_{x131}^+
- [C^{\tt SR,VL}_{\partial a euud}]_{x113}\big) $
\\\hline
\multirow{2}*{$\makecell{n\to \bar\nu_x \, (-) \\
\mbox{\colorbox{gray!15}{$n\to \nu_x \, (+) $}} }$}
& \cellcolor{gray!15}$
- [C^{\tt SR,VL}_{\partial a\nu udd}]_{x122}^- 
-\kappa_2 [C^{\tt VL,SL}_{\partial a \nu ddu}]_{x221} \,\,\, (+)$
& $ \kappa_3\big( [C^{\tt SL,VR}_{\partial a\nu udd}]_{x122}^+
- [C^{\tt SL,VR}_{\partial a\nu ddu}]_{x221} \big)  \,\,\, (-)$   
\\\hhline{~--}
& $ [C^{\tt SL,VR}_{\partial a\nu udd}]_{x122}^-
+ \kappa_2 [C^{\tt VR,SR}_{\partial a\nu ddu}]_{x221} \,\,\, (-)$
& \cellcolor{gray!15}$ -\kappa_3 \big( 
[C^{\tt SR,VL}_{\partial a\nu udd}]_{x122}^+ 
- [C^{\tt SR,VL}_{\partial a\nu ddu}]_{x221} 
\big) \,\,\, (+)$ 
\\\hline
\multirow{4}*{$\makecell{ \Lambda^0 \to \bar\nu_x \, (-)\\
\mbox{\colorbox{gray!15}{$\Lambda^0\to \nu_x \, (+)$}} }$}
& \cellcolor{gray!15}${1 \over \sqrt{6}} \big(
2[C^{\tt SR,VL}_{\partial a  \nu udd}]_{x123}^- 
+ [C^{\tt SR,VL}_{\partial a  \nu udd}]_{x132}^-
- [C^{\tt SR,VL}_{\partial a  \nu ddu}]_{x231}^- \big) $
& $ - \sqrt{3 \over 2} \kappa_3 \big( 
[C^{\tt SL,VR}_{\partial a  \nu udd}]_{x132}^+ 
- [C^{\tt SL,VR}_{\partial a  \nu ddu}]_{x231}^+ 
 \big) \,\,\, (-)$ 
\\
& \cellcolor{gray!15}$
+{\kappa_2\over\sqrt{6}} \big( 
[C^{\tt VL,SL}_{\partial a  \nu ddu}]_{x231} 
+2 [C^{\tt VL,SL}_{\partial a  \nu ddu}]_{x321} \big) \,\,\, (+)$
&
\\\hhline{~--}%
& $- {1\over \sqrt{6}} \big( 
2[C^{\tt SL,VR}_{\partial a  \nu udd}]_{x123}^-
+ [C^{\tt SL,VR}_{\partial a  \nu udd}]_{x132}^- 
-[C^{\tt SL,VR}_{\partial a\nu ddu}]_{x231}^- \big) $
& \cellcolor{gray!15}$ \sqrt{3\over 2} \kappa_3 \big( 
[C^{\tt SR,VL}_{\partial a \nu udd}]_{x132}^+
-[C^{\tt SR,VL}_{\partial a  \nu ddu}]_{x231}^+ \big) \,\,\, (+)$
\\
&$
- {\kappa_2 \over \sqrt{6}} \big( 
[C^{\tt VR,SR}_{\partial a  \nu ddu}]_{x231} 
+2 [C^{\tt VR,SR}_{\partial a  \nu ddu}]_{x321}\big) \,\,\, (-)$
& \cellcolor{gray!15}
\\\hline
\multirow{2}*{$\makecell{ \Sigma^0 \to \bar\nu_x \, (-) \\
\mbox{\colorbox{gray!15}{$\Sigma^0\to \nu_x \, (+)$}} }$}
& \cellcolor{gray!15}$ - {1 \over \sqrt{2}} \big( 
 [C^{\tt SR,VL}_{\partial a\nu udd}]_{x132}^-
+ [C^{\tt SR,VL}_{\partial a  \nu ddu}]_{x231}^- \big)
- {\kappa_2 \over \sqrt{2}} 
[C^{\tt VL,SL}_{\partial a  \nu ddu}]_{x231} \,\,\,(+) $
& $ {\kappa_3 \over \sqrt{2}} \big( 
 2 [C^{\tt SL,VR}_{\partial a  \nu udd}]_{x123}^+
-[C^{\tt SL,VR}_{\partial a  \nu udd}]_{x132}^+
- [C^{\tt SL,VR}_{\partial a \nu ddu}]_{x231}^+  \big) \,\,\,(-) $
\\\hhline{~--}
& $ {1\over \sqrt{2}}\big(
[C^{\tt SL,VR}_{\partial a \nu udd}]_{x132}^-
+[C^{\tt SL,VR}_{\partial a\nu ddu}]_{x231}^-  \big)
 + {\kappa_2\over \sqrt{2}}
[C^{\tt VR,SR}_{\partial a\nu ddu}]_{x231} \,\,\,(-)$
&\cellcolor{gray!15}$ - {\kappa_3\over \sqrt{2}} \big(
2 [C^{\tt SR,VL}_{\partial a\nu udd}]_{x123}^+
-[C^{\tt SR,VL}_{\partial a\nu udd}]_{x132}^+
-[C^{\tt SR,VL}_{\partial a \nu ddu}]_{x231}^+ \big) \,\,\,(+)$
\\\hline
\multirow{2}*{$\makecell{ \Xi^0 \to \bar\nu_x \, (-)\\
\mbox{\colorbox{gray!15}{$\Xi^0 \to \nu_x \, (+)$}} }$}
& \cellcolor{gray!15}$ 
[C^{\tt SR,VL}_{\partial a\nu udd}]_{x133}^-  
+ \kappa_2 [C^{\tt VL,SL}_{\partial a\nu ddu}]_{x331} \,\,\,(+) $
&  $ - \kappa_3 \big( 
[C^{\tt SL,VR}_{\partial a\nu udd}]_{x133}^+ 
- [C^{\tt SL,VR}_{\partial a\nu ddu}]_{x331} \big) \,\,\,(-)$
\\\hhline{~--}
& $  - [C^{\tt SL,VR}_{\partial a\nu udd}]_{x133}^-
- \kappa_2 [C^{\tt VR,SR}_{\partial a\nu ddu}]_{x331} \,\,\,(-)$
& \cellcolor{gray!15}$ \kappa_3\big( 
[C^{\tt SR,VL}_{\partial a\nu udd}]_{x133}^+ 
- [C^{\tt SR,VL}_{\partial a\nu ddu}]_{x331} \big) \,\,\,(+)$
\\\hline
\cellcolor{gray!15}
&  $ -  [C^{\tt SR,VL}_{\partial aeddd}]_{x232}^- 
+ \kappa_2 [C^{\tt VL,SL}_{\partial aeddd}]_{x223} $
& $- \kappa_3  \big( 
 [C^{\tt SL,VR}_{\partial aeddd}]_{x232}^+
 - [C^{\tt SL,VR}_{\partial aeddd}]_{x223} \big) $
\\\hhline{~--}
\multirow{-2}*{\cellcolor{gray!15}$\Sigma^- \to \ell_x^- $}
& $[C^{\tt SL,VR}_{\partial aeddd}]_{x232}^-
- \kappa_2 [C^{\tt VR,SR}_{\partial aeddd}]_{x223} $
& $ \kappa_3 \big(
[C^{\tt SR,VL}_{\partial aeddd}]_{x232}^+ 
- [C^{\tt SR,VL}_{\partial aeddd}]_{x223} \big) $
\\\hline
\cellcolor{gray!15} 
& $ - [C^{\tt SR,VL}_{\partial aeddd}]_{x233}^-
+ \kappa_2 [C^{\tt VL,SL}_{\partial aeddd}]_{x323} $
& $ \kappa_3 \big( 
[C^{\tt SL,VR}_{\partial aeddd}]_{x233}^+
-  [C^{\tt SL,VR}_{\partial aeddd}]_{x332} \big) $
\\\hhline{~--}
\multirow{-2}*{\cellcolor{gray!15}$\Xi^- \to \ell_x^- $} 
& $ [C^{\tt SL,VR}_{\partial aeddd}]_{x233}^-
- \kappa_2 [C^{\tt VR,SR}_{\partial aeddd}]_{x323} $
& $ - \kappa_3 \big( 
[C^{\tt SR,VL}_{\partial aeddd}]_{x233}^+ 
- [C^{\tt SR,VL}_{\partial aeddd}]_{x332} \big) $
\\\hline
\end{tabular}
}
\caption{Specific expressions of coefficients $C_{B\to l}^{1(3)\tL/\tR}$ for each transition mode.}
\label{tab:CB2la}
\end{table}

For the octet baryon two-body decays $B\to l a$ involving an ALP and a lepton in the final state, the single three-point vertex $B{\rm -}l{\rm -}a$ shown in  \cref{fig:2body} can be obtained by replacing spurion fields ${\cal P}$ in the ${\cal P}{\rm -}B$ terms of \cref{eq:LPN} with the corresponding results from \cref{tab:spurion}.
For each decay mode, it is convenient to organize relevant terms in the following general form
\begin{align}
{\cal L}_{B \to l a} & = c_1 \partial^\mu a  
\left[ 
 C^{1\tL}_{B \to l}\big(\overline{l_{\tL}} \gamma_\mu B_{\tL} \big)
+C^{1\tR}_{B \to l}\big(\overline{l_{\tR}} \gamma_\mu B_{\tR} \big) 
+{C^{3\tL}_{B \to l} \over \Lambda_\chi} \big(\overline{l_{\tR}} i\tilde\partial_\mu B_{\tL} \big) 
+{ C^{3\tR}_{B \to l} \over \Lambda_\chi } \big(\overline{l_{\tL}} i\tilde\partial_\mu B_{\tR} \big)
\right], 
\label{eq:LB2la}
\end{align}
where $l$ represents a lepton field ($\ell^\pm, \nu, \bar\nu$) and $\tilde\partial_\mu \equiv \partial_\mu -\gamma_\mu\slashed{\partial}/4$. For a negatively (positively) charged lepton $\ell^- (\ell^+)$, $l_{\tL,\tR} = \ell_{\tL,\tR} (\ell_{\tR,\tL}^\C)$
, and for a neutrino (an antineutrino) field,
$l_{\tL}(l_\tR)=\nu_\tL(\nu_\tL^\C)$.
The first two terms in \cref{eq:LB2la} are associated with the usual $\pmb{8}_{\tL(\tR)} \otimes \pmb{1}_{\tR(\tL)}$ and $\pmb{3}_{\tL(\tR)} \otimes \bar{\pmb{3}}_{\tR(\tL)}$ irreps, while the last two terms are relevant to the $\pmb{6}_{\tL(\tR)} \otimes \pmb{3}_{\tR(\tL)}$ representations.
In terms of the above parametrization, the coefficients $C_{B\to l}^{1(3)\tL/\tR}$ for each transition can be extracted from the chiral Lagrangian and are summarized in \cref{tab:CB2la}.
In this table, we defined two ratios of LECs $\kappa_2\equiv c_2/ c_1, \kappa_3\equiv c_3/c_1$.
From the above Lagrangian, the two-body decay amplitude can be written as
\begin{align}
{\cal M}_{B\to l a} 
= i \overline{u_l}\left[ 
D_{B\to l}^{\tL} P_\tL
+ D_{B\to l}^{\tR} P_\tR \right] u_{B}.
\label{eq:amp2body}
\end{align}
with
\begin{align}
D_{B\to l}^{\tL(\tR)} & \equiv 
c_1 m_B \left(C_{B\to l}^{1\tR(\tL)} 
-{ m_l \over m_B} C_{B\to l}^{1\tL(\tR)}
+ {m_B^2 - 2 m_l^2  + 2 m_a^2\over 4 m_B \Lambda_\chi } C_{B\to l}^{3\tL(\tR)}
+ {m_l \over 4 \Lambda_\chi} C_{B\to l}^{3\tR(\tL)} \right),
\end{align}
where $m_B$, $m_l$, and $m_a$ represent the masses of the baryon, lepton, and ALP, respectively. The spin-averaged and -summed matrix element squared is  
\begin{align}
 \overline{|{\cal M}_{B\to l+a}|^2} = 
 {1\over 2}(m_B^2 + m_l^2- m_a^2)
 \left( |D_{B\to l}^{\tL}|^2 + |D_{B\to l}^{\tR}|^2\right)
 +
 2 m_B m_l \, \Re\left(D_{B\to l}^{\tL}D_{B\to l}^{\tR\,*} \right).
\end{align}
Note that $\overline{u_l}$ in \cref{eq:amp2body} indicates the spinor of the lepton, which should be rewritten as $\overline{v_l^\C}$ when the final-state lepton is an antiparticle. $u$ and $v^\C$ share identical Dirac properties (the equation of motion and completeness relation), and thus, these results remain valid for antiparticles. Finally, the decay width becomes
\begin{align}
\Gamma_{B \to l a} &= { \overline{|{\cal M}_{B\to l+a}|^2} \over 16\pi m_B}\lambda^{1/2}(1,x_l, x_a),
\end{align}
where $\lambda(x,y,z)\equiv x^2+y^2+z^2-2xy-2yz-2zx$ 
represents the triangle function, and
$x_l \equiv m_l^2/m_B^2$, 
$x_a\equiv m_a^2/m_B^2$.
In \cref{app:DW-aLEFT},
we show the complete decay widths for the massless ALP case.
The numerical constraints on the relevant WCs are studied in the next section.

\subsection{Nucleon three-body decays $\texttt{N}\to l+M+a$}
By inserting the explicit expressions of spurion fields from \cref{tab:spurion} into the ${\cal P}{\rm-}{\texttt N}{\rm-}M$ terms in \cref{eq:LPNM}, we obtain effective BNV four-point interactions containing a baryon, a meson, a lepton, and an ALP. These interactions yield the contact contribution to the three-body decays shown in the second diagram of \cref{fig:3body}.
All these local vertices can be written in the general form
\begin{align}
{\cal L}_{\texttt{N} \to l M a} & = 
c_1 \frac{i\partial^\mu a}{F_0}
\left[ 
 C^{1\tL}_{\texttt{N} \to lM}\,\overline{l_{\tL}} \gamma_\mu \texttt{N}_{\tL}
+C^{1\tR}_{\texttt{N} \to lM}\,\overline{l_{\tR}} \gamma_\mu \texttt{N}_{\tR} 
+{ C^{3\tL}_{\texttt{N} \to lM} \over \Lambda_\chi} \overline{l_{\tR}} i\tilde\partial_\mu \texttt{N}_{\tL}
+{ C^{3\tR}_{\texttt{N} \to lM} \over \Lambda_\chi} \overline{l_{\tL}} i\tilde\partial_\mu \texttt{N}_{\tR}
\right] \bar{M}, 
\label{eq:LB2lMa}
\end{align}
where the first two terms are related to irreps $\pmb{8}_{\tL(\tR)} \otimes \pmb{1}_{\tR(\tL)}$ and $\pmb{3}_{\tL(\tR)} \otimes \bar{\pmb{3}}_{\tR(\tL)}$, whereas the last two terms are associated with $\pmb{6}_{\tL(\tR)} \otimes \pmb{3}_{\tR(\tL)}$.
The coefficients $C_{\texttt{N} \to lM}^{1(3)\tL/\tR}$ for all relevant vertices are summarized in \cref{tab:CN2lMa}.

\begin{table}[t]
\center
\resizebox{\linewidth}{!}{
\renewcommand{\arraystretch}{1.3}
\begin{tabular}{|c| c | c |}
\hline
$\texttt{N}\to lM+a$
&  \multicolumn{1}{c|}{ $C^{1\tL}_{\texttt{N} \to lM}$ (upper cell) and $C^{1\tR}_{\texttt{N} \to lM}$ (lower cell)}
& \multicolumn{1}{c|}{ $C^{3\tL}_{\texttt{N} \to lM}$ (upper cell) and $C^{3\tR}_{\texttt{N} \to lM} $ (lower cell)} 
\\\hline
\multirow{2}*{$p \to \ell_x^+ \pi^0 $}
& $ - {1\over 2} \big( 
[C^{\tt SR,VL}_{\partial a  eudu}]_{x121}^-
-\kappa_2 [C^{\tt VL,SL}_{\partial a euud}]_{x112} \big) $ 
& $ {\kappa_3 \over 2} \big( 
[C^{\tt SL,VR}_{\partial a  eudu}]_{x121}^+ 
+ 3 [C^{\tt SL,VR}_{\partial a  euud}]_{x112} \big) $
\\\hhline{~--}
& $ - {1 \over 2} \big( 
[C^{\tt SL,VR}_{\partial a  eudu}]_{x121}^- 
- \kappa_2 [C^{\tt VR,SR}_{\partial a euud}]_{x112}
\big) $
& $ {\kappa_3 \over 2} \big( 
 [C^{\tt SR,VL}_{\partial a  eudu}]_{x121}^+ 
+ 3 [C^{\tt SR,VL}_{\partial a  euud}]_{x112} 
\big) $
\\\hline
\multirow{2}*{$p \to \ell_x^+ \eta $}
& $ {1\over 2\sqrt{3}} \big(
[C^{\tt SR,VL}_{\partial a eudu}]_{x121}^- 
+ 3\kappa_2 
[C^{\tt VL,SL}_{\partial a euud}]_{x112} \big)$ 
& $ - {\kappa_3 \over 2\sqrt{3}} \big( 
[C^{\tt SL,VR}_{\partial a  eudu}]_{x121}^+ 
- [C^{\tt SL,VR}_{\partial a  euud}]_{x112} \big) $
\\\hhline{~--}
& $ { 1\over 2\sqrt{3}} \big( 
[C^{\tt SL,VR}_{\partial a  eudu}]_{x121}^-
+ 3\kappa_2  
[C^{\tt VR,SR}_{\partial a euud}]_{x112} \big)$ 
& $ - {\kappa_3 \over 2\sqrt{3}} \big( 
[C^{\tt SR,VL}_{\partial a  eudu}]_{x121}^+ 
- [C^{\tt SR,VL}_{\partial a  euud}]_{x112} \big) $
\\\hline
\multirow{2}*{$p \to \ell_x^+ K^0 $}
& $ {1\over \sqrt{2}}\big( 
[C^{\tt SR,VL}_{\partial a  eudu}]_{x131} ^-
+ \kappa_2 [C^{\tt VL,SL}_{\partial a euud}]_{x113} \big)$ 
& $ - {\kappa_3 \over \sqrt{2}} \big(
[C^{\tt SL,VR}_{\partial a eudu}]_{x131}^+ 
+ [C^{\tt SL,VR}_{\partial a euud}]_{x113} \big) $
\\\hhline{~--}
& $ {1\over \sqrt{2}} \big( 
[C^{\tt SL,VR}_{\partial a eudu}]_{x131}^- 
+\kappa_2 [C^{\tt VR,SR}_{\partial a euud}]_{x113}  \big)$
& $ - {\kappa_3 \over \sqrt{2}} \big(
[C^{\tt SR,VL}_{\partial a  eudu}]_{x131}^+ 
+ [C^{\tt SR,VL}_{\partial a  euud}]_{x113} \big) $
\\\hline
\multirow{2}*{$n \to \ell_x^+ \pi^-$}
& $-  { 1\over \sqrt{2}} \big( 
[C^{\tt SR,VL}_{\partial a  eudu}]_{x121}^- 
-\kappa_2 [C^{\tt VL,SL}_{\partial a euud}]_{x112} \big)  $
& $ - {\kappa_3 \over \sqrt{2}} 
\big( 
3 [C^{\tt SL,VR}_{\partial a  eudu}]_{x121}^+ - 
[C^{\tt SL,VR}_{\partial a  euud}]_{x112} \big) $
\\\hhline{~--}
& $ - {1 \over \sqrt{2}} \big( 
[C^{\tt SL,VR}_{\partial a  eudu}]_{x121}^- 
- \kappa_2 [C^{\tt VR,SR}_{\partial a euud}]_{x112}\big)  $
& $ - {\kappa_3 \over \sqrt{2}} 
\big(  3 [C^{\tt SR,VL}_{\partial a  eudu}]_{x121}^+ 
- [C^{\tt SR,VL}_{\partial a  euud}]_{x112} \big) $
\\\hline
\multirow{2}*{$\makecell{ p \to \bar\nu_x \pi^+ \, (-)\\
\mbox{ \colorbox{gray!15}{$p \to \nu_x \pi^+ \, (+)$}} }$}
&\cellcolor{gray!15}$ - {1\over \sqrt{2}} \big( 
[C^{\tt SR,VL}_{\partial a  \nu udd}]_{x122}^- 
+ \kappa_2 [C^{\tt VL,SL}_{\partial a \nu ddu}]_{x221}\big) \,\,\,(+)$ 
& $ {\kappa_3 \over \sqrt{2}} 
\big( 3 [C^{\tt SL,VR}_{\partial a  \nu udd}]_{x122}^+
- [C^{\tt SL,VR}_{\partial a  \nu ddu}]_{x221}\big) \,\,\,(-)$
\\\hhline{~--}
& $-  {1 \over \sqrt{2}} \big( 
[C^{\tt SL,VR}_{\partial a  \nu udd}]_{x122}^- 
+ \kappa_2[C^{\tt VR,SR}_{\partial a \nu ddu}]_{x221}\big) \,\,\,(-)$
&\cellcolor{gray!15}$ {\kappa_3 \over \sqrt{2}} 
\big( 3 [C^{\tt SR,VL}_{\partial a  \nu udd}]_{x122}^+ 
-  [C^{\tt SR,VL}_{\partial a  \nu ddu}]_{x221}\big) \,\,\,(+) $
\\\hline
\multirow{2}*{$\makecell{ p \to \bar\nu_x K^+ \, (-)\\
\mbox{ \colorbox{gray!15}{$p \to \nu_x K^+ \, (+)$}} }$}
&\cellcolor{gray!15}$ - {1\over \sqrt{2}} \big( 
[C^{\tt SR,VL}_{\partial a  \nu udd}]_{x123}^-
+[C^{\tt SR,VL}_{\partial a  \nu ddu}]_{x231}^- 
+ \kappa_2  [C^{\tt VL,SL}_{\partial a \nu ddu}]_{x321} \big) \,\,\,(+)$
& $ {\kappa_3 \over \sqrt{2}} \big( 
[C^{\tt SL,VR}_{\partial a  \nu udd}]_{x123}^+
+ 2[C^{\tt SL,VR}_{\partial a  \nu udd}]_{x132}^+
-[C^{\tt SL,VR}_{\partial a  \nu ddu}]_{x231}^+ \big) \,\,\,(-)$
\\\hhline{~--}
& $ - {1\over \sqrt{2}} \big( 
[C^{\tt SL,VR}_{\partial a  \nu udd}]_{x123}^-
+[C^{\tt SL,VR}_{\partial a  \nu ddu}]_{x231}^- 
+\kappa_2 [C^{\tt VR,SR}_{\partial a \nu ddu}]_{x321} \big) \,\,\,(-)$
&\cellcolor{gray!15}$ {\kappa_3 \over \sqrt{2}} \big( 
[C^{\tt SR,VL}_{\partial a  \nu udd}]_{x123}^+
+ 2[C^{\tt SR,VL}_{\partial a  \nu udd}]_{x132}^+
- [C^{\tt SR,VL}_{\partial a  \nu ddu}]_{x231}^+ \big) \,\,\,(+)$
\\\hline
\multirow{2}*{$\makecell{ n \to \bar\nu_x \pi^0 \, (-)\\
\mbox{ \colorbox{gray!15}{$n \to \nu_x \pi^0 \, (+)$}} }$}
&\cellcolor{gray!15}$ {1\over 2} \big( 
[C^{\tt SR,VL}_{\partial a  \nu udd}]_{x122}^- 
+\kappa_2 [C^{\tt VL,SL}_{\partial a \nu ddu}]_{x221}  \big) \,\,\,(+)$
& $ {\kappa_3 \over 2} \big( 
 [C^{\tt SL,VR}_{\partial a  \nu udd}]_{x122}^+ 
+ 3 [C^{\tt SL,VR}_{\partial a  \nu ddu}]_{x221} \big) \,\,\,(-)$
\\\hhline{~--}
& $ {1\over 2} \big( 
[C^{\tt SL,VR}_{\partial a  \nu udd}]_{x122}^- 
+ \kappa_2 [C^{\tt VR,SR}_{\partial a  \nu ddu}]_{x221}\big) \,\,\,(-)$
&\cellcolor{gray!15}$ {\kappa_3 \over 2} \big(
[C^{\tt SR,VL}_{\partial a  \nu udd}]_{x122}^+
+ 3 [C^{\tt SR,VL}_{\partial a  \nu ddu}]_{x221}\big) \,\,\,(+)$
\\\hline
\multirow{2}*{$\makecell{ n \to \bar\nu_x \eta \,(-)\\
\mbox{\colorbox{gray!15}{$n \to \nu_x \eta \, (+)$}} }$}
&\cellcolor{gray!15}$ {1 \over 2\sqrt{3}} \big( 
[C^{\tt SR,VL}_{\partial a  \nu udd}]_{x122}^- 
- 3 \kappa_2 [C^{\tt VL,SL}_{\partial a  \nu ddu}]_{x221} \big) \,\,\,(+)$
& $ {\kappa_3 \over 2\sqrt{3}} \big( 
[C^{\tt SL,VR}_{\partial a  \nu udd}]_{x122}^+
- [C^{\tt SL,VR}_{\partial a  \nu ddu}]_{x221} \big) \,\,\,(-)$
\\\hhline{~--}
& $ {1\over 2\sqrt{3}} \big( 
[C^{\tt SL,VR}_{\partial a  \nu udd}]_{x122}^- 
-3 \kappa_2 [C^{\tt VR,SR}_{\partial a  \nu ddu}]_{x221} \big)  \,\,\,(-)$
&\cellcolor{gray!15}$ {\kappa_3\over 2\sqrt{3}}\big( 
[C^{\tt SR,VL}_{\partial a  \nu udd}]_{x122}^+
-[C^{\tt SR,VL}_{\partial a  \nu ddu}]_{x221} \big) \,\,\,(+)$
\\\hline
\multirow{4}*{$\makecell{ n \to \bar\nu_x K^0 \,(-) \\
\mbox{ \colorbox{gray!15}{$n \to \nu_x K^0 \,(+)$}} }$}
&\cellcolor{gray!15}$ - {1\over \sqrt{2}} \big( 
 [C^{\tt SR,VL}_{\partial a  \nu udd}]_{x123}^- 
 - [C^{\tt SR,VL}_{\partial a  \nu udd}]_{x132}^- \big) $
& $ - {\kappa_3 \over \sqrt{2}} \big( 
[C^{\tt SL,VR}_{\partial a  \nu udd}]_{x123}^+ 
- [C^{\tt SL,VR}_{\partial a  \nu udd}]_{x132}^+
+ 2 [C^{\tt SL,VR}_{\partial a  \nu ddu}]_{x231}^+   \big) \,\,\,(-)$
\\
&\cellcolor{gray!15}$ 
-{\kappa_2 \over \sqrt{2}} \big( 
[C^{\tt VL,SL}_{\partial a  \nu ddu}]_{x231} 
+ [C^{\tt VL,SL}_{\partial a  \nu ddu}]_{x321} \big) \,\,\,(+)$
&
\\\hhline{~--}
& $ - {1 \over \sqrt{2}} \big( 
[C^{\tt SL,VR}_{\partial a \nu udd}]_{x123}^- 
- [C^{\tt SL,VR}_{\partial a  \nu udd}]_{x132}^-\big) $
&\cellcolor{gray!15}$ - {\kappa_3 \over \sqrt{2}} \big( 
[C^{\tt SR,VL}_{\partial a  \nu udd}]_{x123}^+
- [C^{\tt SR,VL}_{\partial a  \nu udd}]_{x132}^+ 
+ 2 [C^{\tt SR,VL}_{\partial a  \nu ddu}]_{x231}^+\big) \,\,\,(+)$
\\
& $ 
-{\kappa_2 \over \sqrt{2}} \big(
[C^{\tt VR,SR}_{\partial a  \nu ddu}]_{x231} 
+ [C^{\tt VR,SR}_{\partial a  \nu ddu}]_{x321} \big) \,\,\,(-)$
& \cellcolor{gray!15}
\\\hline
\cellcolor{gray!15} 
&\cellcolor{gray!15}   ---
&\cellcolor{gray!15}$ \sqrt{2} \kappa_3 
[C^{\tt SL,VR}_{\partial a  eddd}]_{x222} $
\\\hhline{~--}
\multirow{-2}*{\cellcolor{gray!15}$n \to \ell_x^- \pi^+$}
&\cellcolor{gray!15}   ---
&\cellcolor{gray!15}$ \sqrt{2} \kappa_3 
[C^{\tt SR,VL}_{\partial a  eddd}]_{x222} $
\\\hline
\cellcolor{gray!15}  
&\cellcolor{gray!15}$ - {1\over \sqrt{2}} \big(
[C^{\tt SR,VL}_{\partial a  eddd}]_{x232}^-
+ \kappa_2 [C^{\tt VL,SL}_{\partial a eddd}]_{x223}\big)
$ 
&\cellcolor{gray!15}$ {\kappa_3 \over \sqrt{2}} \big( 
[C^{\tt SL,VR}_{\partial a  eddd}]_{x232}^+
+ [C^{\tt SL,VR}_{\partial a  eddd}]_{x223}   \big) $
\\\hhline{~--}
\multirow{-2}*{\cellcolor{gray!15}$n \to \ell_x^- K^+$} 
&\cellcolor{gray!15}$- {1 \over \sqrt{2}} 
\big( [C^{\tt SL,VR}_{\partial a  eddd}]_{x232}^- 
+ \kappa_2 [C^{\tt VR,SR}_{\partial a eddd}]_{x223} \big)$ 
&\cellcolor{gray!15}${\kappa_3 \over \sqrt{2}} \big(
[C^{\tt SR,VL}_{\partial a  eddd}]_{x232}^+
+ [C^{\tt SR,VL}_{\partial a  eddd}]_{x223} \big) $
\\\hline
\end{tabular}
}
\caption{Specific expressions of coefficients $C_{\texttt{N} \to lM}^{1(3)\tL/\tR}$ for each transition mode.}
\label{tab:CN2lMa}
\end{table}

In addition to the BNV interactions given above,
the standard leading-order chiral interactions involving baryons are also required. These enter into the non-contact diagram in \cref{fig:3body} via the three-point vertices involving an octet baryon ($B$), a nucleon ($\tt N$), and a meson ($M$). They are given by \cite{Jenkins:1990jv,Bijnens:1985kj} 
\begin{align}
\label{LofChPT}
{\cal L}_{\tt ChPT}^B & = 
{\rm Tr}[\bar B (i \slashed{D} -M) B] 
+ \frac{D}{2} {\rm Tr}(\bar B \gamma^\mu \gamma_5\{u_\mu,B\}) 
+ \frac{F}{2} {\rm Tr}(\bar B \gamma^\mu \gamma_5 [u_\mu,B]),
\end{align}
where $u_{\mu}=i\left[\xi(\partial_{\mu}-ir_{\mu})\xi^{\dagger}
-\xi^{\dagger}(\partial_{\mu}-il_{\mu})\xi\right]= - i \xi^\dagger (D_\mu\Sigma )\xi^\dagger$.\footnote{We note that the $D$ and $F$ terms differ by a minus sign from those in \cite{Scherer:2002tk}.}
The covariant derivatives of the meson and baryon octets are given by
 $D_\mu \Sigma = \partial_\mu \Sigma - i l_\mu \Sigma +i\Sigma r_\mu$ 
and $D_\mu B = \partial_\mu B + [\Gamma_\mu ,B]-i v_\mu^{(s)}B$, respectively. 
Here, $\Gamma_\mu$ represents the chiral connection,
$\Gamma_{\mu}=\frac{1}{2}
\left[\xi(\partial_{\mu}-ir_{\mu})\xi^{\dagger}+\xi^{\dagger}(\partial_{\mu}-il_{\mu})\xi\right]$, and 
$v_\mu^{(s)}$, $l_{\mu}$, and $r_{\mu}$ are external sources, which can be omitted in our case. 
We use the LECs $ D=0.730(11)$ and $F=0.447^{6}_{7}$ from the recent lattice calculation \cite{Bali:2022qja}. 
By expanding the pseudoscalar meson matrices in \cref{LofChPT} to the linear order, we obtain
\begin{align}
{\cal L}_{\bar{B}\texttt{N}M} & \supset 
\frac{D-F}{2F_0} 
 \left[ 
  \overline{\Sigma^0} \gamma^\mu \gamma_5 p \, \partial_\mu K^-
- \overline{\Sigma^0} \gamma^\mu \gamma_5 n\, \partial_\mu \bar K^0 
+ \sqrt{2}\big(\overline{\Sigma^+}\gamma^\mu \gamma_5 p \, \partial_\mu \bar K^0 
+ \overline{\Sigma^-} \gamma^\mu \gamma_5 n \, \partial_\mu K^- \big) 
\right]
\notag  \\ 
& 
+ \frac{3F-D}{2\sqrt{3}F_0} 
\big(
  \overline{p} \gamma^\mu \gamma_5 p \, \partial_\mu \eta 
+ \overline{n} \gamma^\mu \gamma_5 n \, \partial_\mu \eta \big)
-\frac{D+3F}{2\sqrt{3}F_0} 
\left[ 
  \overline{\Lambda^0} \gamma^\mu \gamma_5 p \, \partial_\mu K^- 
+ \overline{\Lambda^0} \gamma^\mu \gamma_5 n \, \partial_\mu \bar K^0
\right]
\notag  \\ 
&
+\frac{D+F}{2F_0} 
\left[
  \overline{p} \gamma^\mu \gamma_5 p\, \partial_\mu \pi^0
- \overline{n} \gamma^\mu \gamma_5 n\, \partial_\mu \pi^0
+ \sqrt{2} \big(\overline{n} \gamma^\mu \gamma_5 p \, \partial_\mu \pi^- 
+ \overline{p} \gamma^\mu \gamma_5 n \, \partial_\mu \pi^+ \big)
\right].
\label{LBBM}
\end{align}
In general, each term has the form:
\begin{align}
 {\cal L}_{\texttt{N}\to B M} = { C_{\texttt{N}\to BM}\over F_0 }
 \overline{B}\gamma_\mu \gamma_5 \texttt{N} \, \partial^\mu \bar M,   
 \label{eq:N2BM}
\end{align}
where the dimensionless coefficient $C_{\texttt{N}\to BM}$ can be easily read off from \cref{LBBM} for a given configuration of field combinations.
Combining \cref{eq:LB2la,eq:N2BM} will yield the non-contact contribution to the three-body decays shown in the first diagram of \cref{fig:3body}.

After considering both contact and non-contact contributions, the amplitude for a general mode can be parametrized as
\begin{align}
{\cal M}_{\texttt{N}\to l M a} 
= \overline{u_l}\left[
D_{\texttt{N}\to lM}^{{\tt S},\tL} P_\tL 
+ D_{\texttt{N}\to lM}^{{\tt S},\tR} P_\tR 
+ D_{\texttt{N}\to lM}^{{\tt V},\tL} m_\texttt{N}^{-1} \slashed{p}_B P_\tL 
+ D_{\texttt{N}\to lM}^{{\tt V},\tR} m_\texttt{N}^{-1} \slashed{p}_B P_\tR 
\right]u_\texttt{N},
\label{eq:amp3body}
\end{align}
where $B$ represents the baryon in the intermediate state of momentum $p_B$ and mass $m_B$. The expressions for $D_{\texttt{N}\to lM}^{{\tt S},\tL(\tR)}$ and $D_{\texttt{N}\to lM}^{{\tt V},\tL(\tR)}$ are given by 
\begin{subequations}
\label{eq:3coeff}
\begin{align}
{ D_{\texttt{N}\to lM}^{{\tt S},\tL(\tR)}\over c_1 F_0^{-1}}& = 
  m_l  C^{1\tL(\tR)}_{\texttt{N} \to lM}
\pm \Big[ (m_\texttt{N} m_B + s ) \Big( 4 m_l  C^{1\tL(\tR)}_{B \to l}
- { s - 2 m_l^2 + 2m_a^2 \over \Lambda_\chi} C^{3\tL(\tR)}_{B \to l} \Big)
\nonumber
\\
& - (m_\texttt{N} + m_B) s 
\Big( 4 C^{1\tR(\tL)}_{B \to l} + { m_l\over \Lambda_\chi} C^{3\tR(\tL)}_{B \to l} \Big)
\Big]{C_{\texttt{N}\to BM} \over 4(m_B^2 - s)}
\nonumber
\\
&  
- {s+t -m_M^2 - m_l^2 \over 2\Lambda_\chi} C^{3\tL(\tR)}_{\texttt{N} \to lM}
-{m_\texttt{N} m_l \over 4 \Lambda_\chi}C^{3\tR(\tL)}_{\texttt{N} \to lM}, 
\\%
{D_{\texttt{N}\to lM}^{{\tt V},\tL(\tR)}\over c_1 F_0^{-1} m_\texttt{N}} & = 
- C^{1\tL(\tR)}_{\texttt{N} \to lM} 
\pm \Big[
(m_\texttt{N} + m_B) \Big( 4 m_l C^{1\tR(\tL)}_{B \to l} 
- {s-2 m_l^2 + 2 m_a^2 \over \Lambda_\chi} C^{3\tR(\tL)}_{B \to l} \Big)
\nonumber
\\
& -  ( m_\texttt{N} m_B + s) \Big( 4 C^{1\tL(\tR)}_{B \to l}
+ {m_l\over \Lambda_\chi} C^{3\tL(\tR)}_{B \to l} \Big) \Big]{C_{\texttt{N}\to BM} \over 4(m_B^2 - s)}
+ {m_\texttt{N} \over 4 \Lambda_\chi}  C^{3\tR(\tL)}_{\texttt{N} \to lM}, 
\end{align}
\end{subequations} 
where $m_\texttt{N}$ represents the nucleon mass, $s\equiv p_B^2 = (p_l + p_a)^2$, and
$t \equiv (p-p_l)^2 = (p_M +p_a)^2$. 
In the above equations, a summation over the virtual baryon $B$ is implied. \cref{LBBM} clearly indicates that decay modes $p \to \nu_x(\bar{\nu}_x) K^+ a$ and $n \to \nu_x(\bar{\nu}_x) K^0 a$ receive non-contact contributions from both the $\Sigma^0$ and $\Lambda^0$ intermediate states as these baryons share the same quark content. For all other decay modes ${\tt N} \to M+l+a$, the contributing virtual baryon $B$ is uniquely determined by vertices $C_{{\tt N} \to B M}$ in \cref{LBBM}. 
The process $n \to \ell^- \pi^+ a$ changes the isospin by $3/2$ units, so that the required local vertex has to contain three down quarks. This flavor condition is uniquely satisfied by operators $\calO_{\partial a eddd}^{\tt SL,VR(SR,VL)}$ in the irreps $\pmb{6}_{\tL(\tR)} \otimes \pmb{3}_{\tR(\tL)}$, 
making only $C_{\texttt{N} \to lM}^{3\tL/\tR}$ nonzero.
Meanwhile, the non-contact contribution is absent because of the lack of a $C_{n\to B \pi^-}$ vertex from \cref{LBBM}.

Based on \cref{eq:amp3body}, the spin-averaged and -summed matrix element squared can be expressed compactly as 
\begin{align}
 \overline{|{\cal M}_{\texttt{N}\to lMa}|^2} & = 
 {1\over 2}(m_\texttt{N}^2 + m_l^2 - t)
\left(|D_{\texttt{N}\to lM}^{{\tt S},\tL}|^2 + |D_{\texttt{N}\to lM}^{{\tt S},\tR}|^2\right)
\nonumber
\\
& + {1\over 2 m_\texttt{N}^2} \left[(m_\texttt{N}^2 - m_M^2)(m_l^2 - m_a^2) 
+ s (s+t - m_M^2 - m_a^2) \right] 
\left(|D_{\texttt{N}\to lM}^{{\tt V},\tL}|^2 
+ |D_{\texttt{N}\to lM}^{{\tt V},\tR}|^2\right)
\nonumber
\\
& + 2 m_l m_\texttt{N} \, \Re\left(
D_{\texttt{N}\to lM}^{{\tt S},\tL}D_{\texttt{N}\to lM}^{{\tt S},\tR\,*}  \right)
+ 2 { m_l \over m_\texttt{N} } s \, \Re \left( D_{\texttt{N}\to lM}^{{\tt V},\tL}D_{\texttt{N}\to lM}^{{\tt V},\tR\,*} \right)
\nonumber
\\
& + { m_l \over m_\texttt{N} } (s + m_\texttt{N}^2 - m_M^2)\,\Re\left(
D_{\texttt{N}\to lM}^{{\tt S},\tL}D_{\texttt{N}\to lM}^{{\tt V},\tL\,*} 
+ D_{\texttt{N}\to lM}^{{\tt S},\tR}D_{\texttt{N}\to lM}^{{\tt V},\tR\,*}
\right) 
\nonumber
\\
& + (s+ m_l^2 - m_a^2) \,\Re\left(
D_{\texttt{N}\to lM}^{{\tt S},\tL}D_{\texttt{N}\to lM}^{{\tt V},\tR\,*} 
+ D_{\texttt{N}\to lM}^{{\tt S},\tR} D_{\texttt{N}\to lM}^{{\tt V},\tL\,*} \right). 
\end{align}
Finally, the decay width is
\begin{align}
\Gamma_{\texttt{N}\to lMa}=\frac{1}{256\pi^3 m_{\texttt N}^3}  \int d s\int_{t_-}^{t_+} d t \overline{|{\cal M}_{\texttt{N}\to lMa}|^2},
\end{align}
where integration domains are 
\begin{align}
& (m_l +m_a)^2 \leq s \leq (m_{\texttt N} - m_M)^2,
\notag\\
& t_\pm = (E_2^* + E_3^*)^2 - \Big(\sqrt{E_2^{*2} - m_a^2} \mp \sqrt{E_3^{*2} - m_M^2} \Big)^2, 
\notag\\
& E_2^* \equiv \frac{s - m_l^2 + m_a^2}{2\sqrt{s} },\,
E_3^* \equiv \frac{ m_{\texttt N}^2 - s - m_M^2}{2\sqrt{s}}.
\end{align}
Using the above formalism, each decay width can be parametrized in terms of aLEFT WCs after performing the full phase space integration for a given ALP mass. In  \cref{app:DW-aLEFT}, we present complete numerical results for all considered baryon and nucleon BNV decay modes with an ALP, assuming the ALP mass is negligible.    

\subsection{Normalized distribution}

Experimentally, the event numbers are usually binned against some kinetic variables such as momentum, missing energy, and invariant mass. 
These distributions encode rich information about the underlying dynamics and kinematic properties of involved particles.
We analyze the momentum distribution of the three-body nucleon decays to extract information from these distributions.  

For each mode, we define the normalized differential decay width against the momentum of either the charged lepton or octet meson via 
\begin{align}
\frac{d\tilde \Gamma}{d |\pmb{p}_\ell|} \equiv \frac{1}{\Gamma} \frac{d\Gamma}{d |\pmb{p}_\ell|}, \quad 
\frac{d\tilde \Gamma}{d |\pmb{p}_M|} \equiv \frac{1}{\Gamma}  \frac{d\Gamma}{d |\pmb{p}_M|}, 
\end{align}
where $\pmb{p}_\ell$ and $\pmb{p}_M$ represent the three-momenta of the charged lepton and octet meson in the final state, respectively.
Since both neutrino and ALP are invisible to the detector, their momentum distributions are not considered.
For illustration, we focus on proton and neutron decay modes that involve both a charged lepton ($e,\mu$) and a pion, as well as proton decay modes involving $\eta$ meson. All of these are related to operators containing two up quarks and one down quark.
We assume that only one WC is nonzero at a time and consider only half of the operators with ``{\tt VL,SL}'' and ``{\tt SR,VL}'' chiral structures because the other chirality-flipped counterparts (``{\tt VR,SR}'' and ``{\tt SL,VR}'') yield the same distributions. 

\begin{figure}
\centering
\includegraphics[width=6cm]{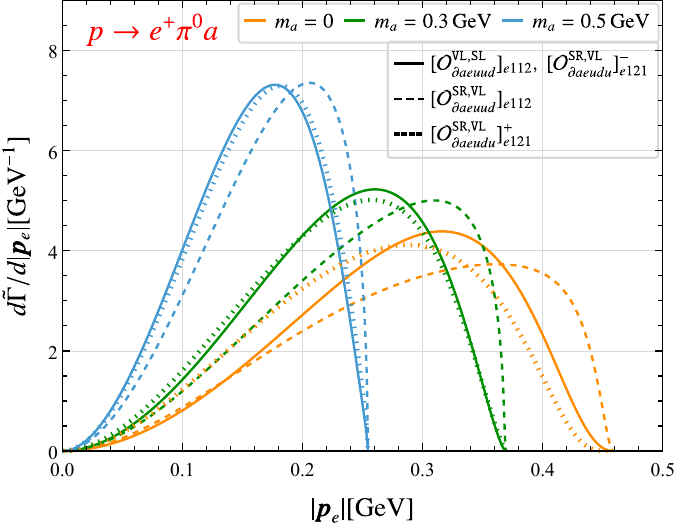}\qquad
\includegraphics[width=6cm]{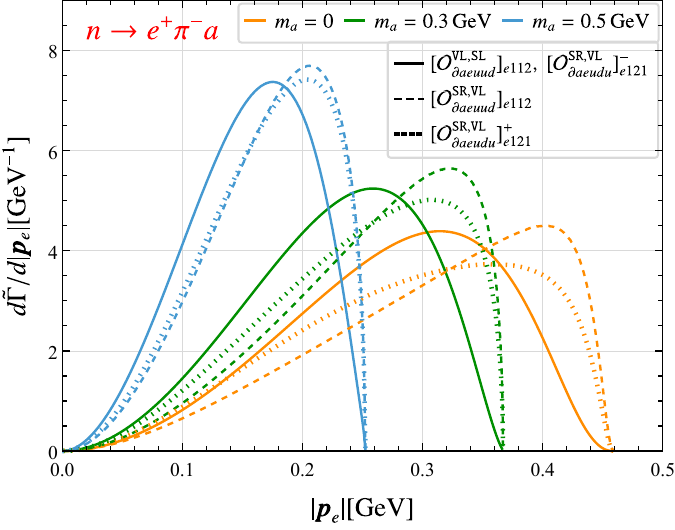}
\\[3pt]
\includegraphics[width=6cm]{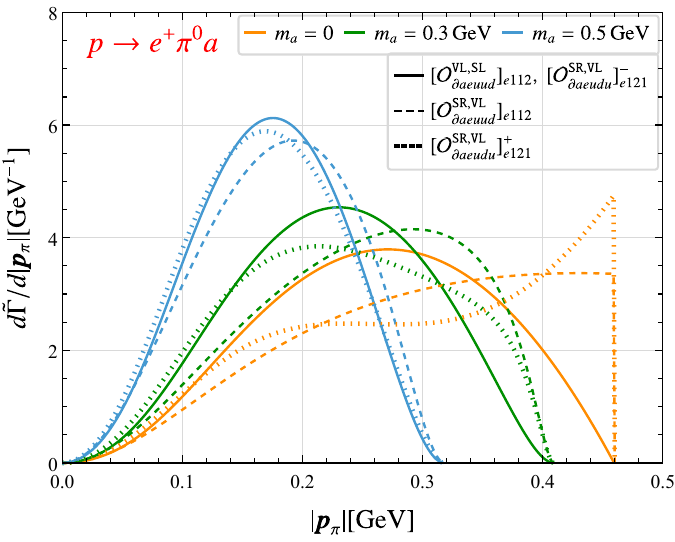}\qquad
\includegraphics[width=6cm]{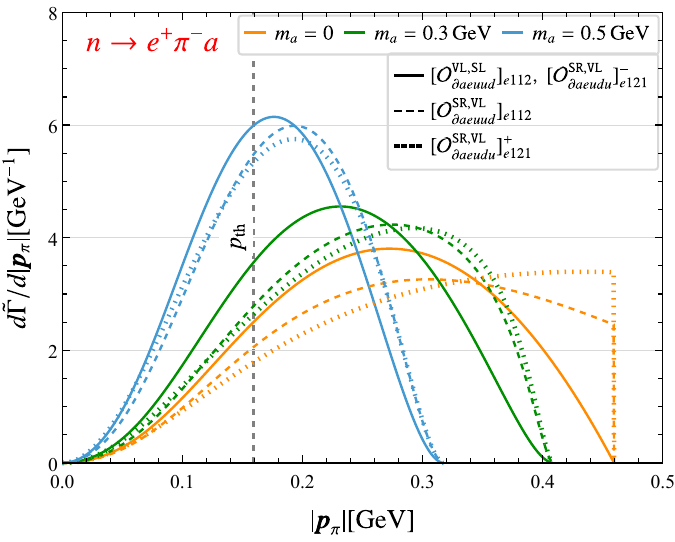}
\\[3pt]
\includegraphics[width=6cm]{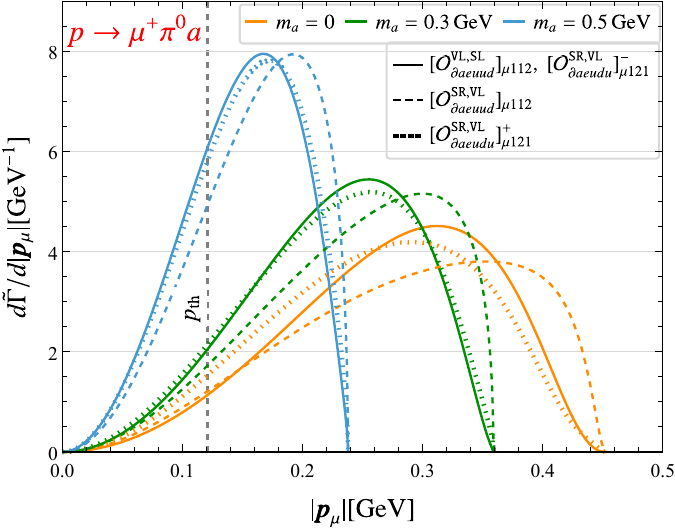}\qquad
\includegraphics[width=6cm]{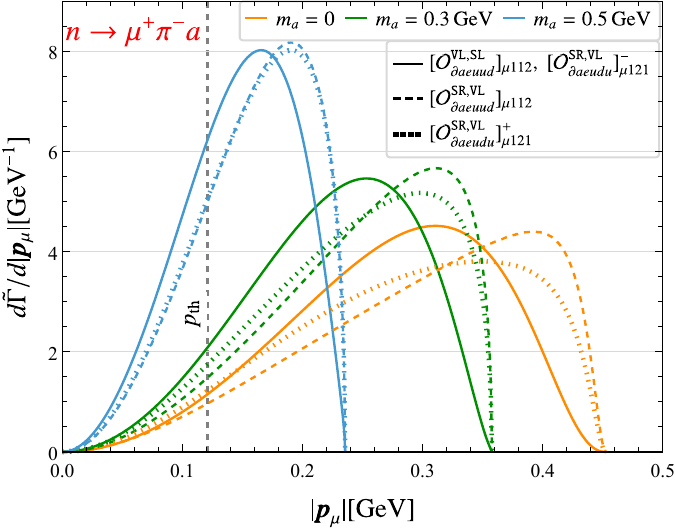}
\\[3pt]
\includegraphics[width=6cm]{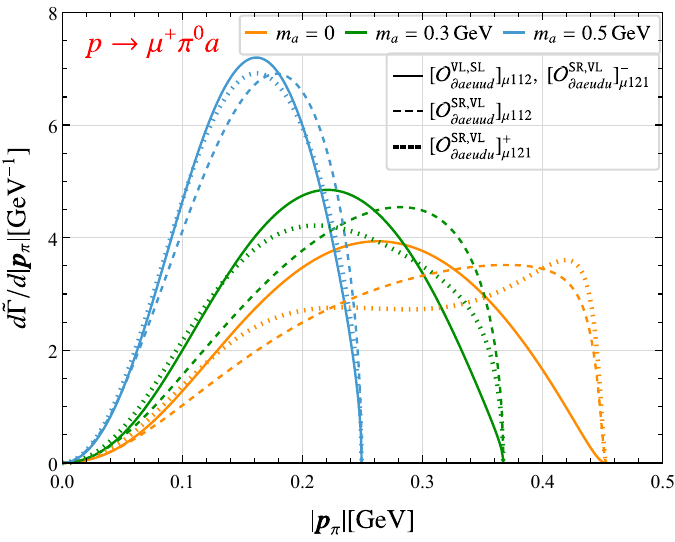}\qquad
\includegraphics[width=6cm]{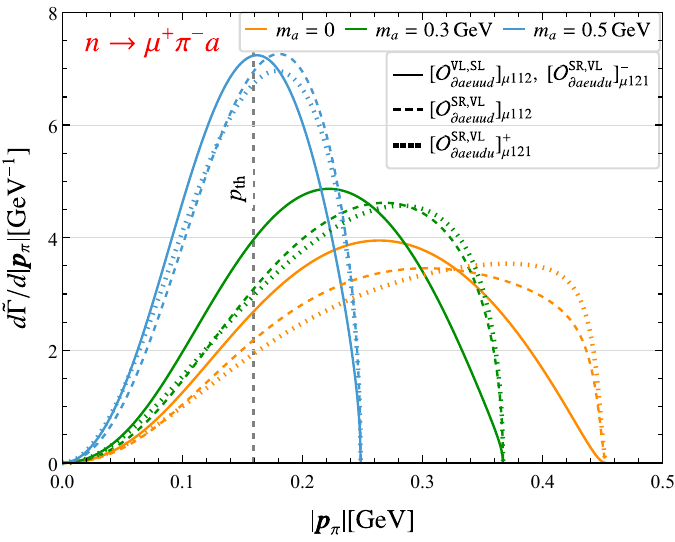}
\caption{
Normalized momentum distributions for $p \to \ell^+ \pi^0 a$ and $n \to \ell^+ \pi^- a$ resulting from the insertion of various operators.
All distributions are evaluated for three benchmark ALP masses: 
$m_a=\{0,~0.3,~0.5\}\,\rm GeV$.
The vertical gray dashed lines indicate the Cherenkov thresholds: $121 \, \MeV$ for $\mu^+$ and $159 \, \MeV$ for $\pi^-$ \cite{Heeck:2019kgr}.
$[{\cal O}^{\tt SR,VL}_{\partial a eudu}]_{x1y1}^\pm \equiv 
[{\cal O}^{\tt SR,VL}_{\partial a eudu}]_{x1y1} 
\pm [{\cal O}^{\tt SR,VL}_{\partial a eduu}]_{xy11}$
correspond to operators associated with WCs $[C^{\tt SR,VL}_{\partial a eudu}]_{x1y1}^\pm$, defined through the relationship $[C^{\tt SR,VL}_{\partial a eudu}]_{x1y1}^+ [{\cal O}^{\tt SR,VL}_{\partial a eudu}]_{x1y1}^+
+ [C^{\tt SR,VL}_{\partial a eudu}]_{x1y1}^- [{\cal O}^{\tt SR,VL}_{\partial a eudu}]_{x1y1}^-
=[C^{\tt SR,VL}_{\partial a eudu}]_{x1y1}
[{\cal O}^{\tt SR,VL}_{\partial a eudu}]_{x1y1}
+[C^{\tt SR,VL}_{\partial a eduu}]_{xy11}
[{\cal O}^{\tt SR,VL}_{\partial a eduu}]_{xy11}$.} 
\label{fig:p2pi}
\end{figure}

\cref{fig:p2pi} shows normalized differential distributions for the processes $p \to \ell^+ \pi^0a$ (left panels) and $n \to \ell^+ \pi^- a$ (right panels) from the insertion of various operators.
Given that the mass of the ALP spans a wide range, we consider three benchmark points with $m_a=\{0,~0.3,~0.5\}$ GeV. 
In the plots, the solid curves represent contributions from the usual chiral irrep operators $[{\cal O}^{\tt VL,SL}_{\partial a euud}]_{x112}\in \pmb{8}_{\tL} \otimes \pmb{1}_{\tR}$ and $[{\cal O}^{\tt SR,VL}_{\partial a eudu}]^-_{x121} \in \pmb{3}_{\tL} \otimes \bar{\pmb{3}}_{\tR}$, while the dashed and dotted curves correspond to contributions from operators $[{\cal O}^{\tt SR,VL}_{\partial a euud}]_{x112}$
and $[{\cal O}^{\tt SR,VL}_{\partial a eudu}]^+_{x121}$, respectively,  which belong to the new chiral irrep $\pmb{3}_{\tL} \otimes \pmb{6}_{\tR}$. 
As shown in \cref{fig:p2pi}, 
for each ALP mass scenario, the operators $[{\cal O}^{\tt VL,SL}_{\partial a euud}]_{x112}$ and $[{\cal O}^{\tt SR,VL}_{\partial a eudu}]^-_{x121}$ lead to the same lepton/pion-momentum distributions. These operators have a definite isospin change of $\Delta I=1/2$, and therefore, they yield approximately the same distributions for the decay modes $p \to \ell^+ \pi^0 a$ and $n \to \ell^+ \pi^- a$ in each case.
In contrast, operators belonging to the new chiral irrep $\pmb{3}_{\tL} \otimes \pmb{6}_{\tR}$ exhibit markedly different 
$|\pmb{p}_i|$-distributions across each ALP mass point and between the two decay modes because of the presence of both $\Delta I=1/2$ and $\Delta I=3/2$ isospin components in these operators.
In each $|\pmb{p}_\pi|$-distribution panel, distributions from two $\pmb{3}_{\tL} \otimes \pmb{6}_{\tR}$ operators tend to be concentrated at higher $|\pmb{p}_\pi|$ values compared to those of the two usual operators. These distinct patterns in the distributions can help disentangle operator structures and determine the ALP mass in future experimental searches.

\begin{figure}
\centering
\includegraphics[width=6cm]{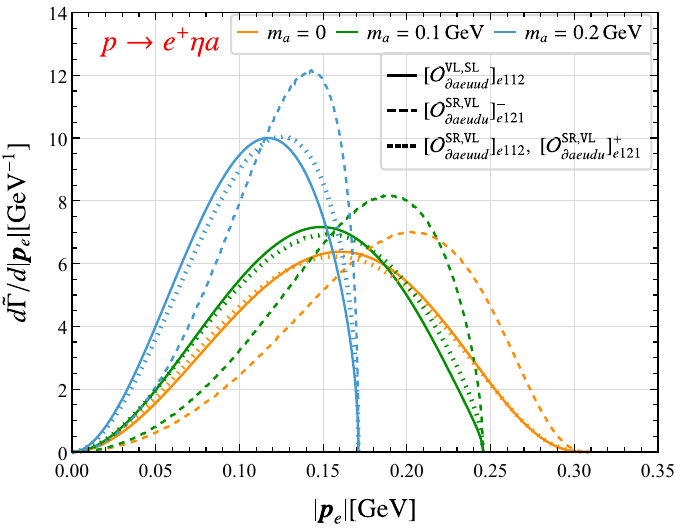}\qquad
\includegraphics[width=6cm]{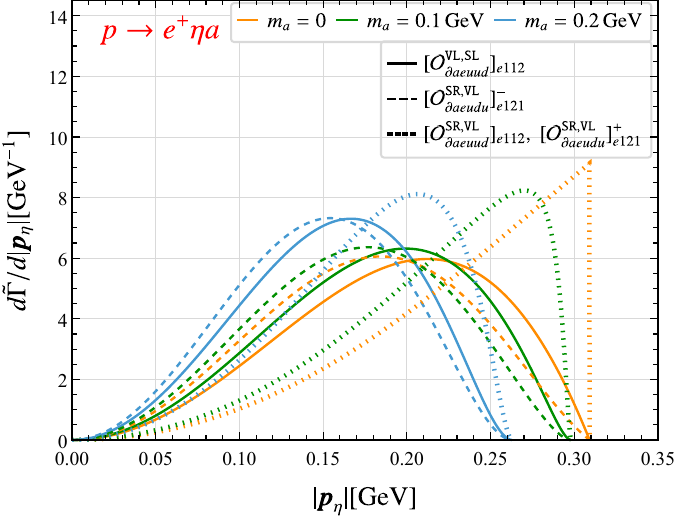}
\\[3pt]
\includegraphics[width=6cm]{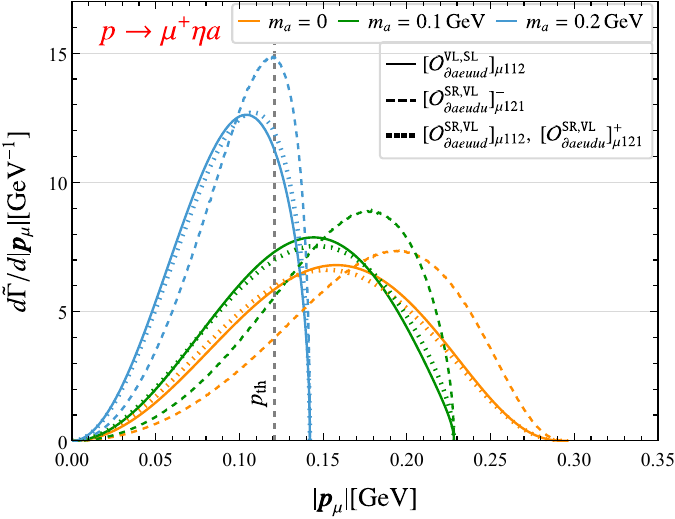}\qquad
\includegraphics[width=6cm]{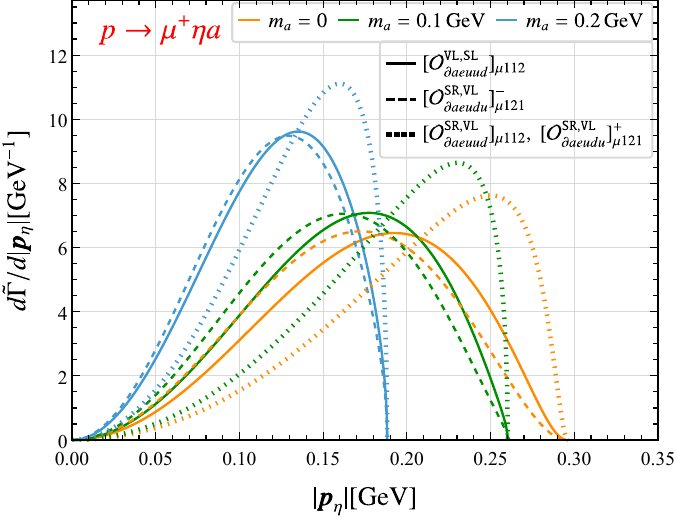}
\caption{Same as \cref{fig:p2pi} but for the decay mode $p \to \ell^+ \eta a$ with $m_a=\{0,~0.1,~0.2\}\, \GeV$.}
\label{fig:p2eta}
\end{figure}  

\cref{fig:p2eta} shows $|\pmb{p}_i|$-distributions for the decay mode  $p\to \ell^+ \eta a$, which involves the same set of operators as the pion modes.
Unlike the previous cases, the distributions from the two usual irrep operators 
$[{\cal O}^{\tt VL,SL}_{\partial a euud}]_{x112}$ (solid) and 
$[{\cal O}^{\tt SR,VL}_{\partial a eudu}]^-_{x121}$ (dashed) exhibit clear differences across all three ALP mass values and lepton/eta-momentum scenarios.
However, the two operators $[{\cal O}^{\tt SR,VL}_{\partial a euud}]_{x112}$ and $[{\cal O}^{\tt SR,VL}_{\partial a eudu}]^+_{x121}$ in the new chiral irrep share the same behavior in all cases, which is in contrast to the scenario in process $p\to \ell^+ \pi^0a$. 
This distribution characteristic can be understood from isospin considerations. 
For the decay $p\to \ell^+ \eta a$, only the $\Delta I=1/2$ components of the two operators contribute, which results in the coefficients of $[C^{\tt SR,VL}_{\partial a euud}]_{x112}$ and $[C^{\tt SR,VL}_{\partial a eudu}]^+_{x121}$ sharing the same relative ratio in both the contact ($C_{p \to \ell \eta}^{3\tR}$) and non-contact ($C_{p\to \ell}^{3\tR}$) contributions. 
For each operator, the behavior of the $|\pmb{p}_\eta|$-distribution differs from that of the $\pmb{p}_{e,\mu}$-distribution. 
The $|\pmb{p}_\eta|$-distributions from the two new irrep operators are enhanced at higher $|\pmb{p}_\eta|$ values, whereas the two usual operators peak at intermediate $|\pmb{p}_\eta|$ values. Complementary to the pion modes, these new features can be employed to distinguish among various scenarios in future experimental searches. 

Similar distributions can be studied for the other three-body decay modes.
For example, $|\pmb{p}_\pi|$-distributions in the process $p\to \nu_x(\bar\nu_x)\pi^+ a$ closely resemble those of $n\to e^+\pi^- a$ for corresponding operators with the simultaneous exchange of $u\leftrightarrow d$ and $e\leftrightarrow \nu(\nu^\C)$.   
This is physically understandable;
on the one hand, nucleon-to-pion matrix elements are connected with each other through the exchange of $u \leftrightarrow d$, 
and on the other hand, the masses of both the neutrino and electron are negligible. All these results can be understood analytically.
Given amplitude formulas in \cref{eq:amp3body,eq:3coeff} and neglecting lepton masses in both processes, a direct comparison of their corresponding coefficients ($C_{B\to l}^{1(3)\tL/\tR}$, $C_{\texttt{N} \to lM}^{1(3)\tL/\tR}$, and $C_{\texttt{N}\to BM}$) suffices to determine the relationship between their amplitudes and further their distributions.
Accordingly, such similarity in distributions is anticipated between $p \to e^+ \pi^0 (\eta) a$ and $n \to \nu_x (\bar{\nu}_x) \pi^0 (\eta) a$, $p \to \nu_x (\bar{\nu}_x) K^+ a$ and $n \to \nu_x (\bar{\nu}_x) K^0 a$,
as well as $p \to \ell^+ K^0 a$ and $n \to \ell^- K^+ a$ for corresponding operators with the simultaneous exchange of $u\leftrightarrow d$ and involved leptons.

\section{Constraints and implications}
\label{sec:constraints}

After establishing the theoretical framework for nucleon decay with an ALP in the final state,
we investigate constraints on relevant WCs from available experimental data. Exotic nucleon decays involving new light invisible particles have been largely overlooked in the past experimental searches, thereby resulting in a lack of direct constraints on their occurrence. The only exceptions are the two-body proton decay modes $p\to e^+X$ and $\mu^+X$. The Super-K experiment has set stringent lower bounds on their inverse decay widths in the limit of $m_X\to 0$: $\Gamma^{-1}({p \to e^+ + X}) > 7.9 \times 10^{32}$ yr and
$\Gamma^{-1}(p \to \mu^+ + X)> 4.1 \times 10^{32}$ yr  \cite{Super-Kamiokande:2015pys}.  
However, given the non-observation of any BNV nucleon decays, existing experimental data from searches for decay modes involving only SM final states can be used to constrain exotic modes involving new light particles. In the following subsections, we discuss how to reinterpret the existing experimental data into bounds on these new channels, and then, we use these bounds to set limits on the WCs of the effective operators.

\subsection{Recasting existing data into constraints on nucleon decay with an ALP}

First, the bounds on several inclusive decay modes can be directly applied to analogous processes involving an invisible ALP. These include the neutron invisible decay ($n\to{\rm invisible}$) and  proton or neutron decays into a charged $e^+$ or $\mu^+$ plus any other particles ($N\to e^+/\mu^++{\rm anything}$, where $N=p,n$). Due to the invisible nature of the ALP, the constraint on $n\to{\rm invisible}$ is directly applicable to the two-body modes $n\to\bar\nu(\nu) a$. 
For the decay $n\to{\rm invisible}$, we adopt the most recent partial lifetime bound reported by the SNO$+$ experiment, $\Gamma^{-1}(n\to\text{invisible})> 9.0\times 10^{29}\,{\rm yr}$ \cite{SNO:2022trz}.\footnote{Note that the old KamLAND result $\Gamma^{-1}(n\to\text{invisible})> 5.8\times 10^{29}\,{\rm yr}$ \cite{KamLAND:2005pen} was used by Ref.~\cite{Li:2024liy} in their analysis.}
This limit is expected to be further improved by two orders of magnitude in the JUNO experiment \cite{JUNO:2024pur}. 
For the nucleon decays involving a positively charged lepton, the
inclusive searches around 1980s also provide partial lifetime limits with $\Gamma^{-1}(N\to e^+ + \text{anything})>0.6\times 10^{30}$ yr \cite{Learned:1979gp} and $\Gamma^{-1}(N\to \mu^+ + \text{anything})>12\times 10^{30}$ yr \cite{Cherry:1981uq}, based on detecting both prompt and secondary charged leptons. These limits are expected to be broadly applicable to the decay channels involving the same charged leptons in the final state considered in this work, provided Cherenkov threshold effects are properly accounted for.

Although there are no dedicated searches for nucleon decays involving new light particles besides the aforementioned $p \to e^+(\mu^+) X$ channels with a massless $X$, existing experimental data and background information from conventional channels can be reinterpreted to constrain such processes. We focus on the Super-K experiment, a water-Cherenkov detector, for our analysis because it provides the most stringent bound.
Charged particles traversing the detector produce Cherenkov radiation, forming characteristic rings classified as either showering $(e^\pm,\gamma)$ or non-showing $(\mu^\pm,\pi^\pm)$ types based on their topological features. 
Each molecule (${\rm H_2 O}$) contains two free protons (from hydrogen atoms) and eight bound protons (from the oxygen nucleus).
In this preliminary attempt, we consider free proton decays in the hydrogen atom in our simulation to derive conservative bounds on relevant modes.
A more complete treatment of bound nucleon decays requires accounting for nuclear effects such as Fermi motion, nucleon correlations, and meson-nucleon interactions. Consequently, we restrict ourselves to the analysis of proton decay channels in this study, deferring the detailed investigations of these bound nucleon decays to our future work. 

The simulated channels include the two-body modes $p\to\ell^+a$ and the three-body modes $p\to\nu(\bar\nu)\pi^+ a$, $p\to\ell^+\pi^0 a$, $p\to\ell^+\eta a$, and $p\to \mu^+ K^0 a$, with $\ell=e,\mu$.
For $p\to\ell^+a$, we recast the spectral search results for $p\rightarrow e^+X$ and $p\rightarrow \mu^+X$ with a massless $X$, as reported in \cite{Super-Kamiokande:2015pys}. We employ the reconstructed momentum distributions of the charged lepton $\ell^+$ shown in Fig.\,1 of that reference to obtain the bounds on $p\to \ell^+ a$.
Similarly, for the channels $p\to\nu(\bar\nu)\pi^+ a$ in which only the charged pion is visible in the final state, we derive constraints using the reconstructed momentum distributions of the $\pi^+$, as provided in Fig.\,3 of \cite{Super-Kamiokande:2013rwg}. 
For $p\to\ell^+\pi^0 a$ and $p\to\ell^+\eta a$, we use the reconstructed invariant mass distributions of the charged lepton and neutral meson. The $\pi^0$ predominantly decays into two photons ($\textrm{Br}\simeq 98.8\%$), producing Cherenkov rings. The $\eta$ meson is reconstructed through the two-photon decay mode $\eta\to 2 \gamma$ ($\textrm{Br}\simeq39\%$), while reconstruction through $\eta\to3\pi^0$ is less effective because of the ring-counting algorithm’s limitation of identifying at most five Cherenkov rings. 
For the $p\to \mu^+ K^0 a$ channel, we utilize the dominant decay channel of $K^0_S\to\pi^+\pi^-$($\textrm{Br}\simeq69\%$) to reconstruct the $K^0$ meson.
We simulate aforementioned proton decay processes using analytical expressions and momentum distributions presented in previous sections, considering one operator at a time, and we use \texttt{Pythia8}~\cite{Sjostrand:2006za,Bierlich:2022pfr} to model the subsequent decays of the mesons. 
The momentum resolution of the detector is implemented according to $\sigma_e=(0.6+2.6/\sqrt{p_e [{\rm GeV}]})\%$ and $\sigma_\mu=(1.7+0.7/\sqrt{p_\mu [{\rm GeV}]})\%$ \cite{Super-Kamiokande:2005mbp}. 
To constrain lower limits on the partial lifetime $\Gamma^{-1}$, we require the expected signal to not exceed the observed number of events by more than $2\sigma$ in any bin, i.e., $N_s^i+N_b^i < N_o^i+2 \sigma^i$, where $N_s^i$, $N_b^i$, and $N_o^i$ represent the number of signals, background, and observed events in the $i$-th bin, respectively, and $\sigma^i$ represents the corresponding error bar. 
We neglected the angular resolution effects in our analysis, which were estimated to be $3.0^{\circ}$ for single-ring $e$-like events  and $1.8^{\circ}$ for $\mu$-like events \cite{Super-Kamiokande:2005mbp}, respectively.

\begin{figure}
\centering
\includegraphics[width=0.4\linewidth]{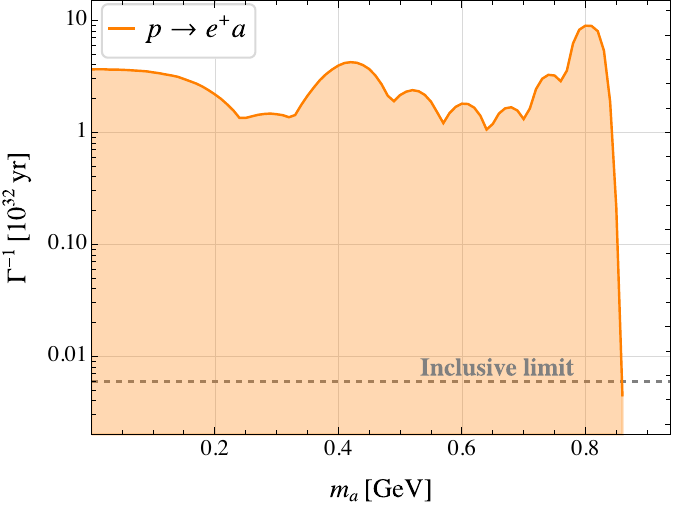}\qquad
\includegraphics[width=0.4\linewidth]{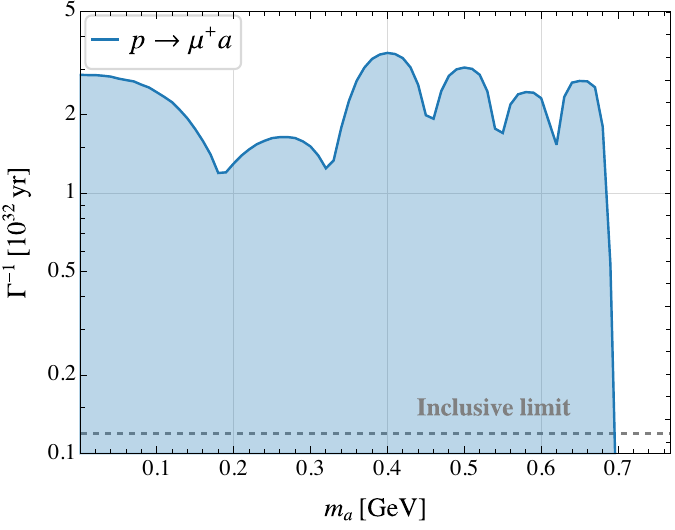}
\\
\includegraphics[width=0.4\linewidth]{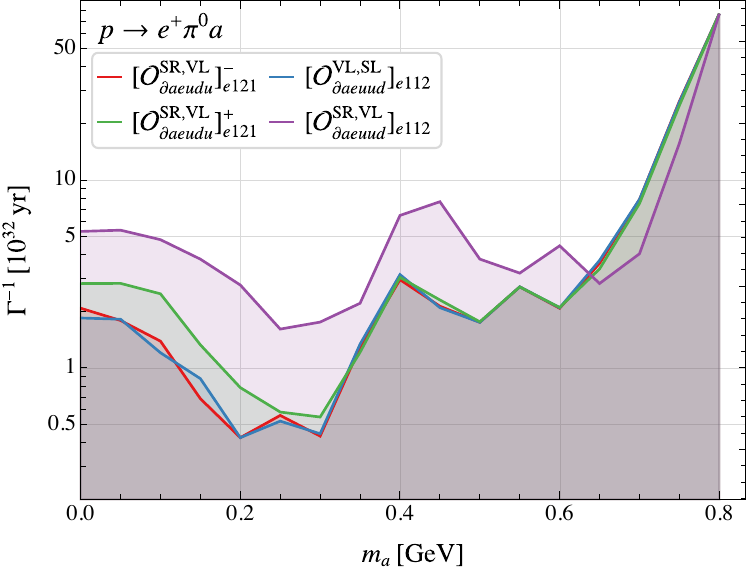}\qquad
\includegraphics[width=0.4\linewidth]{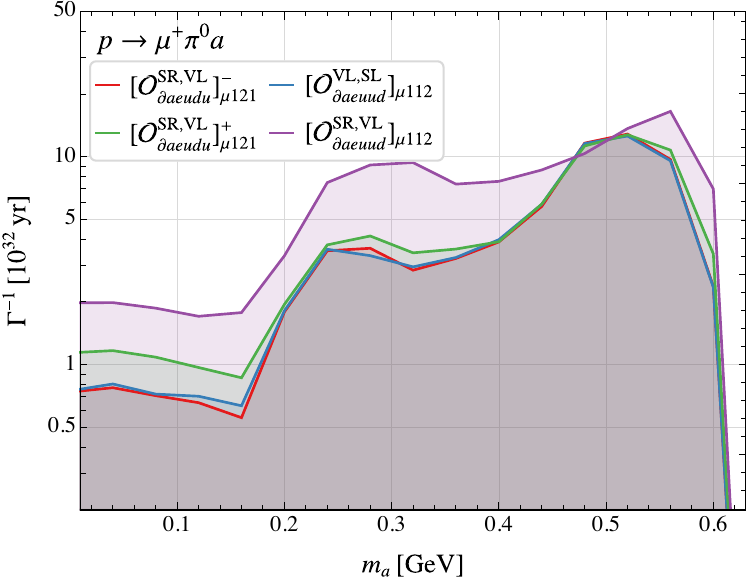}
\\
\includegraphics[width=0.4\linewidth]{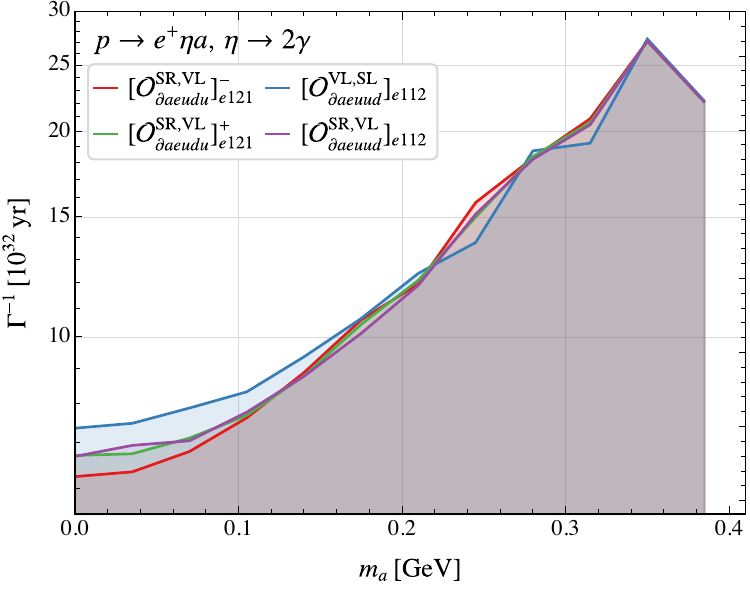}\qquad
\includegraphics[width=0.4\linewidth]{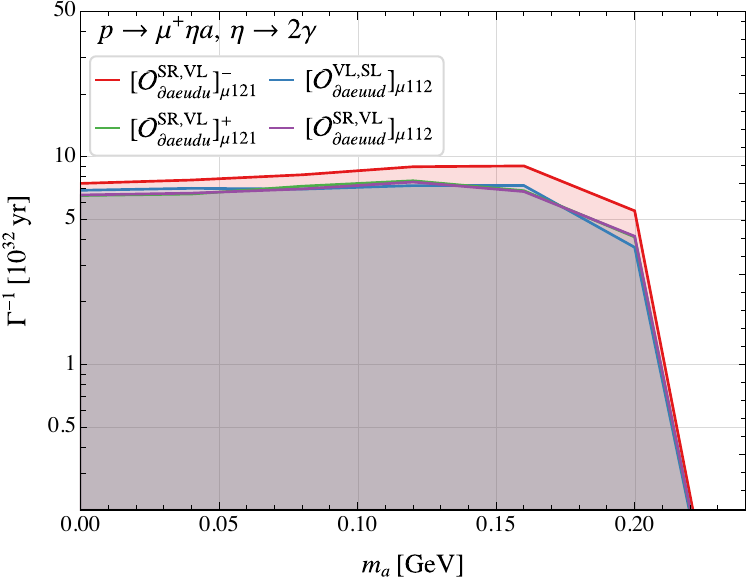}
\\
\includegraphics[width=0.4\linewidth]{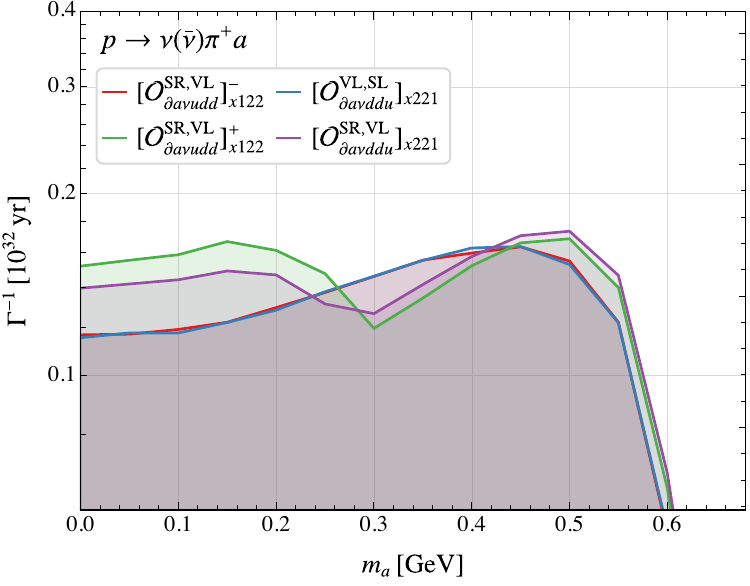}\qquad
\includegraphics[width=0.4\linewidth]{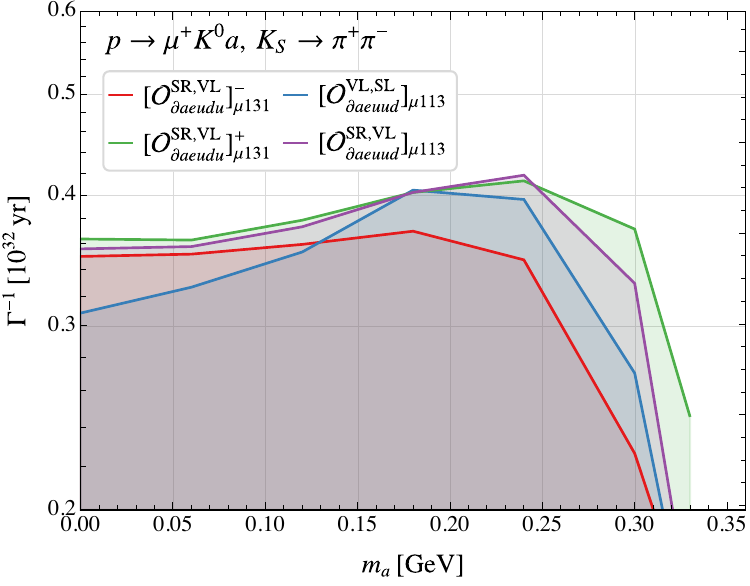}
\caption{New constraints on various proton decay channels as a function of the ALP mass $m_a$ obtained from our simulation based on Super-K experimental data.
The plots for $p\to \ell^+ a$ are obtained by recasting the analysis in \cite{Super-Kamiokande:2015pys}.
The plots for $p\to \ell^+\pi^0 a$ are based on data provided in Fig.\,4 of \cite{Super-Kamiokande:2016exg}.
The plots for $p\to \ell^+\eta a$ are based on the data provided in Fig.\,5 of \cite{Super-Kamiokande:2024qbv}, where the $\eta$ meson is reconstructed by the decay channel $\eta\to 2 \gamma$. The results for $p\to \nu(\bar\nu)\pi^+ a$ and $p \to \mu^+ K^0 a $ are obtained by reanalyzing data provided in Fig.\,3 of \cite{Super-Kamiokande:2013rwg} and Fig.\,6 of \cite{Super-Kamiokande:2022egr}, respectively.
The results are the same for the chirality-flipped operators with $\tL\leftrightarrow\tR$. The peaks or fluctuations observed in resulting constraints arise from the simplified statistical analysis and the neglect of nuclear effects.
} 
\label{fig:newbound}
\end{figure}

The recasting results for various proton decay channels are presented in \cref{fig:newbound}. Owing to the fixed kinematics in two-body decays, the obtained limits on the modes $p\to e^+(\mu^+) a$ are independent of the underlying interaction structures. In contrast, momentum distributions in three-body decays strongly depend on interaction structures; therefore, the limits for all three-body modes are presented on an operator-by-operator basis. 
The plots indicate that recasting constraints on modes containing a charged lepton are stronger than the corresponding inclusive limits by one to four orders of magnitude with the exact improvement  depending on the decay mode and ALP mass. For the two-body channels $p\to\ell^+a$, the inclusive limits cover a slightly larger ALP mass range compared to the recasting bounds caused by experimental cuts of $|\pmb{p}_e|>100$ MeV and $|\pmb{p}_\mu|>200$ MeV in the search \cite{Super-Kamiokande:2015pys}. Stronger bounds for heavier ALPs arise in $e^+$-related channels because a sharper and more localized signal peak appears at lower reconstructed invariant masses, improving single bin sensitivity. However, this effect is diminished by the reduced detection efficiency for low-momentum $\mu^+$ or $\pi^+$ produced in association with heavier ALPs, causing the bounds on channels involving $\mu^+$ or $\pi^+$ to weaken as the ALP approaches the threshold.

We did not perform a similar analysis for the $p\to e^+ K^0 a$ channel because the most recent search for $p\to e^+ K^0$ by Super-K \cite{Super-Kamiokande:2005lev} does not provide the required kinematic distributions. Therefore, we adopt the inclusive limit to set $\Gamma^{-1}(p\to e^+ K^0 a)>0.6\times 10^{30}$ yr. 
For channels with the positively charged kaon, $p\to \bar\nu(\nu) K^+ a$, $K^+$ carries momentum below its Cherenkov threshold ($563$ MeV). Thus, it cannot be detected directly in water Cherenkov detectors. Instead, $K^+$ is identified via its decay products. Owing to its low momentum and short lifetime, $K^+$ decays at rest. The Super-K experiment utilizes the dominant two-body decay modes $K\to\mu^+\nu_\mu\,(64\%)$ and $K\to\pi^+\pi^0\,(21\%)$ for its detection. Further, for nucleon decays occurring within the oxygen nucleus, the residual nucleus may be left in an excited state, which can promptly de-excite via gamma-ray emission, thereby offering an additional signal. 
Given the similarity in final-state signatures, existing searches for $p\to\nu K^+$ can be directly reinterpreted as constraints on $p\to \bar\nu(\nu) K^+ a$. Consequently, we adopt the current Super-K limit \cite{Super-Kamiokande:2014otb} to set $\Gamma^{-1}(p \to \bar\nu(\nu) K^+ a) > 5.9 \times 10^{33}$ yr. 

Although we focused on the current Super-K experiment, we anticipate that constraints will be further improved in the upcoming Hyper-K experiment, which will have a fiducial mass approximately eight times that of the Super-K. In addition, JUNO and DUNE experiments, which utilize tracking detectors, are expected to offer enhanced capabilities for probing the mass of the ALP near the threshold of nucleon decays because of their excellent low-energy thresholds.

\subsection{Comprehensive constraints on the BNV aLEFT interactions}

\begin{figure}
\centering
\includegraphics[width=0.4\linewidth]{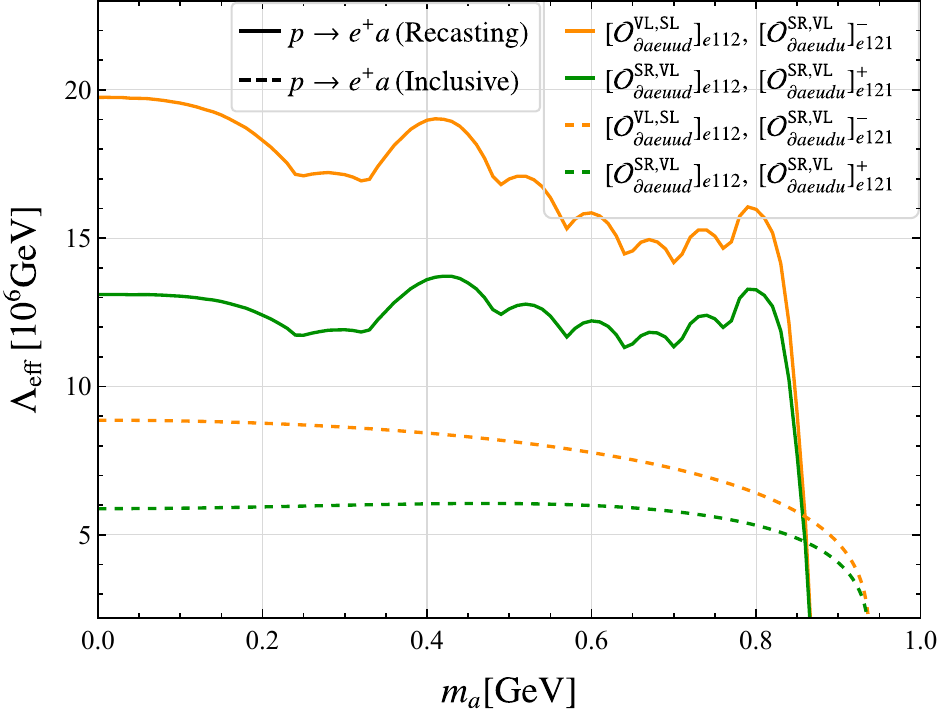}\qquad
\includegraphics[width=0.4\linewidth]{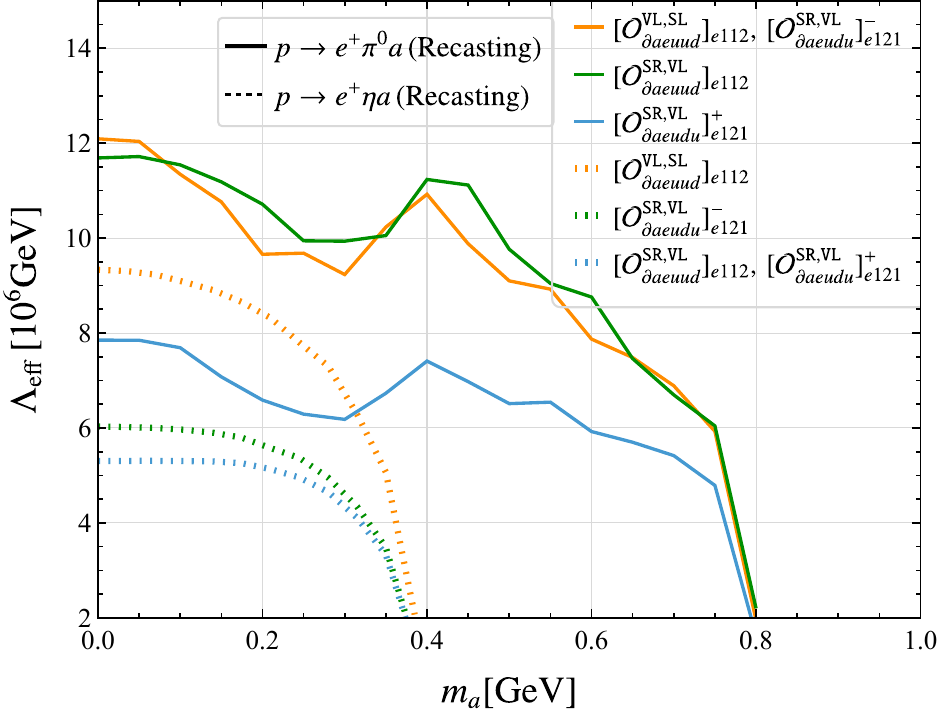}
\\
\includegraphics[width=0.4\linewidth]{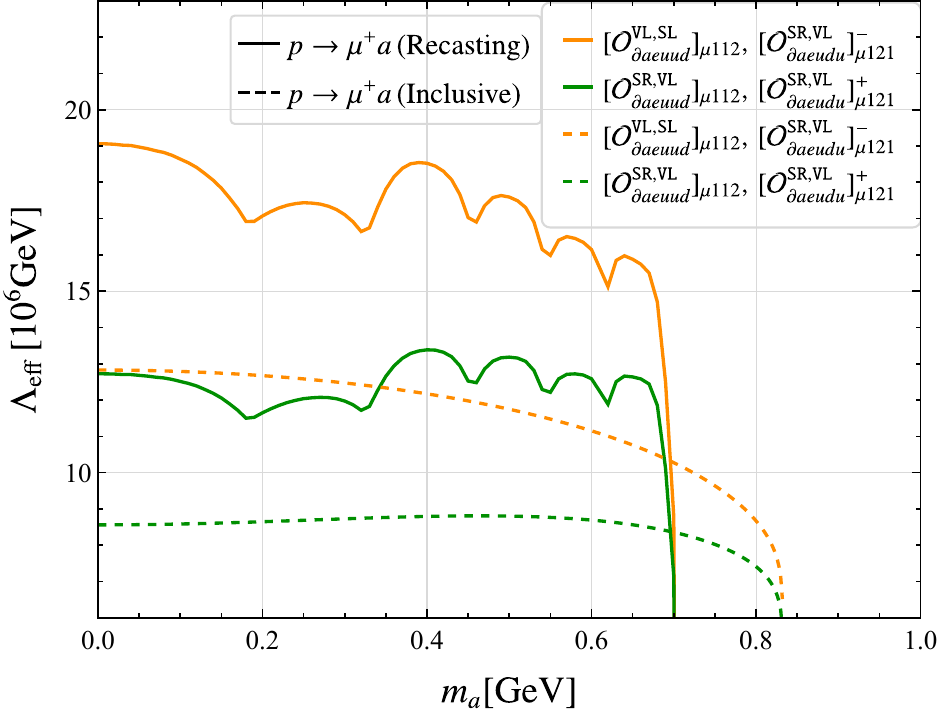}\qquad
\includegraphics[width=0.4\linewidth]{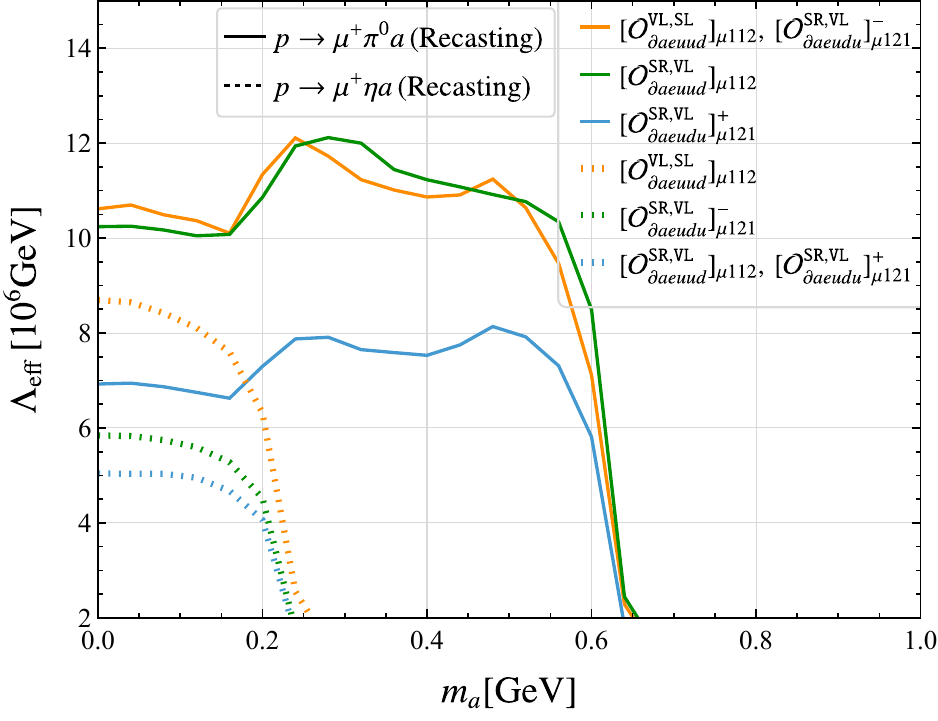}
\\
\includegraphics[width=0.4\linewidth]{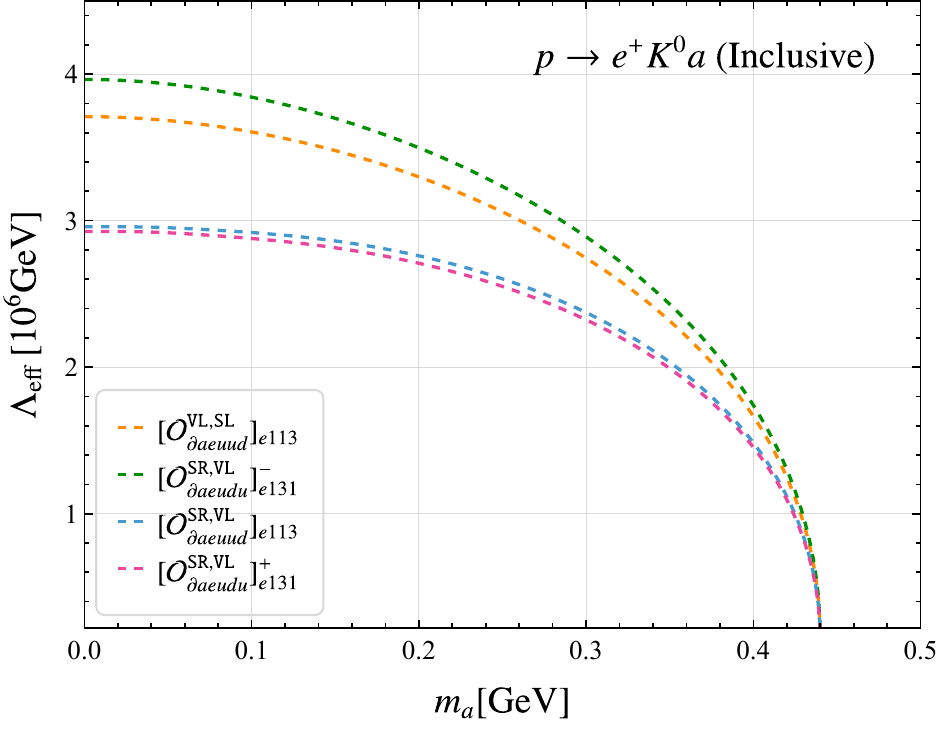}\qquad
\includegraphics[width=0.4\linewidth]{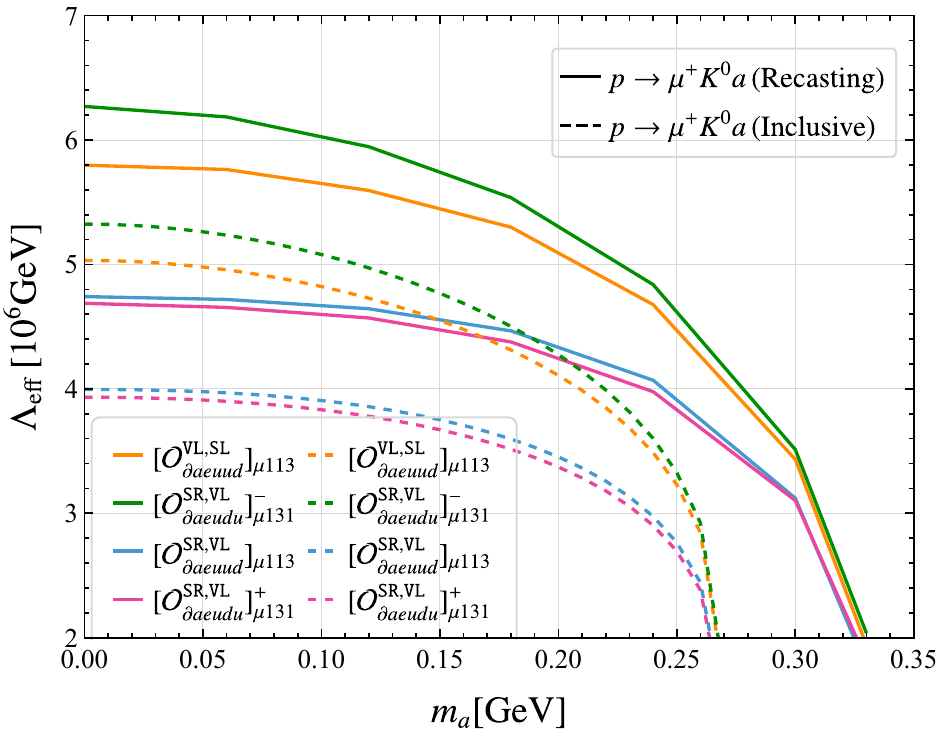}
\\
\includegraphics[width=0.4\linewidth]{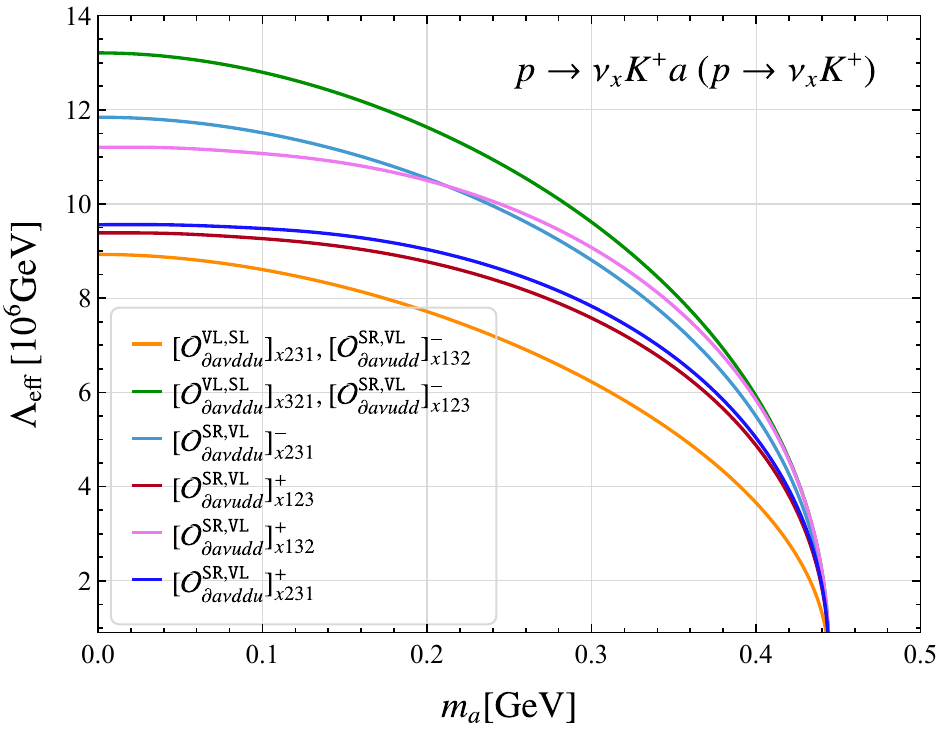}\qquad
\includegraphics[width=0.4\linewidth]{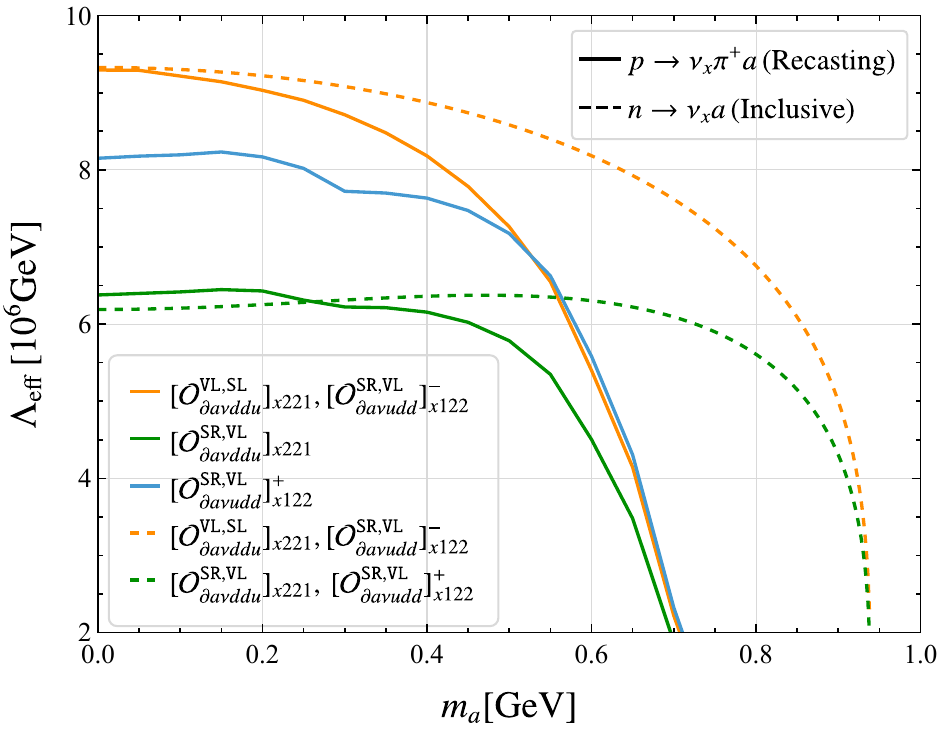}
\caption{Constraints on the effective new physics scale associated with the dim-8 aLEFT operators as a function of the ALP mass $m_a$. 
We have set $\kappa_3=1$ in the numerical analysis, and the results for the operators belonging to the irreps $\pmb{3}_{\tt L(R)}\otimes \pmb{6}_{\tt R(L)}$ can be easily rescaled if $\kappa_3$ is found to differ from 1. The peaks observed in certain solid curves originate from the corresponding peaks in recasting bounds presented in \cref{fig:newbound}.}  
\label{fig:Lam-bounds}
\end{figure}

With bounds on the inverse decay widths for  decay modes summarized in the previous subsection, we constrain the aLEFT WCs. For the WC $C_i$ of an aLEFT operator $\calO_i$, we define an associated effective scale $\Lambda_{\tt eff}$ via $\Lambda_{\tt eff}\equiv |C_i^{-1/4}|$, and study the constraint on $\Lambda_{\tt eff}$ as a function of the ALP mass $m_a$. 
To establish the bound, we deal with one operator at a time and require the theoretical decay width to be less than the limit given in the previous subsection. 

Our final results are presented in \cref{fig:Lam-bounds}. 
Only half of the operators with ``{\tt VL,SL}'' and ``{\tt SR,VL}'' chiral structures are presented because their chirality-flipped counterparts (``{\tt VR,SR}'' and ``{\tt SL,VR}'') receive identical constraints. 
For each operator, besides the constraint derived from our recast analysis (solid or dotted curves), we present the limit from inclusive searches (dashed curves) in some channels for comparison. 
Operators with $ddd$, $dds$,  $uss$, $dss$, and $sss$ quark configurations cannot be constrained because of the lack of corresponding experimental data; we neglected them in this work. Both panels in the top row of \cref{fig:Lam-bounds} feature the same set of operators with the $a\text{-}e\text{-}u\text{-}u\text{-}d$ field content. The left panel shows constraints from the two-body mode $p\to e^+ a$ based on both the recasting and inclusive results, while the right panel presents only recasting results from the three-body modes $p\to e^+ \pi^0(\eta)a$.
The two-body mode based on the recasting results provides the most stringent constraints over a wide range of ALP masses. 
A similar behavior is observed in the muonic case with the $a\text{-}\mu\text{-}u\text{-}u\text{-}d$ field content, as shown in both panels of the second row. The results for the operators with the $u\text{-}u\text{-}s$ quark content are shown in the third row, with only inclusive limits provided for the $e^+$-related operators caused by the lack of experimental distribution data.
The last row presents results for operators with the  $u\text{-}d\text{-}s$ quark content in the left panel and $u\text{-}d\text{-}d$ quark content in the right panel.
For $u\text{-}d\text{-}d$-type operators, unlike the other cases, the inclusive bounds are stronger than the recasting ones for several operators across a wide range of ALP masses.
These inclusive searches combined with the recasting analysis of available experimental data enable us to constrain a broader set of operators and explore a larger ALP mass range.

\begin{table}[t]
\center
\resizebox{\linewidth}{!}{
\renewcommand{\arraystretch}{1.}
\begin{tabular}{|c|c|c|c|c|c|}
	\hline
	WCs/Process & \multicolumn{4}{c|}{Derived bound on the effective scale $\Lambda_{\tt eff}\equiv [C_i]^{-\frac{1}{4}}~(\text{GeV})$} &\makecell[c]{ Exp. lower bound \\ on 
    $\Gamma^{-1}~(10^{30}~\text{yr})$}
    \\
    \hline\hline 
	WCs & \cellcolor{gray!50} $[C_{\partial a euud}^{\tt VL,SL}]_{e112}$
	& \cellcolor{gray!50} $ [C_{\partial a eudu}^{\tt SR,VL}]_{e121}^-$
	& $ [C_{\partial a eudu}^{\tt SR,VL}]_{e121}^+$
	& $ [C_{\partial a euud}^{\tt SR,VL}]_{e112}$ 
	& 
	\\
	\hline
	$p \to e^+ a$ 
	& $ 2.2 \times 10^7 $  
	& $ 2.2 \times 10^7 $
	& $ 1.4 \times 10^7 $  
	& $ 1.4 \times 10^7 $
	& 790 (Exclusive)~\cite{Super-Kamiokande:2015pys}
	\\
	$p \to e^+ a$
    & $ 2.0 \times 10^7 $  
    & $ 2.0 \times 10^7 $ 
    & $ 1.3 \times 10^7 $ 
    & $ 1.3 \times 10^7 $
    & 364
    \\
	$p \to e^+ \pi^0 a$ 
	& $ 1.2 \times 10^7 $ 
	& $ 1.2 \times 10^7 $
	& $ 7.8 \times 10^6 $
	& $ 1.1 \times 10^7 $
    & $184~\big|~208~\big|~281~\big|~532$
	\\
	$p \to e^+ \eta a$ 
	& $ 9.3 \times 10^6 $  
	& $ 6.0 \times 10^6 $ 
	& $ 5.3 \times 10^6 $
	& $ 5.3 \times 10^6 $
    & $735~\big|~624~\big|~670~\big|~668$
	\\
	$n \to e^+ \pi^- a$ 
	& $ 6.4 \times 10^6 $ 
	& $ 6.4 \times 10^6 $ 
	& $ 5.4 \times 10^6 $ 
	& $ 4.3 \times 10^6 $ 
	& 0.6 (Inclusive)~\cite{ParticleDataGroup:2024cfk}
	\\
	\hline\hline
	WCs 
    & \cellcolor{gray!50} $ [C_{\partial a euud}^{\tt VL,SL}]_{\mu 112}  $ 
	& \cellcolor{gray!50} $ [C_{\partial a eudu}^{\tt SR,VL}]_{\mu 121}^-$
	& $ [C_{\partial a eudu}^{\tt SR,VL}]_{\mu 121}^+$
	& $ [C_{\partial a euud}^{\tt SR,VL}]_{\mu 112}$ 
	& 
	\\
	\hline
	$p \to \mu^+ a$ 
	& $ 2.0 \times 10^7 $ 
	& $ 2.0 \times 10^7 $
	& $ 1.3 \times 10^7 $ 
	& $ 1.3 \times 10^7 $
	&  410 (Exclusive)~\cite{Super-Kamiokande:2015pys}
    \\
	$p \to \mu^+ a$
    & $ 1.9 \times 10^7 $
    & $ 1.9 \times 10^7 $
    & $ 1.3 \times 10^7 $
    & $ 1.3 \times 10^7 $
    & 285
	\\
	$p \to \mu^+ \pi^0 a$ 
	& $ 1.1 \times 10^7 $ 
	& $ 1.1 \times 10^7 $ 
	& $ 6.9 \times 10^6 $
	& $ 1.0 \times 10^7 $
    & $~75~\big|~~74~\big|~114~\big|~198$
	\\
	$p \to \mu^+ \eta a$ 
	& $ 8.7 \times 10^6 $
	& $ 5.8 \times 10^6 $ 
	& $ 5.0 \times 10^6 $
	& $ 5.0 \times 10^6 $
    & $688~\big|~743~\big|~650~\big|~654$
	\\
	$n \to \mu^+ \pi^- a$ 
	& $ 9.2 \times 10^6 $ 
	& $ 9.2 \times 10^6 $ 
	& $ 7.8 \times 10^6 $ 
	& $ 6.2 \times 10^6 $ 
	& 12 (Inclusive)~\cite{ParticleDataGroup:2024cfk}
	\\
	\hline\hline
	WCs
    & $ [C_{\partial a euud}^{\tt VL,SL}]_{e113} $
	& $ [C_{\partial a eudu}^{\tt SR,VL}]_{e131}^-  $
	& $ [C_{\partial a eudu}^{\tt SR,VL}]_{e131}^+  $
	& $ [C_{\partial a euud}^{\tt SR,VL}]_{e113} $
	& 
	\\
	\hline
	$p \to e^+ K^0 a$ 
	& $ 3.7 \times 10^6 $
	& $ 4.0 \times 10^6 $ 
	& $ 3.0 \times 10^6 $ 
	& $ 3.0 \times 10^6 $
	&  0.6 (Inclusive)~\cite{ParticleDataGroup:2024cfk}
	\\
	\hline\hline
	WCs
    & $ [C_{\partial a euud}^{\tt VL,SL}]_{\mu 113}  $
	& $ [C_{\partial a eudu}^{\tt SR,VL}]_{\mu 131}^- $
	& $ [C_{\partial a eudu}^{\tt SR,VL}]_{\mu 131}^+ $
	& $ [C_{\partial a euud}^{\tt SR,VL}]_{\mu 113}$
	& 
	\\
	\hline
	$p \to \mu^+ K^0 a$ 
	& $ 5.8 \times 10^6 $ 
	& $ 6.3 \times 10^6 $ 
	& $ 4.7 \times 10^6 $ 
	& $ 4.7 \times 10^6 $
	& $31~~\big|~~35~~\big|~~36~~\big|~~36$
	\\
	\hline\hline
	WCs
    & \cellcolor{gray!50} $ [C_{\partial a \nu ddu}^{\tt VL,SL}]_{x221} $ 
	& \cellcolor{gray!50} $ [C_{\partial a \nu udd}^{\tt SR,VL}]_{x122}^-  $
	& $ [C_{\partial a \nu udd}^{\tt SR,VL}]_{x122}^+  $
	& $ [C_{\partial a \nu ddu}^{\tt SR,VL}]_{x221}  $
	&
	\\
	\hline
	$p \to \nu_x \pi^+ a$ 
	& $ 9.3 \times 10^6 $ 
	& $ 9.3 \times 10^6 $ 
	& $ 8.1 \times 10^6 $ 
	& $ 6.4 \times 10^6 $
	& $12~~\big|~~12~~\big|~~15~~\big|~~14$
	\\
	\hline\hline
	WCs
    & \cellcolor{gray!50} $ [C_{\partial a \nu ddu}^{\tt VL,SL}]_{x321} $ 
    & \cellcolor{gray!50} $ [C_{\partial a \nu udd}^{\tt SR,VL}]_{x132}^-  $
    & $ [C_{\partial a \nu udd}^{\tt SR,VL}]_{x132}^+  $
	& $ [C_{\partial a \nu ddu}^{\tt SR,VL}]_{x231}^+  $
	& \multirow{4}*{  5900 ($p \to \nu_x K^+$) \cite{Super-Kamiokande:2014otb} }
	\\
	\cline{1-5}
	$p \to \nu_x K^+ a$ 
	& $ 1.3 \times 10^7 $ 
	& $ 8.5 \times 10^6 $ 
	  & $ 1.1 \times 10^7 $
	& $ 9.3 \times 10^6 $ 
	&  
	\\
	\cline{1-5}
	WCs
    & \cellcolor{gray!50} $ [C_{\partial a \nu ddu}^{\tt VL,SL}]_{x231} $
    & \cellcolor{gray!50} $ [C_{\partial a \nu udd}^{\tt SR,VL}]_{x123}^-  $
    & $ [C_{\partial a \nu ddu}^{\tt SR,VL}]_{x231}^-  $
    & $ [C_{\partial a \nu udd}^{\tt SR,VL}]_{x123}^+  $
	&
	\\
	\cline{1-5}
	$p \to \nu_x K^+ a$ 
    & $ 8.5 \times 10^6 $
    & $ 1.3 \times 10^7 $ 
    & $ 1.2 \times 10^7 $
	& $ 9.3 \times 10^6 $
	& 
	\\
	\hline
\end{tabular}
}
\caption{Lower bounds on the effective scale of relevant dim-8 aLEFT operators in the massless ALP limit. The limits are identical for corresponding chirality-flipped operators with $\tL \leftrightarrow \tR$, and therefore, we omitted them for brevity. 
Experimental lower bounds without a reference in the last column correspond to the recasting bounds shown in \cref{fig:newbound} in the limit of $m_a\to0$. 
}
\label{tab:bound_massless_a}
\end{table}

To compare with the results in \cite{Li:2024liy},
\cref{tab:bound_massless_a} presents the lower bound on $\Lambda_{\tt eff}$ when $m_a\to 0$. WCs highlighted in dark gray correspond to those considered in that paper. The last column shows the experimental lower bounds on the inverse decay widths used to reach constraints on WCs listed in the front cells, including recasting bounds obtained in \cref{fig:newbound}.
All WCs in the fourth and fifth columns (except for the last one in the fourth column) correspond to operators belonging to the new chiral irrep $\pmb{3}_{\tL} \otimes \pmb{6}_{\tR}$. 
The derived energy scales associated with these operators are essentially of the same order as those for operators in the usual chiral irreps $\pmb{3}_{\tL} \otimes \bar{\pmb{3}}_{\tR}$ and $\pmb{8}_{\tL} \otimes \pmb{1}_{\tR}$.
The most stringent constraints for $uud$-type operators arise from the search for nucleon decays $N \to \ell^+ X$ in the Super-K experiment \cite{Super-Kamiokande:2015pys} and are consistent with those obtained in \cite{Li:2024liy}.
For effective scales associated with operators containing $u\text{-}d\text{-}s$ quarks, our results are five-to-six orders of magnitude stronger than those obtained in that paper based on the $\Lambda^0$-hyperon invisible search performed by BESIII \cite{BESIII:2021slv}. 
Derived bounds on $\Lambda_{\tt eff}$ associated with $[C^{\tt VL,SL}_{\partial a \nu ddu}]_{x221}$ and $[C^{\tt SR,VL}_{\partial a \nu udd}]^-_{x122}$ are slightly stronger.

\begin{table}[t]
\centering
\resizebox{\linewidth}{!}{
\renewcommand{\arraystretch}{1.1}
\begin{tabular}{|l|c|c|l|c|c|}
    \hline 
    \multirow{2}*{~~~~Mode}  
	&\multicolumn{2}{c|}{ Derived bounds on branching ratios }
	&\multirow{2}*{~~~~Mode}
	&\multicolumn{2}{c|}{ Derived bounds on inverse decay widths }
	\\\cline{2-3}\cline{5-6}
	& Br & aLEFT operator
	&
	& $\Gamma^{-1} (\rm yr) $ & aLEFT operator
	\\
	\hline%
	& $9.3\times 10^{-45}$
	& $[\calO_{\partial a euud}^{\tt VL,SL}]_{e113}$
	&
	& $2.2\times 10^{31}$
	& $[\calO_{\partial a \nu ddu}^{\tt VL,SL}]_{x221}, \, [\calO_{\partial a \nu udd}^{\tt SR,VL}]^-_{x122}$
	\\
	$~\Sigma^+ \to e^+ a$ 
	& $4.9\times 10^{-45}$
	& $[\calO_{\partial a eudu}^{\tt SR,VL}]^-_{e131}$
	& $~n \to \nu_x \pi^0 a$
	& $4.2\times 10^{30}$
	& $[\calO_{\partial a \nu ddu}^{\tt SR,VL}]_{x221}$
	\\
	& $3.0\times 10^{-45}$
	& $[\calO_{\partial a eudu}^{\tt SR,VL}]^+_{e131}, \, [\calO_{\partial a euud}^{\tt SR,VL}]_{e113} $
	&
	& $3.5\times 10^{32}$
	& $[\calO_{\partial a \nu udd}^{\tt SR,VL}]^+_{x122}$
	\\
	\hline%
	& $2.5\times 10^{-46}$
	& $[\calO_{\partial a euud}^{\tt VL,SL}]_{\mu 113}$
	&
	& $6.9\times 10^{32}$
	& $[\calO_{\partial a \nu ddu}^{\tt VL,SL}]_{x221}$
	\\
	$~\Sigma^+ \to \mu^+ a$ 
	& $1.3\times 10^{-46}$
	& $[\calO_{\partial a eudu}^{\tt SR,VL}]^-_{\mu 131}$
	& $~n \to \nu_x \eta a$ 
	& $1.9\times 10^{34}$
	& $[\calO_{\partial a \nu udd}^{\tt SR,VL}]^-_{x122}, \, [\calO_{\partial a \nu udd}^{\tt SR,VL}]^+_{x122} $
	\\
	& $8.2\times 10^{-47}$
	& $[\calO_{\partial a eudu}^{\tt SR,VL}]^+_{\mu 131}, \, [\calO_{\partial a euud}^{\tt SR,VL}]_{\mu 113} $
	& 
	& $2.9\times 10^{33}$
	& $[\calO_{\partial a \nu ddu}^{\tt SR,VL}]_{x221} $
	\\
	\hline%
	& $6.9\times 10^{-49}$
	& $[\calO_{\partial a \nu ddu}^{\tt VL,SL}]_{x321},\, [\calO_{\partial a \nu udd}^{\tt SR,VL}]^-_{x123}$
	&
	& $5.3\times 10^{33}$
	& $[\calO_{\partial a \nu ddu}^{\tt VL,SL}]_{x321},\, [\calO_{\partial a \nu udd}^{\tt SR,VL}]^-_{x123}$
	\\
	& $5.6\times 10^{-48}$
	& $[\calO_{\partial a \nu ddu}^{\tt VL,SL}]_{x231},\, [\calO_{\partial a \nu udd}^{\tt SR,VL}]^-_{x132}$
	&
	& $2.7\times 10^{32}$
	& $[\calO_{\partial a \nu ddu}^{\tt VL,SL}]_{x231}$
	\\
	$~\Lambda^0 \to \nu_x a$ 
	& $3.5\times 10^{-49}$
	& $[\calO_{\partial a \nu ddu}^{\tt SR,VL}]^-_{x231}$
	&
	& $4.4\times 10^{32}$
	& $[\calO_{\partial a \nu udd}^{\tt SR,VL}]^-_{x132}$
	\\
	& $1.3\times 10^{-48}$
	& $[\calO_{\partial a \nu ddu}^{\tt SR,VL}]^+_{x231}$
	& $~n \to \nu_x K^0 a$
	& $9.0\times 10^{34}$
	& $[\calO_{\partial a \nu ddu}^{\tt SR,VL}]^-_{x231}$
	\\
	& $3.3\times 10^{-49}$
	& $[\calO_{\partial a \nu udd}^{\tt SR,VL}]^+_{x132}$
	&
	& $1.3\times 10^{33}$
	& $[\calO_{\partial a \nu ddu}^{\tt SR,VL}]^+_{x231}$
	\\
	\cline{1-3}
	& $5.6\times 10^{-57}$
	& $[\calO_{\partial a \nu ddu}^{\tt VL,SL}]_{x231},\, [\calO_{\partial a \nu udd}^{\tt SR,VL}]^-_{x132}$
	&
	& $2.2\times 10^{34}$
	& $[\calO_{\partial a \nu udd}^{\tt SR,VL}]^+_{x132}$
	\\
	& $3.5\times 10^{-58}$
	& $[\calO_{\partial a \nu ddu}^{\tt SR,VL}]^-_{x231}$
	&
	& $5.8\times 10^{33}$
	& $[\calO_{\partial a \nu udd}^{\tt SR,VL}]^+_{x123}$
	\\ 
	\cline{4-6}
	$~\Sigma^0 \to \nu_x a$
	& $1.7\times 10^{-58}$
	& $[\calO_{\partial a \nu ddu}^{\tt SR,VL}]^+_{x231}$
	&
	&
	&
	\\
	& $6.7\times 10^{-58}$
	& $[\calO_{\partial a \nu udd}^{\tt SR,VL}]^+_{x123}$
	&
	&
	&
	\\
	& $4.3\times 10^{-59}$
	& $[\calO_{\partial a \nu udd}^{\tt SR,VL}]^+_{x132}$
	&
	&
	&
	\\
	\hline%
\end{tabular} }
\caption{Derived bounds on the hyperon two-body decay modes and neutron three-body decay modes involving a neutrino and a massless ALP based on constraints provided in \cref{tab:bound_massless_a}.
}
\label{tab:bound_hyperon}
\end{table}	

Finally, we utilize the constraints in \cref{tab:bound_massless_a} to derive new bounds on several hyperon and neutron BNV decay modes. In \cref{tab:bound_hyperon}, we present the derived bounds along with the corresponding operators whose constraints are used to obtain them. As shown in the table, the constraints obtained in \cref{tab:bound_massless_a} impose very stringent limits on the occurrence of these hyperon decay modes. However, the bounds on neutron decay modes are comparable to those for similar modes with the SM final states, making them promising targets for future experiments. 

\section{Summary}
\label{sec:Conclusion}

In this study, we systematically investigated two- and three-body BNV nucleon decays involving an invisible ALP. These decays are described within the framework of aLEFT, i.e., the LEFT framework extended by an ALP. By imposing shift symmetry on the ALP field, we identified relevant BNV aLEFT operators, which first appear at dimension 8 at the leading order. Then, we performed a chiral decomposition of these aLEFT operators under the QCD chiral group $\rm SU(3)_\tL\otimes SU(3)_\tR$ and identified corresponding spurion fields that enter the recently developed chiral perturbation theory for nucleon decays. With these spurion fields at hand, we conducted chiral matching for these BNV ALP interactions and derived general expressions for nucleon decays involving an ALP in terms of the associated Wilson coefficients and hadronic parameters in the chiral framework. Compared to the previous study in the literature, we considered the complete set of aLEFT BNV operators involving light $u,~d,~s$ quarks. We found that 12 aLEFT operators belonging to the new chiral irreps $\pmb{6}_{\tL(\tR)} \otimes \pmb{3}_{\tR(\tL)}$ contribute at the same chiral order as operators in the usual irreps $\pmb{8}_{\tL(\tR)} \otimes \pmb{1}_{\tR(\tL)}$ 
and $\bar{\pmb{3}}_{\tL(\tR)} \otimes \pmb{3}_{\tR(\tL)}$. Processes that change isospin by $3/2$ units, such as $n\to\pi^+\ell^- a$, can only be induced by operators in new irreps. In addition, we analyzed the momentum distributions of the charged lepton and mesons in three-body decay modes. Our results confirm that the distinct distribution behavior may help distinguish underlying operator structures and potentially determine the ALP mass in future experimental searches.

Based on this comprehensive theoretical framework, we simulated proton decay processes involving an ALP and recast existing Super-K data to set bounds on the inverse decay widths of these exotic modes. 
Although our treatment of the experimental recasts is simplified and conservative, we still obtain meaningful results because of the large number of protons in the experiment.
These results are complementary to current searches for nucleon decay modes involving only SM final states. The bounds we derived not only significantly improve upon inclusive limits set by  experiments nearly four decades ago, but also extend across a broad range of ALP masses. Using these results, we established conservative lower limits on the effective scales $\Lambda_{\tt eff}$ associated with the relevant Wilson coefficients. We found that the lower bounds on $\Lambda_{\tt eff}$ for operators corresponding to the new chiral irreps are of the same order as those for operators in the usual irreps. 
Finally, by employing these limits on $\Lambda_{\tt eff}$ in the massless ALP limit, we predicted new bounds on the occurrence of some BNV neutron and hyperon decay modes. Our results imposed very stringent limits on the branching ratios of exotic hyperon decay modes, while the projected bounds on neutron decays were within the reach of future neutrino experiments such as JUNO and DUNE.

\acknowledgments
This work was supported 
by the Grants 
No.\,NSFC-12035008, 
No.\,NSFC-12247151, 
and No.\,NSFC-12305110.  

\appendix

\section{General chiral terms}
\label{app:ChiL}

By expanding the pseudoscalar matrix in \cref{eq:chiB} to the zeroth order in the meson fields, we obtain the following general vertices  involving a baryon without any pseudoscalar mesons:
\begin{align}
{\cal L}_{{\cal P}B}
& = 
\Big\{ 
\big[ c_1 {\cal P}_{uud}^{\tL\tR} + c_2 {\cal P}_{uud}^{\tL\tL} 
+ c_3 \Lambda_\chi^{-1} ( {\cal P}_{uud}^{\tL\tR,\mu} - {\cal P}_{udu}^{\tL\tR,\mu} ) 
i\tilde{\partial}_\mu \big]p_\tL
 \notag\\
&
+ \big[ c_1 {\cal P}_{dud}^{\tL\tR} + c_2 {\cal P}_{dud}^{\tL\tL} 
- c_3 \Lambda_\chi^{-1} ({\cal P}_{ddu}^{\tL\tR,\mu} -{\cal P}_{udd}^{\tL\tR,\mu} ) 
i\tilde{\partial}_\mu \big] n_\tL
 \notag\\
&+ {1\over \sqrt{6} } \big[
  c_1 ( {\cal P}_{uds}^{\tL\tR} + {\cal P}_{dsu}^{\tL\tR}-2{\cal P}_{sud}^{\tL\tR})
+ c_2 ({\cal P}_{dsu}^{\tL\tL} - 2 {\cal P}_{sud}^{\tL\tL}) 
- 3 c_3 \Lambda_\chi^{-1} ({\cal P}_{usd}^{\tL\tR,\mu} - {\cal P}_{dsu}^{\tL\tR,\mu})
i \tilde{\partial}_\mu 
\big] \Lambda^0_\tL
\notag\\
& 
+{1\over \sqrt{2}} \big[ 
  c_1 ( {\cal P}_{uds}^{\tL\tR} - {\cal P}_{dsu}^{\tL\tR})
- c_2 {\cal P}_{dsu}^{\tL\tL} 
+ c_3 \Lambda_\chi^{-1} ( 2 {\cal P}_{uds}^{\tL\tR,\mu} - {\cal P}_{usd}^{\tL\tR,\mu}- {\cal P}_{dsu}^{\tL\tR,\mu} ) 
i\tilde{\partial}_\mu \big] \Sigma^0_\tL 
 \notag\\
& + \big[ c_1 {\cal P}_{usu}^{\tL\tR} + c_2 {\cal P}_{usu}^{\tL\tL} 
- c_3 \Lambda_\chi^{-1} ( {\cal P}_{uus}^{\tL\tR,\mu} - {\cal P}_{usu}^{\tL\tR,\mu} ) i\tilde{\partial}_\mu \big] \Sigma^+_\tL 
 \notag\\
& + \big[ c_1 {\cal P}_{dds}^{\tL\tR} + c_2 {\cal P}_{dds}^{\tL\tL} 
+ c_3 \Lambda_\chi^{-1} ({\cal P}_{dds}^{\tL\tR,\mu}-{\cal P}_{dsd}^{\tL\tR,\mu})
i\tilde{\partial}_\mu \big] 
\Sigma^-_\tL 
 \notag\\
& +\big[ c_1 {\cal P}_{ssu}^{\tL\tR} + c_2 {\cal P}_{ssu}^{\tL\tL} 
+ c_3 \Lambda_\chi^{-1} ( {\cal P}_{ssu}^{\tL\tR,\mu} - {\cal P}_{uss}^{\tL\tR,\mu} ) i\tilde{\partial}_\mu \big]\Xi^0_\tL 
 \notag\\
& + \big[ c_1 {\cal P}_{sds}^{\tL\tR} + c_2 {\cal P}_{sds}^{\tL\tL}  
- c_3 \Lambda_\chi^{-1} ( {\cal P}_{ssd}^{\tL\tR,\mu} - {\cal P}_{dss}^{\tL\tR,\mu} ) 
i\tilde{\partial}_\mu \big] \Xi^-_\tL \Big\} 
- \tL \leftrightarrow \tR. 
\label{eq:LPN}
\end{align} 
Similarly, expanding the pseudoscalar matrix in \cref{eq:chiB} to the first order in the meson fields yields the following vertices containing a nucleon and a meson:
\begin{align}
{\cal L}_{{\cal P}{\texttt N}M} 
& = {i \over F_0}\Big\{ 
  - \sqrt{2} \pi^+ c_3 \Lambda_\chi^{-1} {\cal P}_{uuu}^{\tL\tR,\mu} i \tilde{\partial}_\mu p_\tL 
  + \sqrt{2} \pi^- c_3 \Lambda_\chi^{-1} {\cal P}_{ddd}^{\tL\tR,\mu} i \tilde{\partial}_\mu n_\tL
\notag\\
&+  \frac{1}{\sqrt{2}} \pi^- \big[ 
c_1 {\cal P}_{dud}^{\tL\tR} 
+ c_2 {\cal P}_{dud}^{\tL\tL} 
- c_3 \Lambda_\chi^{-1} ({\cal P}_{ddu}^{\tL\tR,\mu}-3{\cal P}_{udd}^{\tL\tR,\mu})i\tilde{\partial}_\mu
\big] p_\tL    
\notag\\
&+  \frac{1}{\sqrt{2}} \pi^+ \big[ 
c_1 {\cal P}_{uud}^{\tL\tR} 
+ c_2 {\cal P}_{uud}^{\tL\tL} 
+ c_3 \Lambda_\chi^{-1} ({\cal P}_{uud}^{\tL\tR,\mu}-3{\cal P}_{udu}^{\tL\tR,\mu})i\tilde{\partial}_\mu
\big] n_\tL
\notag\\
& + \frac{1}{2} \pi^0 \big[ 
c_1 {\cal P}_{uud}^{\tL\tR} 
+ c_2 {\cal P}_{uud}^{\tL\tL} 
+ c_3 \Lambda_\chi^{-1} ({\cal P}_{udu}^{\tL\tR,\mu} + 3 {\cal P}_{uud}^{\tL\tR,\mu})i\tilde{\partial}_\mu   
\big] p_\tL    
\notag\\
& - \frac{1}{2} \pi^0 \big[
c_1 {\cal P}_{dud}^{\tL\tR} 
+ c_2 {\cal P}_{dud}^{\tL\tL} 
- c_3 \Lambda_\chi^{-1} (3{\cal P}_{ddu}^{\tL\tR,\mu} + {\cal P}_{udd}^{\tL\tR,\mu})i\tilde{\partial}_\mu   
\big]n_\tL
\notag\\
&-  \frac{1}{2\sqrt{3}} \eta \big[
c_1 {\cal P}_{uud}^{\tL\tR} 
-3 c_2 {\cal P}_{uud}^{\tL\tL} 
+ c_3 \Lambda_\chi^{-1} ({\cal P}_{udu}^{\tL\tR,\mu} - {\cal P}_{uud}^{\tL\tR,\mu}) i \tilde{\partial}_\mu
\big]p_\tL      
\notag\\
&-  \frac{1}{2\sqrt{3}} \eta \big[ 
c_1 {\cal P}_{dud}^{\tL\tR} 
-3 c_2 {\cal P}_{dud}^{\tL\tL} 
+ c_3 \Lambda_\chi^{-1} ({\cal P}_{ddu}^{\tL\tR,\mu} - {\cal P}_{udd}^{\tL\tR,\mu}) i \tilde{\partial}_\mu 
\big] n_\tL
\notag\\
&+ \frac{1}{\sqrt{2}} \bar{K}^0 \big[ 
c_1 {\cal P}_{usu}^{\tL\tR} 
- c_2 {\cal P}_{usu}^{\tL\tL} 
- c_3 \Lambda_\chi^{-1}({\cal P}_{usu}^{\tL\tR,\mu} + {\cal P}_{uus}^{\tL\tR,\mu})i\tilde{\partial}_\mu
\big]p_\tL 
\notag\\
& + \frac{1}{\sqrt{2}} K^- \big[
c_1 {\cal P}_{dds}^{\tL\tR} 
- c_2 {\cal P}_{dds}^{\tL\tL} 
+ c_3 \Lambda_\chi^{-1}({\cal P}_{dds}^{\tL\tR,\mu} + {\cal P}_{dsd}^{\tL\tR,\mu})i\tilde{\partial}_\mu
\big]n_\tL    
\notag\\
&+  \frac{1}{\sqrt{2}} K^- \big[ 
c_1 ( {\cal P}_{sud}^{\tL\tR} + {\cal P}_{uds}^{\tL\tR}) 
+ c_2 {\cal P}_{sud}^{\tL\tL} 
- c_3 \Lambda_\chi^{-1} ( {\cal P}_{dsu}^{\tL\tR,\mu}-{\cal P}_{uds}^{\tL\tR,\mu}
-2{\cal P}_{usd}^{\tL\tR,\mu}) i \tilde{\partial}_\mu 
\big] p_\tL 
\notag\\
&+ \frac{ 1}{\sqrt{2}} \bar{K}^0 
\big[ 
c_1 ( {\cal P}_{dsu}^{\tL\tR} + {\cal P}_{sud}^{\tL\tR} ) 
+ c_2 ( {\cal P}_{sud}^{\tL\tL} - {\cal P}_{dsu}^{\tL\tL} )
\notag\\
&- c_3 \Lambda_\chi^{-1} (2{\cal P}_{dsu}^{\tL\tR,\mu} + {\cal P}_{uds}^{\tL\tR,\mu} 
- {\cal P}_{usd}^{\tL\tR,\mu} ) i \tilde{\partial}_\mu  
\big]n_\tL \Big\}
+ \tL \leftrightarrow \tR.
\label{eq:LPNM}    
\end{align}
The absence of $K^+\texttt{N}$ and $K^0\texttt{N}$ terms is caused by the requirement of an anti-strange quark.

\section{Complete expressions for decay widths in the aLEFT}
\label{app:DW-aLEFT}

In this Appendix, we summarize our complete numerical results for decay widths expressed in terms of the WCs in the aLEFT. 
We denote the specific flavors in each WC by subscripts, with the first letter ($e,~\mu,~\tau$) indicating lepton flavors, and the other three numbers indicating quark flavors ($1,~2,~3$ represent $u,~d,~s$, respectively).  
For modes involving a neutrino, the subscript $x$ can be either $e$, $\mu$, or $\tau$. 
We neglect the ALP mass and numerically integrate all phase space factors associated with each WC squared. 
$c_3$ remains undetermined now, and therefore, we retain its explicit dependence through $\kappa_3(\equiv c_3/c_1)$ in these results.
To present the results more compactly, we removed the prefix ``$\partial a$'' from the subscripts of all relevant WCs.

For two-body decay processes, the complete results are
{\footnotesize
\begin{subequations}	
\begin{align}
{\Gamma_{p\to e^+ a} \over (0.1\rm GeV)^9} & =   
	1300 |[C_{ euud}^{\tt VL,SL}]_{e112}|^2
	+ 1300 |[C_{ eudu}^{\tt SL,VR}]^-_{e121}|^2 
	+ 50 \kappa_3^2 \big( |[C_{ eudu}^{\tt SL,VR}]^+_{e121}|^2
	+ |[C_{ euud}^{\tt SL,VR}]_{e112}|^2 \big) 
    \notag\\
	&+2600 \Re\big( [C_{ eudu}^{\tt SL,VR}]^-_{e121} [C_{ euud}^{\tt  VR,SR}]_{e112}^*  \big)
    -510 \kappa_3 \Re\big( ([C_{ eudu}^{\tt SL,VR}]^+_{e121} 
	-[C_{ euud}^{\tt SL,VR}]_{e112} )
	[C_{ euud}^{\tt VR,SR}]_{e112}^*  \big)
	\notag\\
	&- 510 \kappa_3 \Re\big( ([C_{ eudu}^{\tt SL,VR}]^+_{e121} 
	-[C_{ euud}^{\tt SL,VR}]_{e112} )
	[C_{ eudu}^{\tt SL,VR}]_{e121}^{-*}  \big)
    -99 \kappa_3^2 \Re\big( [C_{ eudu}^{\tt SL,VR}]^+_{e121} [C_{ euud}^{\tt  SL,VR}]_{e112}^*  \big) 
	\notag\\
	&- 0.6 \kappa_3 \Re\big( ([C_{ euud}^{\tt SL,VR}]_{e112} 
	-[C_{ eudu}^{\tt SL,VR}]_{e121}^+ )
	[C_{ euud}^{\tt VL,SL}]_{e112}^{*}  \big)
    + 0.2 \kappa_3^2 \Re\big( [C_{ eudu}^{\tt SL,VR}]_{e121}^+ 
	[C_{ euud}^{\tt SR,VL}]_{e112}^{*}  \big) 
	\notag\\
	&- 0.5 \kappa_3 \Re\big( ([C_{ euud}^{\tt SL,VR}]_{e112} 
	-[C_{ eudu}^{\tt SL,VR}]_{e121}^+ )
	[C_{ eudu}^{\tt SR,VL}]_{e121}^{-*}  \big)
	- 0.1 \kappa_3^2 \Re\big( [C_{ eudu}^{\tt SL,VR}]_{e121}^+ 
	[C_{ eudu}^{\tt SR,VL}]_{e121}^{+*}\big)
    \notag\\
	&- 0.1 \kappa_3^2 \Re\big( [C_{ euud}^{\tt SL,VR}]_{e112}
	[C_{ euud}^{\tt SR,VL}]_{e112}^{*}\big)+ \tL \leftrightarrow \tR,
\\
	{\Gamma_{p\to \mu^+ a} \over (0.1\rm GeV)^9} & =
	1300 |[C_{ euud}^{\tt VL,SL}]_{\mu 112}|^2
	+ 1200 |[C_{ eudu}^{\tt SL,VR}]^-_{\mu 121}|^2
    + 50 \kappa_3^2 \big( |[C_{ eudu}^{\tt SL,VR}]^+_{\mu 121}|^2
	+ |[C_{ euud}^{\tt SL,VR}]_{\mu 112}|^2 \big) 
	\notag\\
	&+ 2500 \Re\big( [C_{ eudu}^{\tt SL,VR}]^-_{\mu 121} [C_{ euud}^{\tt  VR,SR}]_{\mu 112}^*  \big)
	- 490 \kappa_3 \Re\big( ([C_{ eudu}^{\tt SL,VR}]^+_{\mu 121} 
	-[C_{ euud}^{\tt SL,VR}]_{\mu 112} )
	[C_{ euud}^{\tt VR,SR}]_{\mu 112}^*  \big)
	\notag\\
	&- 490 \kappa_3 \Re\big( ([C_{ eudu}^{\tt SL,VR}]^+_{\mu 121} 
	-[C_{ euud}^{\tt SL,VR}]_{\mu 112} )
	[C_{ eudu}^{\tt SL,VR}]_{\mu 121}^{-*}  \big)
    -100 \kappa_3^2 \Re\big( [C_{ eudu}^{\tt SL,VR}]^+_{\mu 121} [C_{ euud}^{\tt  SL,VR}]_{\mu112}^*  \big)
	\notag\\
	&- 110 \kappa_3 \Re\big( ([C_{ euud}^{\tt SL,VR}]_{\mu 112} 
	-[C_{ eudu}^{\tt SL,VR}]^+_{\mu 121} )
	[C_{ euud}^{\tt VL,SL}]_{\mu 112}^{*}  \big)
    + 43 \kappa_3^2 \Re\big( [C_{ eudu}^{\tt SL,VR}]_{\mu 121}^+ 
	[C_{ euud}^{\tt SR,VL}]_{\mu 112}^{*}  \big)
	\notag\\
	&- 110 \kappa_3 \Re\big( ([C_{ euud}^{\tt SL,VR}]_{\mu 112} 
	-[C_{ eudu}^{\tt SL,VR}]_{\mu 121}^+ )
	[C_{ eudu}^{\tt SR,VL}]_{\mu 121}^{-*}  \big)
    - 22 \kappa_3^2 \Re\big( [C_{ eudu}^{\tt SL,VR}]_{\mu 121}^+ 
	[C_{ eudu}^{\tt SR,VL}]_{\mu 121}^{+*} \big)
	\notag\\
	&- 22 \kappa_3^2 \Re\big( [C_{ euud}^{\tt SL,VR}]_{\mu 112}
	[C_{ euud}^{\tt SR,VL}]_{\mu 112}^{*}  \big)    + \tL \leftrightarrow \tR,
\\%
	{\Gamma_{\Sigma^+ \to e^+ a} \over (0.1\rm GeV)^9} & = 
	2700 |[C_{ euud}^{\tt VL,SL}]_{e113}|^2
	+ 2600 |[C_{ eudu}^{\tt SL,VR}]^-_{e131}|^2
	+ 160 \kappa_3^2 \big( |[C_{ eudu}^{\tt SL,VR}]^+_{e131}|^2
	+ |[C_{ euud}^{\tt SL,VR}]_{e113}|^2  \big) 
	\notag\\
	&+5300 \Re\big( [C_{ eudu}^{\tt SL,VR}]^-_{e131} [C_{ euud}^{\tt  VR,SR}]_{e113}^* \big)
    - 1300 \kappa_3 \Re\big( ([C_{ eudu}^{\tt SL,VR}]^+_{e131} 
	-[C_{ euud}^{\tt SL,VR}]_{e113} )
	[C_{ euud}^{\tt VR,SR}]_{e113}^*  \big)
	\notag\\
	&- 1300 \kappa_3 \Re\big( ([C_{ eudu}^{\tt SL,VR}]^+_{e131} 
	-[C_{ euud}^{\tt SL,VR}]_{e113} )
	[C_{ eudu}^{\tt SL,VR}]_{e131}^{-*}  \big)
    -330 \kappa_3^2 \Re\big( [C_{ eudu}^{\tt SL,VR}]^+_{e131} [C_{ euud}^{\tt  SL,VR}]_{e113}^* \big)
	\notag\\
	&- 1.1 \kappa_3 \Re\big( ([C_{ euud}^{\tt SL,VR}]_{e113} 
	-[C_{ eudu}^{\tt SL,VR}]_{e131}^+ )
	([C_{ eudu}^{\tt SR,VL}]_{e131}^{-*}
    + [C_{ euud}^{\tt VL,SL}]_{e113}^{*} )  \big)
	\notag\\
	&+ 0.6 \kappa_3^2 \Re\big( [C_{ eudu}^{\tt SL,VR}]_{e131}^+ 
	[C_{ euud}^{\tt SR,VL}]_{e113}^{*}  \big)
    - 0.3 \kappa_3^2 \Re\big( [C_{ eudu}^{\tt SL,VR}]_{e131}^+ 
	[C_{ eudu}^{\tt SR,VL}]_{e131}^{+*}\big)
	\notag\\
	&- 0.3 \kappa_3^2 \Re\big([C_{ euud}^{\tt SL,VR}]_{e113}
	[C_{ euud}^{\tt SR,VL}]_{e113}^{*}  \big) + \tL \leftrightarrow \tR,
\\%
	{\Gamma_{\Sigma^+ \to \mu^+ a} \over (0.1\rm GeV)^9} &=
	2600 |[C_{ euud}^{\tt VL,SL}]_{\mu 113}|^2
	+ 2600 |[C_{ eudu}^{\tt SL,VR}]^-_{\mu 131}|^2 
	+ 160 \kappa_3^2 \big( |[C_{ eudu}^{\tt SL,VR}]^+_{\mu 131}|^2
	+ |[C_{ euud}^{\tt SL,VR}]_{\mu 113}|^2 \big) 
	\notag\\
	&+ 5200 \Re\big( [C_{ eudu}^{\tt SL,VR}]^-_{\mu 131} [C_{ euud}^{\tt  VR,SR}]_{\mu 113}^* \big)
    - 1300 \kappa_3 \Re\big( ([C_{ eudu}^{\tt SL,VR}]^+_{\mu 131} 
	-[C_{ euud}^{\tt SL,VR}]_{\mu 113} )
	[C_{ euud}^{\tt VR,SR}]_{\mu 113}^*  \big)
	\notag\\
	&- 1300 \kappa_3 \Re\big( ([C_{ eudu}^{\tt SL,VR}]^+_{\mu 131} 
	-[C_{ euud}^{\tt SL,VR}]_{\mu 113} )
	[C_{ eudu}^{\tt SL,VR}]_{\mu 131}^{-*}  \big)
	- 330 \kappa_3^2 \Re\big( [C_{ eudu}^{\tt SL,VR}]^+_{\mu 131} 
    [C_{ euud}^{\tt  SL,VR}]_{\mu 113}^* \big)
     \notag\\
	&- 230 \kappa_3 \Re\big( ([C_{ euud}^{\tt SL,VR}]_{\mu 113} 
	-[C_{ eudu}^{\tt SL,VR}]_{\mu 131}^+ )
	[C_{ euud}^{\tt VL,SL}]_{\mu 113}^{*}  \big)
	+ 110 \kappa_3^2 \Re\big( [C_{ eudu}^{\tt SL,VR}]_{\mu 131}^+ 
	[C_{ euud}^{\tt SR,VL}]_{\mu 113}^{*}  \big)
    \notag\\
	&- 230 \kappa_3 \Re\big( ([C_{ euud}^{\tt SL,VR}]_{\mu 113} 
	-[C_{ eudu}^{\tt SL,VR}]_{\mu 131}^+ )
	[C_{ eudu}^{\tt SR,VL}]_{\mu 131}^{-*}  \big)
    - 56 \kappa_3^2 \Re\big( [C_{ eudu}^{\tt SL,VR}]_{\mu 131}^+ 
	[C_{ eudu}^{\tt SR,VL}]_{\mu 131}^{+*}\big)
	\notag\\
	&- 56 \kappa_3^2 \Re\big(
	[C_{ euud}^{\tt SL,VR}]_{\mu 113}
	[C_{ euud}^{\tt SR,VL}]_{\mu 113}^{*}  \big) + \tL \leftrightarrow \tR,
\\%
	{\Gamma_{n \to \nu_x a} \over (0.1\rm GeV)^9} &=
	1300 |[C_{ \nu ddu}^{\tt VL,SL}]_{x221}|^2
	+ 1300 |[C_{ \nu udd}^{\tt SR,VL}]^-_{x122}|^2
	+ 50 \kappa_3^2 \big( |[C_{ \nu ddu}^{\tt SR,VL}]_{x221}|^2
	+ |[C_{ \nu udd}^{\tt SR,VL}]^+_{x122}|^2 \big)
	\notag\\
	&- 2600 \Re\big( [C_{ \nu udd}^{\tt SR,VL}]_{x122}^-
	[C_{ \nu ddu}^{\tt VL,SL}]_{x221}^{*} \big)
	- 520 \kappa_3 \Re\big( ([C_{ \nu udd}^{\tt SR,VL}]^+_{x122} 
	-[C_{ \nu ddu}^{\tt SR,VL}]_{x221} )
	[C_{ \nu ddu}^{\tt VL,SL}]_{x221}^{*}  \big)
	\notag\\
	&- 510 \kappa_3 \Re\big( ([C_{ \nu ddu}^{\tt SR,VL}]_{x221} 
	-[C_{ \nu udd}^{\tt SR,VL}]^+_{x122} )
	[C_{ \nu udd}^{\tt SR,VL}]_{x122}^{-*}  \big)
    - 100 \kappa_3^2 \Re\big( [C_{ \nu udd}^{\tt SR,VL}]_{x122}^+
	[C_{ \nu ddu}^{\tt SR,VL}]_{x221}^{*} \big),
\\%
	{\Gamma_{\Lambda^0 \to \nu_x a} \over (0.1\rm GeV)^9} & =
	1500 |[C_{ \nu ddu}^{\tt VL,SL}]_{x321}|^2
	+ 1500 |[C_{ \nu udd}^{\tt SR,VL}]^-_{x123}|^2
	+ 370 |[C_{ \nu ddu}^{\tt VL,SL}]_{x231}|^2
	\notag\\
	&+ 360 \big( |[C_{ \nu ddu}^{\tt SR,VL}]^-_{x231}|^2
	+ |[C_{ \nu udd}^{\tt SR,VL}]^-_{x132}|^2 \big)
	+ 180 \kappa_3^2 \big( |[C_{ \nu ddu}^{\tt SR,VL}]^+_{x231}|^2
	+ |[C_{ \nu udd}^{\tt SR,VL}]^+_{x132}|^2 \big)
	\notag\\
	&- 2900 \Re\big( [C_{ \nu udd}^{\tt SR,VL}]_{x123}^-
	[C_{ \nu ddu}^{\tt VL,SL}]_{x321}^{*} \big)
	+ 1500 \Re\big( [C_{ \nu ddu}^{\tt VL,SL}]_{x231}
	[C_{ \nu ddu}^{\tt VL,SL}]_{x321}^{*} \big)
	\notag\\
	&+ 1500 \Re\big( ([C_{ \nu ddu}^{\tt SR,VL}]^-_{x231} 
	-[C_{ \nu udd}^{\tt SR,VL}]^-_{x132} )
	[C_{ \nu ddu}^{\tt VL,SL}]_{x321}^{*}  \big)
	- 1500 \Re\big( [C_{ \nu udd}^{\tt SR,VL}]^-_{x123} 
	[C_{ \nu ddu}^{\tt VL,SL}]_{x231}^{*}  \big)
	\notag\\
	&+ 1500 \Re\big( ([C_{ \nu udd}^{\tt SR,VL}]^-_{x132} 
	-[C_{ \nu ddu}^{\tt SR,VL}]^-_{x231} )
	[C_{ \nu udd}^{\tt SR,VL}]_{x123}^{-*}  \big)
	\notag\\
	&- 1000 \kappa_3 \Re\big( ([C_{ \nu udd}^{\tt SR,VL}]^+_{x132} 
	-[C_{ \nu ddu}^{\tt SR,VL}]^+_{x231} )
	[C_{ \nu ddu}^{\tt VL,SL}]_{x321}^{*}  \big)
	\notag\\
	&- 1000 \kappa_3 \Re\big( ([C_{ \nu ddu}^{\tt SR,VL}]^+_{x231} 
	-[C_{ \nu udd}^{\tt SR,VL}]^+_{x132} )
	[C_{ \nu udd}^{\tt SR,VL}]_{x123}^{-*}  \big)
	\notag\\
	&+ 730 \Re\big( ([C_{ \nu ddu}^{\tt SR,VL}]^-_{x231} 
	-[C_{ \nu udd}^{\tt SR,VL}]^-_{x132} )
	[C_{ \nu ddu}^{\tt VL,SL}]_{x231}^{*}  \big)
	- 730 \Re\big( [C_{ \nu ddu}^{\tt SR,VL}]^-_{x231} 
	[C_{ \nu udd}^{\tt SR,VL}]_{x132}^{-*}  \big)
	\notag\\
	&- 510 \kappa_3 \Re\big( ([C_{ \nu udd}^{\tt SR,VL}]^+_{x132} 
	-[C_{ \nu ddu}^{\tt SR,VL}]^+_{x231} )
	[C_{ \nu ddu}^{\tt VL,SL}]_{x231}^{*}  \big)
	- 350 \kappa_3^2 \Re\big( [C_{ \nu ddu}^{\tt SR,VL}]_{x231}^+
	[C_{ \nu udd}^{\tt SR,VL}]_{x132}^{+*} \big)
    \notag\\
	&- 510 \kappa_3 \Re\big( ([C_{ \nu udd}^{\tt SR,VL}]^+_{x132} 
	-[C_{ \nu ddu}^{\tt SR,VL}]^+_{x231} )
	([C_{ \nu ddu}^{\tt SR,VL}]_{x231}^{-*}
	-[C_{ \nu udd}^{\tt SR,VL}]_{x132}^{-*} )  \big),
\\%
	{\Gamma_{\Sigma^0 \to \nu_x a} \over (0.1\rm GeV)^9} &=
	1400 |[C_{ \nu ddu}^{\tt VL,SL}]_{x231}|^2
	+ 1300 \big( |[C_{ \nu ddu}^{\tt SR,VL}]^-_{x231}|^2
	+ |[C_{ \nu udd}^{\tt SR,VL}]^-_{x132}|^2 \big)
	\notag\\
	&+ 300 \kappa_3^2 |[C_{ \nu udd}^{\tt SR,VL}]_{x123}^+|^2
	+ 82 \kappa_3^2 \big( |[C_{ \nu ddu}^{\tt SR,VL}]^+_{x231}|^2
	+ |[C_{ \nu udd}^{\tt SR,VL}]^+_{x132}|^2 \big)
	\notag\\
	&- 2700 \Re\big( ([C_{ \nu ddu}^{\tt SR,VL}]^-_{x231} 
	+[C_{ \nu udd}^{\tt SR,VL}]^-_{x132} )
	[C_{ \nu ddu}^{\tt VL,SL}]_{x231}^{*}  \big)
	+ 2700 \Re\big( [C_{ \nu ddu}^{\tt SR,VL}]^-_{x231} 
	[C_{ \nu udd}^{\tt SR,VL}]_{x132}^{-*}  \big)
	\notag\\
	&- 1300 \kappa_3 \Re\big( [C_{ \nu udd}^{\tt SR,VL}]^+_{x123} 
	[C_{ \nu ddu}^{\tt VL,SL}]_{x231}^{*}  \big)
	+ 1300 \kappa_3 \Re\big( ([C_{ \nu ddu}^{\tt SR,VL}]^-_{x231} 
	+[C_{ \nu udd}^{\tt SR,VL}]^-_{x132} )
	[C_{ \nu udd}^{\tt SR,VL}]_{x123}^{+*}  \big)
	\notag\\
	&+ 670 \kappa_3 \Re\big( ([C_{ \nu ddu}^{\tt SR,VL}]^+_{x231} 
	+[C_{ \nu udd}^{\tt SR,VL}]^+_{x132} )
	[C_{ \nu ddu}^{\tt VL,SL}]_{x231}^{*}  \big)
	+ 160 \kappa_3^2 \Re\big( [C_{ \nu ddu}^{\tt SR,VL}]_{x231}^+
	[C_{ \nu udd}^{\tt SR,VL}]_{x132}^{+*} \big)
    \notag\\
	&- 660 \kappa_3 \Re\big( ([C_{ \nu udd}^{\tt SR,VL}]^+_{x132} 
	+[C_{ \nu ddu}^{\tt SR,VL}]^+_{x231} )
	([C_{ \nu ddu}^{\tt SR,VL}]_{x231}^{-*}
	+[C_{ \nu udd}^{\tt SR,VL}]_{x132}^{-*} )  \big)
	\notag\\
	&- 330 \kappa_3^2 \Re\big( ([C_{ \nu ddu}^{\tt SR,VL}]^+_{x231} 
	+[C_{ \nu udd}^{\tt SR,VL}]^+_{x132} )
	[C_{ \nu udd}^{\tt SR,VL}]_{x123}^{+*}  \big),
\\%
	{\Gamma_{\Xi^0 \to \nu_x a} \over (0.1\rm GeV)^9} &=
	3600 |[C_{ \nu ddu}^{\tt VL,SL}]_{x331}|^2
	+ 3600 |[C_{ \nu udd}^{\tt SR,VL}]^-_{x133}|^2
	+ 270 \kappa_3^2 \big( |[C_{ \nu ddu}^{\tt SR,VL}]_{x331}|^2
	+ |[C_{ \nu udd}^{\tt SR,VL}]^+_{x133}|^2 \big)
	\notag\\
	&- 7200 \Re\big( [C_{ \nu udd}^{\tt SR,VL}]_{x133}^-
	[C_{ \nu ddu}^{\tt VL,SL}]_{x331}^{*} \big)
    - 2000 \kappa_3 \Re\big( ([C_{ \nu udd}^{\tt SR,VL}]^+_{x133} 
	-[C_{ \nu ddu}^{\tt SR,VL}]_{x331} )
	[C_{ \nu ddu}^{\tt VL,SL}]_{x331}^{*}  \big)
	\notag\\
	&- 2000 \kappa_3 \Re\big( ([C_{ \nu ddu}^{\tt SR,VL}]_{x331} 
	-[C_{ \nu udd}^{\tt SR,VL}]^+_{x133} )
	[C_{ \nu udd}^{\tt SR,VL}]_{x133}^{-*}  \big)
    - 540 \kappa_3^2 \Re\big( [C_{ \nu udd}^{\tt SR,VL}]_{x133}^+
	[C_{ \nu ddu}^{\tt SR,VL}]_{x331}^{*} \big),
\\%
	{\Gamma_{\Sigma^- \to e^- a} \over (0.1\rm GeV)^9} &=
	2800 |[C_{ eddd}^{\tt VL,SL}]_{e223}|^2
	+ 2700 |[C_{ eddd}^{\tt SL,VR}]^-_{e232}|^2
	+ 170 \kappa_3^2 \big( |[C_{ eddd}^{\tt SL,VR}]^+_{e232}|^2
	+ |[C_{ eddd}^{\tt SL,VR}]_{e223}|^2 \big) 
	\notag\\
	&+ 5400 \Re\big( [C_{ eddd}^{\tt SL,VR}]^-_{e232} [C_{ eddd}^{\tt  VR,SR}]_{e223}^* \big)
    - 1400 \kappa_3 \Re\big( ([C_{ eddd}^{\tt SL,VR}]^+_{e232} 
	-[C_{ eddd}^{\tt SL,VR}]_{e223} )
	[C_{ eddd}^{\tt VR,SR}]_{e223}^*   \big)
	\notag\\
	&- 1300 \kappa_3 \Re\big( ([C_{ eddd}^{\tt SL,VR}]^+_{e232} 
	-[C_{ eddd}^{\tt SL,VR}]_{e223} )
	[C_{ eddd}^{\tt SL,VR}]_{e232}^{-*}  \big)
    - 340 \kappa_3^2 \Re\big( [C_{ eddd}^{\tt SL,VR}]^+_{e232} [C_{ eddd}^{\tt  SL,VR}]_{e223}^* \big)
    \notag\\
	&- 1.2 \kappa_3 \Re\big( ([C_{ eddd}^{\tt SL,VR}]_{e223} 
	-[C_{ eddd}^{\tt SL,VR}]_{e232}^+ )
	[C_{ eddd}^{\tt VL,SL}]_{e223}^{*}  \big)
	+ 0.6 \kappa_3^2 \Re\big( [C_{ eddd}^{\tt SL,VR}]_{e223} 
	[C_{ eddd}^{\tt SR,VL}]_{e232}^{+*}  \big)
    \notag\\
	&- 1.1 \kappa_3 \Re\big( ([C_{ eddd}^{\tt SL,VR}]_{e223} 
	-[C_{ eddd}^{\tt SL,VR}]_{e232}^+ )
	[C_{ eddd}^{\tt SR,VL}]_{e232}^{-*}  \big)
	- 0.3 \kappa_3^2 \Re\big( [C_{ eddd}^{\tt SL,VR}]_{e223} 
	[C_{ eddd}^{\tt SR,VL}]_{e223}^{*}\big)
	\notag\\
	&- 0.3 \kappa_3^2 \Re\big([C_{ eddd}^{\tt SL,VR}]_{e232}^+ 
	[C_{ eddd}^{\tt SR,VL}]_{e232}^{+*}  \big)+ \tL \leftrightarrow \tR ,
\\%
	{\Gamma_{\Sigma^- \to \mu^- a} \over (0.1\rm GeV)^9} &=
	2700 |[C_{ eddd}^{\tt VL,SL}]_{\mu 223}|^2
	+ 2600 |[C_{ eddd}^{\tt SL,VR}]^-_{\mu 232}|^2
	+ 170 \kappa_3^2 \big( |[C_{ eddd}^{\tt SL,VR}]^+_{\mu 232}|^2
	+ |[C_{ eddd}^{\tt SL,VR}]_{\mu 223}|^2 \big) 
	\notag\\
	&+ 5300 \Re\big( [C_{ eddd}^{\tt SL,VR}]^-_{\mu 232} [C_{ eddd}^{\tt  VR,SR}]_{\mu 223}^* \big)
	- 1300 \kappa_3 \Re\big( ([C_{ eddd}^{\tt SL,VR}]^+_{\mu 232} 
	-[C_{ eddd}^{\tt SL,VR}]_{\mu 223} )
	[C_{ eddd}^{\tt VR,SR}]_{\mu 223}^*  \big)
	\notag\\
	&- 1300 \kappa_3 \Re\big( ([C_{ eddd}^{\tt SL,VR}]^+_{\mu 232} 
	-[C_{ eddd}^{\tt SL,VR}]_{\mu 223} )
	[C_{ eddd}^{\tt SL,VR}]_{\mu 232}^{-*}  \big)
	- 340 \kappa_3^2 \Re\big( [C_{ eddd}^{\tt SL,VR}]^+_{\mu 232} 
    [C_{ eddd}^{\tt  SL,VR}]_{\mu 223}^* \big)
    \notag\\
	&- 230 \kappa_3 \Re\big( ([C_{ eddd}^{\tt SL,VR}]_{\mu 223} 
	-[C_{ eddd}^{\tt SL,VR}]_{\mu 232}^+ )
	[C_{ eddd}^{\tt VL,SL}]_{\mu 223}^{*}  \big)
	+ 120 \kappa_3^2 \Re\big( [C_{ eddd}^{\tt SL,VR}]_{\mu 223} 
	[C_{ eddd}^{\tt SR,VL}]_{\mu 232}^{+*} \big)
    \notag\\
	&- 230 \kappa_3  \Re\big( ([C_{ eddd}^{\tt SL,VR}]_{\mu 223} 
	-[C_{ eddd}^{\tt SL,VR}]_{\mu 232}^+ )
	[C_{ eddd}^{\tt SR,VL}]_{\mu 232}^{-*}  \big)
	- 58 \kappa_3^2 \Re\big( [C_{ eddd}^{\tt SL,VR}]_{\mu 223} 
	[C_{ eddd}^{\tt SR,VL}]_{\mu 223}^{*}\big) 
	\notag\\
	&- 58 \kappa_3^2 \Re\big( [C_{ eddd}^{\tt SL,VR}]_{\mu 232}^+ 
	[C_{ eddd}^{\tt SR,VL}]_{\mu 232}^{+*}  \big)  + \tL \leftrightarrow \tR ,
\\%
	{\Gamma_{\Xi^- \to e^- a} \over (0.1\rm GeV)^9} &=
	3700  |[C_{ eddd}^{\tt VL,SL}]_{e323}|^2
	+ 3600  |[C_{ eddd}^{\tt SL,VR}]^-_{e233}|^2
	+ 270 \kappa_3^2 \big( |[C_{ eddd}^{\tt SL,VR}]^+_{e233}|^2
	+ |[C_{ eddd}^{\tt SL,VR}]_{e332}|^2 \big) 
	\notag\\
	&+7300 \Re\big( [C_{ eddd}^{\tt SL,VR}]^-_{e233} [C_{ eddd}^{\tt  VR,SR}]_{e323}^* \big)
    - 2000 \kappa_3 \Re\big( ([C_{ eddd}^{\tt SL,VR}]_{e332} 
	- [C_{ eddd}^{\tt SL,VR}]^+_{e233} )
	[C_{ eddd}^{\tt VR,SR}]_{e323}^*  \big)
	\notag\\
	&- 2000 \kappa_3 \Re\big( ( [C_{ eddd}^{\tt SL,VR}]_{e332} 
	- [C_{ eddd}^{\tt SL,VR}]^+_{e233} )
	[C_{ eddd}^{\tt SL,VR}]_{e233}^{-*}  \big)
	- 550 \kappa_3^2 \Re\big( [C_{ eddd}^{\tt SL,VR}]^+_{e233} [C_{ eddd}^{\tt  SL,VR}]_{e332}^* \big)
    \notag\\
	&- 1.6 \kappa_3 \Re\big( ([C_{ eddd}^{\tt SL,VR}]_{e233}^+
	- [C_{ eddd}^{\tt SL,VR}]_{e332}  )
	[C_{ eddd}^{\tt VL,SL}]_{e323}^{*}  \big)
	+ 0.8 \kappa_3^2 \Re\big( [C_{ eddd}^{\tt SL,VR}]_{e332} 
	[C_{ eddd}^{\tt SR,VL}]_{e233}^{+*}  \big)
    \notag\\
	&- 1.5 \kappa_3 \Re\big( ([C_{ eddd}^{\tt SL,VR}]_{e233}^+
	- [C_{ eddd}^{\tt SL,VR}]_{e332}  )
	[C_{ eddd}^{\tt SR,VL}]_{e233}^{-*}  \big)
	- 0.4 \kappa_3^2 \Re\big( [C_{ eddd}^{\tt SL,VR}]_{e332} 
	[C_{ eddd}^{\tt SR,VL}]_{e332}^{*}\big) 
    \notag\\
	&- 0.4 \kappa_3^2 \Re\big(
	[C_{ eddd}^{\tt SL,VR}]_{e233}^+ 
	[C_{ eddd}^{\tt SR,VL}]_{e233}^{+*}  \big)+ \tL \leftrightarrow \tR  ,
\\%
	{\Gamma_{\Xi^- \to \mu^- a} \over (0.1\rm GeV)^9} &=
	3600 |[C_{ eddd}^{\tt VL,SL}]_{\mu 323}|^2
	+ 3600 |[C_{ eddd}^{\tt SL,VR}]^-_{\mu 233}|^2
	+ 280 \kappa_3^2 \big( |[C_{ eddd}^{\tt SL,VR}]^+_{\mu 233}|^2
	+ |[C_{ eddd}^{\tt SL,VR}]_{\mu 332}|^2 \big) 
	\notag\\
	&+ 7200 \Re\big( [C_{ eddd}^{\tt SL,VR}]^-_{\mu 233} [C_{ eddd}^{\tt  VR,SR}]_{\mu 323}^* \big)
    - 2000 \kappa_3 \Re\big( ([C_{ eddd}^{\tt SL,VR}]_{\mu 332} 
	- [C_{ eddd}^{\tt SL,VR}]^+_{\mu 233} )
	[C_{ eddd}^{\tt VR,SR}]_{\mu 323}^*  \big)
	\notag\\
	&- 2000 \kappa_3 \Re\big( ( [C_{ eddd}^{\tt SL,VR}]_{\mu 332} 
	- [C_{ eddd}^{\tt SL,VR}]^+_{\mu 233} )
	[C_{ eddd}^{\tt SL,VR}]_{\mu 233}^{-*}  \big)
	- 550 \kappa_3^2 \Re\big( [C_{ eddd}^{\tt SL,VR}]^+_{\mu 233} 
    [C_{ eddd}^{\tt  SL,VR}]_{\mu 332}^* \big)\notag\\
	&- 320 \kappa_3 \Re\big( ([C_{ eddd}^{\tt SL,VR}]_{\mu 233}^+
	- [C_{ eddd}^{\tt SL,VR}]_{\mu 332}  )
	[C_{ eddd}^{\tt VL,SL}]_{\mu 323}^{*}  \big)
    + 170 \kappa_3^2 \Re\big( [C_{ eddd}^{\tt SL,VR}]_{\mu 332} 
	[C_{ eddd}^{\tt SR,VL}]_{\mu 233}^{+*}  \big) 
	\notag\\
	&- 310 \kappa_3 \Re\big( ([C_{ eddd}^{\tt SL,VR}]_{\mu 233}^+
	- [C_{ eddd}^{\tt SL,VR}]_{\mu 332}  )
	[C_{ eddd}^{\tt SR,VL}]_{\mu 233}^{-*}  \big)
    - 86 \kappa_3^2 \Re\big( [C_{ eddd}^{\tt SL,VR}]_{\mu 332} 
	[C_{ eddd}^{\tt SR,VL}]_{\mu 332}^{*}\big)
	\notag\\
	&- 86 \kappa_3^2 \Re\big([C_{ eddd}^{\tt SL,VR}]_{\mu 233}^+ 
	[C_{ eddd}^{\tt SR,VL}]_{\mu 233}^{+*}  \big) + \tL \leftrightarrow \tR.
\end{align}
\end{subequations} }%
For processes with an antineutrino in the final state, their decay widths can be obtained by exchanging the chiral labels $\tL$ and $\tR$ in the WCs of the corresponding neutrino cases.

For the three-body nucleon decay processes, the final results are summarized as
{\footnotesize
\begin{subequations}
\begin{align}
{\Gamma_{p\to e^+ \pi^0 a} \over (0.1\rm GeV)^9} &=
	52 |[C_{euud}^{\tt VL,SL}]_{e112}|^2
	+ 51 |[C_{eudu}^{\tt SL,VR}]^-_{e121}|^2
	+ 14 \kappa_3^2 |[C_{euud}^{\tt SL,VR}]_{e112}|^2 
	+ \kappa_3^2 |[C_{eudu}^{\tt SL,VR}]^+_{e121}|^2
	\notag\\
	&+ 102 \Re\big( [C_{eudu}^{\tt SL,VR}]^-_{e121} 
	[C_{ euud}^{\tt  VR,SR}]_{e112}^* \big)
	- 14 \kappa_3 \Re\big( ([C_{ euud}^{\tt VR,SR}]_{e112} 
	+[C_{ eudu}^{\tt SL,VR}]^-_{e121} )
	[C_{euud}^{\tt SL,VR}]_{e112}^*   \big)
	\notag\\
	&- 9.6 \kappa_3 \Re\big( ([C_{euud}^{\tt VR,SR}]_{e112} 
	+[C_{eudu}^{\tt SL,VR}]^-_{e121} )
	[C_{ eudu}^{\tt SL,VR}]_{e121}^{+*}  \big)
	+ 6.3 \kappa_3^2 \Re\big( [C_{ eudu}^{\tt SL,VR}]^+_{e121} 
	[C_{ euud}^{\tt  SL,VR}]_{e112}^* \big)
	\notag\\
	&- 0.1 \kappa_3 \Re\big( ([C_{ euud}^{\tt VL,SL}]_{e112} 
	+[C_{ eudu}^{\tt SR,VL}]_{e121}^- )
	[C_{ euud}^{\tt SL,VR}]_{e112}^*  \big)
	 + 0.03 \kappa_3^2 \Re \big( [C_{ euud}^{\tt SL,VR}]_{e112} 
	[C_{ euud}^{\tt SR,VL}]_{e112}^* \big)
	\notag\\
	& -0.02 \kappa_3 \Re\big( ([C_{euud}^{\tt VR,SR}]_{e112} 
	+[C_{eudu}^{\tt SL,VR}]^-_{e121} )
	[C_{eudu}^{\tt SR,VL}]_{e121}^{+*}  \big)
	  + 0.02 \kappa_3^2 \Re \big( [C_{eudu}^{\tt SL,VR}]^+_{e121} 
	[C_{euud}^{\tt SR,VL}]_{e112}^*  \big)\notag\\
	&
    + 0.003 \kappa_3^2 \Re \big( [C_{ eudu}^{\tt SL,VR}]_{e121}^+ 
	[C_{ eudu}^{\tt SR,VL}]_{e121}^{+*} \big)+ \tL \leftrightarrow \tR,
\\%
	{\Gamma_{p\to \mu^+ \pi^0 a} \over (0.1\rm GeV)^9} & = 
	45  |[C_{ euud}^{\tt VL,SL}]_{\mu 112}|^2
	+ 44 |[C_{ eudu}^{\tt SL,VR}]^-_{\mu 121}|^2
	+ 13 \kappa_3^2  |[C_{ euud}^{\tt SL,VR}]_{\mu 112}|^2 
	+ \kappa_3^2 |[C_{ eudu}^{\tt SL,VR}]^+_{\mu 121}|^2
	\notag\\
	&+ 89 \Re\big( [C_{ eudu}^{\tt SL,VR}]^-_{\mu 121} 
	[C_{ euud}^{\tt  VR,SR}]_{\mu 112}^* \big)
	- 16 \kappa_3 \Re\big( ([C_{ euud}^{\tt VR,SR}]_{\mu 112} 
	+[C_{ eudu}^{\tt SL,VR}]^-_{\mu 121} )
	[C_{ euud}^{\tt SR,VL}]_{\mu 112}^*   \big)
	\notag\\
	&- 13 \kappa_3 \Re\big( ([C_{ euud}^{\tt VR,SR}]_{\mu 112} 
	+[C_{ eudu}^{\tt SL,VR}]^-_{\mu 121} )
	[C_{ euud}^{\tt SL,VR}]_{\mu 112}^{*}  \big)
	+ 5.7 \kappa_3^2 \Re \big( [C_{ eudu}^{\tt SL,VR}]^+_{\mu 121} 
	[C_{ euud}^{\tt SL,VR}]_{\mu 112}^*  \big)
    \notag\\
	&- 8.6 \kappa_3 \Re\big( ([C_{ euud}^{\tt VR,SR}]_{\mu 112} 
	+[C_{ eudu}^{\tt SL,VR}]^-_{\mu 121} )
	[C_{ eudu}^{\tt SL,VR}]_{\mu 121}^{+*}  \big)
	  + 4.6 \kappa_3^2 \Re \big( [C_{ euud}^{\tt SL,VR}]_{\mu 112} 
	[C_{ euud}^{\tt SR,VL}]_{\mu 112}^* \big) 
    \notag\\
	& -2.6 \kappa_3 \Re\big( ([C_{ euud}^{\tt VR,SR}]_{\mu 112} 
	+[C_{ eudu}^{\tt SL,VR}]^-_{\mu 121} )
	[C_{ eudu}^{\tt SR,VL}]_{\mu 121}^{+*}  \big)
    + 3.6 \kappa_3^2 \Re \big( [C_{ eudu}^{\tt SL,VR}]^+_{\mu 121} 
	[C_{ euud}^{\tt SR,VL}]_{\mu 112}^*  \big)
	\notag\\
	&+ 0.45 \kappa_3^2 \Re \big( [C_{ eudu}^{\tt SL,VR}]_{\mu 121}^+ 
	[C_{ eudu}^{\tt SR,VL}]_{\mu 121}^{+*} \big)+ \tL \leftrightarrow \tR  ,
\\%
	{\Gamma_{p\to e^+ \eta a} \over (0.1\rm GeV)^9} &=
	1.6 |[C_{ euud}^{\tt VL,SL}]_{e112}|^2
	+ 0.06 |[C_{ eudu}^{\tt SL,VR}]^-_{e121}|^2
	+ 0.02 \kappa_3^2 \big( |[C_{ euud}^{\tt SL,VR}]_{e112}|^2  
	+ |[C_{ eudu}^{\tt SL,VR}]^+_{e121}|^2\big)
	\notag\\
	&- 0.6 \Re\big( [C_{ eudu}^{\tt SL,VR}]^-_{e121} 
	[C_{ euud}^{\tt  VR,SR}]_{e112}^* \big)
    + 0.2 \kappa_3 \Re\big( ([C_{ eudu}^{\tt SL,VR}]^+_{e121} 
	- [C_{ euud}^{\tt SL,VR}]_{e112} )
	[C_{ euud}^{\tt VR,SR}]_{e112}^*   \big)
	\notag\\
	&+ 0.05 \kappa_3 \Re\big( ([C_{ euud}^{\tt SL,VR}]_{e112} 
	- [C_{ eudu}^{\tt SL,VR}]^+_{e121} )
	[C_{ eudu}^{\tt SL,VR}]_{e121}^{-*}   \big)
	- 0.04 \kappa_3^2 \Re\big( [C_{ eudu}^{\tt SL,VR}]^+_{e121} 
	[C_{ euud}^{\tt  SL,VR}]_{e112}^* \big)
    \notag\\
	&+ 10^{-3} \kappa_3 \Re\big( ([C_{ eudu}^{\tt SR,VL}]^+_{e121} 
	- [C_{ euud}^{\tt SR,VL}]_{e112} )
	[C_{ euud}^{\tt VR,SR}]_{e112}^*   \big)
	- 10^{-4} \kappa_3^2 \Re \big( [C_{ eudu}^{\tt SL,VR}]^+_{e121} 
	[C_{ euud}^{\tt SR,VL}]_{e112}^*  \big)
    \notag\\
	&+ 2\cdot10^{-4} \kappa_3 \Re\big( ([C_{ euud}^{\tt SR,VL}]_{e112} 
	- [C_{ eudu}^{\tt SR,VL}]^+_{e121} )
	[C_{ eudu}^{\tt SL,VR}]_{e121}^{-*}   \big)
	\notag\\
	&+ 5\cdot10^{-5} \kappa_3^2 \Re \big( [C_{ eudu}^{\tt SL,VR}]_{e121}^+ 
	[C_{ eudu}^{\tt SR,VL}]_{e121}^{+*} 
	+ [C_{ euud}^{\tt SL,VR}]_{e112} 
	[C_{ euud}^{\tt SR,VL}]_{e112}^*   \big)    + \tL \leftrightarrow \tR,
\\%
	{\Gamma_{p\to \mu^+ \eta a} \over (0.1\rm GeV)^9} &=
	|[C_{ euud}^{\tt VL,SL}]_{\mu 112}|^2
	+ 0.04 |[C_{ eudu}^{\tt SL,VR}]^-_{\mu 121}|^2
    + 0.01 \kappa_3^2 \big( |[C_{ euud}^{\tt SL,VR}]_{\mu 112}|^2 
	+ |[C_{ eudu}^{\tt SL,VR}]^+_{\mu 121}|^2 \big)
	\notag\\
	& - 0.4 \Re\big( [C_{ eudu}^{\tt SL,VR}]^-_{\mu 121} 
	[C_{ euud}^{\tt  VR,SR}]_{\mu 112}^* \big)
	+ 0.1 \kappa_3 \Re\big( ([C_{ eudu}^{\tt SR,VL}]^+_{\mu 121} 
	- [C_{ euud}^{\tt SR,VL}]_{\mu 112} )
	[C_{ euud}^{\tt VR,SR}]_{\mu 112}^*   \big)
	\notag\\
	&+ 0.1 \kappa_3 \Re\big( ([C_{ eudu}^{\tt SL,VR}]^+_{\mu 121} 
	- [C_{ euud}^{\tt SL,VR}]_{\mu 112} )
	[C_{ euud}^{\tt VR,SR}]_{\mu 112}^*   \big)
	- 0.03 \kappa_3^2 \Re\big( [C_{ eudu}^{\tt SL,VR}]^+_{\mu 121} 
	[C_{ euud}^{\tt  SL,VR}]_{\mu 112}^* \big)
    \notag\\
	&+ 0.03 \kappa_3 \Re\big( ([C_{ euud}^{\tt SL,VR}]_{\mu 112} 
	- [C_{ eudu}^{\tt SL,VR}]^+_{\mu 121} )
	[C_{ eudu}^{\tt SL,VR}]_{\mu 121}^{-*}   \big)
	- 0.01 \kappa_3^2 \Re \big( [C_{ eudu}^{\tt SL,VR}]^+_{\mu 121} 
	[C_{ euud}^{\tt SR,VL}]_{\mu 112}^* \big)
    \notag\\
	&+ 0.02 \kappa_3 \Re\big( ([C_{ euud}^{\tt SR,VL}]_{\mu 112} 
	- [C_{ eudu}^{\tt SR,VL}]^+_{\mu 121} )
	[C_{ eudu}^{\tt SL,VR}]_{\mu 121}^{-*}   \big)
    \notag\\
	&+ 0.005 \kappa_3^2 \Re \big( [C_{ eudu}^{\tt SL,VR}]_{\mu 121}^+ 
	[C_{ eudu}^{\tt SR,VL}]_{\mu 121}^{+*} 
	+ [C_{ euud}^{\tt SL,VR}]_{\mu 112} 
	[C_{ euud}^{\tt SR,VL}]_{\mu 112}^*   \big)    + \tL \leftrightarrow \tR ,
\\%
	{\Gamma_{p\to e^+ K^0 a} \over (0.1\rm GeV)^9} &=
	2.1 |[C_{ eudu}^{\tt SL,VR}]^-_{e131}|^2
	+ 1.2 |[C_{ euud}^{\tt VL,SL}]_{e113}|^2
	+ 0.2 \kappa_3^2 \big( |[C_{ euud}^{\tt SL,VR}]_{e113}|^2 
	+ |[C_{ eudu}^{\tt SL,VR}]^+_{e131}|^2\big)
	\notag\\
	&- 3.2 \Re\big( [C_{ eudu}^{\tt SL,VR}]^-_{e131} 
	[C_{ euud}^{\tt  VR,SR}]_{e113}^* \big)
    - 0.7 \kappa_3 \Re\big( ([C_{ euud}^{\tt SL,VR}]_{e113} 
	+ [C_{ eudu}^{\tt SL,VR}]^+_{e131} )
	[C_{ eudu}^{\tt SL,VR}]_{e131}^{-*}   \big)
	\notag\\
	&+ 0.6 \kappa_3 \Re\big( ([C_{ eudu}^{\tt SL,VR}]^+_{e131} 
	+ [C_{ euud}^{\tt SL,VR}]_{e113} )
	[C_{ euud}^{\tt VR,SR}]_{e113}^*   \big)
	+ 0.4 \kappa_3^2 \Re\big( [C_{ eudu}^{\tt SL,VR}]^+_{e131} 
	[C_{ euud}^{\tt  SL,VR}]_{e113}^* \big)
    \notag\\
	&- 0.004 \kappa_3 \Re\big( ([C_{ euud}^{\tt SR,VL}]_{e113} 
	+ [C_{ eudu}^{\tt SR,VL}]^+_{e131} )
	[C_{ eudu}^{\tt SL,VR}]_{e131}^{-*}   \big)
	+ 0.001 \kappa_3^2 \Re \big( [C_{ eudu}^{\tt SL,VR}]^+_{e131} 
	[C_{ euud}^{\tt SR,VL}]_{e113}^*  \big)
    \notag\\ 
	&+ 0.003 \kappa_3 \Re\big( ([C_{ eudu}^{\tt SR,VL}]^+_{e131} 
	+ [C_{ euud}^{\tt SR,VL}]_{e113} )
	[C_{ euud}^{\tt VR,SR}]_{e113}^*   \big)
	\notag\\
	&+ 5\cdot10^{-4} \kappa_3^2 \Re \big( [C_{ eudu}^{\tt SL,VR}]_{e131}^+ 
	[C_{ eudu}^{\tt SR,VL}]_{e131}^{+*} 
	+ [C_{ euud}^{\tt SL,VR}]_{e113} 
	[C_{ euud}^{\tt SR,VL}]_{e113}^*   \big)  + \tL \leftrightarrow \tR,
\\%
	{\Gamma_{p\to \mu^+ K^0 a} \over (0.1\rm GeV)^9} &=
	1.4 |[C_{ eudu}^{\tt SL,VR}]^-_{\mu 131}|^2
	+ 0.9 |[C_{ euud}^{\tt VL,SL}]_{\mu 113}|^2
	+ 0.1 \kappa_3^2 \big( |[C_{ euud}^{\tt SL,VR}]_{\mu 113}|^2 
	+ |[C_{ eudu}^{\tt SL,VR}]^+_{\mu 131}|^2 \big)
	\notag\\
	& - 2.2 \Re\big( [C_{ eudu}^{\tt SL,VR}]^-_{\mu 131} 
	[C_{ euud}^{\tt  VR,SR}]_{\mu 113}^* \big)
	- 0.5 \kappa_3 \Re\big( ([C_{ euud}^{\tt SL,VR}]_{\mu 113} 
	+ [C_{ eudu}^{\tt SL,VR}]^+_{\mu 131} )
	[C_{ eudu}^{\tt SL,VR}]_{\mu 131}^{-*}   \big)
	\notag\\
	&- 0.4 \kappa_3 \Re\big( ([C_{ euud}^{\tt SR,VL}]_{\mu 113} 
	+ [C_{ eudu}^{\tt SR,VL}]^+_{\mu 131} )
	[C_{ eudu}^{\tt SL,VR}]_{\mu 131}^{-*}   \big)
	+ 0.3 \kappa_3^2 \Re\big( [C_{ eudu}^{\tt SL,VR}]^+_{\mu 131} 
	[C_{ euud}^{\tt  SL,VR}]_{\mu 113}^* \big)
    \notag\\ 
	&+ 0.4 \kappa_3 \Re\big( ([C_{ eudu}^{\tt SL,VR}]^+_{\mu 131} 
	+ [C_{ euud}^{\tt SL,VR}]_{\mu 113} )
	[C_{ euud}^{\tt VR,SR}]_{\mu 113}^*   \big)
	+ 0.1 \kappa_3^2 \Re \big( [C_{ eudu}^{\tt SL,VR}]^+_{\mu 131} 
	[C_{ euud}^{\tt SR,VL}]_{\mu 113}^*  \big)
    \notag\\
	&+ 0.3 \kappa_3 \Re\big( ([C_{ eudu}^{\tt SR,VL}]^+_{\mu 131} 
	+ [C_{ euud}^{\tt SR,VL}]_{\mu 113} )
	[C_{ euud}^{\tt VR,SR}]_{\mu 113}^*   \big)
    \notag\\
	&+ 0.05 \kappa_3^2 \Re \big( [C_{ eudu}^{\tt SL,VR}]_{\mu 131}^+ 
	[C_{ eudu}^{\tt SR,VL}]_{\mu 131}^{+*} 
	+ [C_{ euud}^{\tt SL,VR}]_{\mu 113} 
	[C_{ euud}^{\tt SR,VL}]_{\mu 113}^*   \big)+ \tL \leftrightarrow \tR ,
\\%
	{\Gamma_{n \to e^+ \pi^- a} \over (0.1\rm GeV)^9} &=
	102 |[C_{ euud}^{\tt VL,SL}]_{e112}|^2
	+ 100|[C_{ eudu}^{\tt SL,VR}]^-_{e121}|^2
	+ 27 \kappa_3^2 |[C_{ eudu}^{\tt SL,VR}]^+_{e121}|^2
	+ 4.2 \kappa_3^2 |[C_{ euud}^{\tt SL,VR}]_{e112}|^2 
	\notag\\
	&+ 202 \Re\big( [C_{ eudu}^{\tt SL,VR}]^-_{e121} 
	[C_{ euud}^{\tt  VR,SR}]_{e112}^* \big)
	  + 29 \kappa_3 \Re\big( ([C_{ euud}^{\tt VR,SR}]_{e112} 
	+[C_{ eudu}^{\tt SL,VR}]^-_{e121} )
	[C_{ eudu}^{\tt SL,VR}]_{e121}^{+*}  \big)
	\notag\\
	& - 5 \kappa_3 \Re\big( ([C_{ euud}^{\tt VR,SR}]_{e112} 
	+[C_{ eudu}^{\tt SL,VR}]^-_{e121} )
	[C_{ euud}^{\tt SL,VR}]_{e112}^*   \big)
	- 21 \kappa_3^2 \Re\big( [C_{ eudu}^{\tt SL,VR}]^+_{e121} 
	[C_{ euud}^{\tt  SL,VR}]_{e112}^* \big)
    \notag\\
	&+ 0.2 \kappa_3 \Re\big( ([C_{ euud}^{\tt VR,SR}]_{e112} 
	+[C_{ eudu}^{\tt SL,VR}]^-_{e121} )
	[C_{ eudu}^{\tt SR,VL}]_{e121}^{+*}  \big)
	- 0.03 \kappa_3^2 \Re \big( [C_{ eudu}^{\tt SL,VR}]^+_{e121} 
	[C_{ euud}^{\tt SR,VL}]_{e112}^*  \big)
    \notag\\
	&- 0.08 \kappa_3 \Re\big( ([C_{ euud}^{\tt VL,SL}]_{e112} 
	+[C_{ eudu}^{\tt SR,VL}]_{e121}^- )
	[C_{ euud}^{\tt SL,VR}]_{e112}^*  \big)
	+ 0.05 \kappa_3^2 \Re \big( [C_{ eudu}^{\tt SL,VR}]_{e121}^+ 
	[C_{ eudu}^{\tt SR,VL}]_{e121}^{+*} \big)
    \notag\\
	&+ 0.005 \kappa_3^2 \Re \big( [C_{ euud}^{\tt SL,VR}]_{e112} 
	[C_{ euud}^{\tt SR,VL}]_{e112}^* \big)+ \tL \leftrightarrow \tR,
\\%
	{\Gamma_{n \to \mu^+ \pi^- a} \over (0.1\rm GeV)^9} &=
	89 |[C_{ euud}^{\tt VL,SL}]_{\mu 112}|^2
	+ 87 |[C_{ eudu}^{\tt SL,VR}]^-_{\mu 121}|^2
	+ 25 \kappa_3^2 |[C_{ eudu}^{\tt SL,VR}]^+_{\mu 121}|^2
	+ 4 \kappa_3^2 |[C_{ euud}^{\tt SL,VR}]_{\mu 112}|^2 
	\notag\\
	&+ 176 \Re\big( [C_{ eudu}^{\tt SL,VR}]^-_{\mu 121} 
	[C_{ euud}^{\tt  VR,SR}]_{\mu 112}^*  \big)
	+ 31 \kappa_3 \Re\big( ([C_{ euud}^{\tt VR,SR}]_{\mu 112} 
	+[C_{ eudu}^{\tt SL,VR}]^-_{\mu 121} )
	[C_{ eudu}^{\tt SR,VL}]_{\mu 121}^{+*}  \big)
	\notag\\
	&+ 26 \kappa_3 \Re\big( ([C_{ euud}^{\tt VR,SR}]_{\mu 112} 
	+[C_{ eudu}^{\tt SL,VR}]^-_{\mu 121} )
	[C_{ eudu}^{\tt SL,VR}]_{\mu 121}^{+*}  \big)
	- 20 \kappa_3^2 \Re\big( [C_{ eudu}^{\tt SL,VR}]^+_{\mu 121} 
	[C_{ euud}^{\tt  SL,VR}]_{\mu 112}^* \big)
	\notag\\
	&- 13 \kappa_3 \Re\big( ([C_{ euud}^{\tt VL,SL}]_{\mu 112} 
	+[C_{ eudu}^{\tt SR,VL}]_{\mu 121}^- )
	[C_{ euud}^{\tt SL,VR}]_{\mu 112}^*  \big)
	- 5.5 \kappa_3^2 \Re \big( [C_{ eudu}^{\tt SL,VR}]^+_{\mu 121} 
	[C_{ euud}^{\tt SR,VL}]_{\mu 112}^*  \big)
	\notag\\
	&- 4.4 \kappa_3 \Re\big( ([C_{ euud}^{\tt VR,SR}]_{\mu 112} 
	+[C_{ eudu}^{\tt SL,VR}]^-_{\mu 121} )
	[C_{ euud}^{\tt SL,VR}]_{\mu 112}^*   \big)
	+ 9 \kappa_3^2 \Re \big( [C_{ eudu}^{\tt SL,VR}]_{\mu 121}^+ 
	[C_{ eudu}^{\tt SR,VL}]_{\mu 121}^{+*} \big)
    \notag\\
	& + 0.7 \kappa_3^2 \Re \big( [C_{ euud}^{\tt SL,VR}]_{\mu 112} 
	[C_{ euud}^{\tt SR,VL}]_{\mu 112}^* \big)+ \tL \leftrightarrow \tR,
\\%
	{\Gamma_{p \to \nu_x \pi^+ a} \over (0.1\rm GeV)^9} &=
	101 |[C_{ \nu ddu}^{\tt VL,SL}]_{x221}|^2 
	+ 99 |[C_{ \nu udd}^{\tt SR,VL}]^-_{x122}|^2
	+ 27 \kappa_3^2 |[C_{ \nu udd}^{\tt SR,VL}]^+_{x122}|^2
	+ 4.1 \kappa_3^2 |[C_{ \nu ddu}^{\tt SR,VL}]_{x221}|^2
	\notag\\
	&- 199 \Re \big( [C_{ \nu udd}^{\tt SR,VL}]^-_{x122} 
	[C_{ \nu ddu}^{\tt VL,SL}]_{x221}^*  \big)
    + 29 \kappa_3 \Re\big( ([C_{ \nu ddu}^{\tt VL,SL}]_{x221} 
	- [C_{ \nu udd}^{\tt SR,VL}]^-_{x122} )
	[C_{ \nu udd}^{\tt SR,VL}]_{x122}^{+*}  \big)
	\notag\\
	&- 21 \kappa_3^2 \Re\big( [C_{ \nu udd}^{\tt SR,VL}]^+_{x122} 
	[C_{ \nu ddu}^{\tt  SR,VL}]_{x221}^*  \big)
    + 5.1 \kappa_3 \Re\big( ([C_{ \nu udd}^{\tt SR,VL}]^-_{x122}
	- [C_{ \nu ddu}^{\tt VL,SL}]_{x221} )
	[C_{ \nu ddu}^{\tt SR,VL}]_{x221}^* \big),
\\%
	{\Gamma_{p \to \nu_x K^+ a} \over (0.1\rm GeV)^9} &=
	3.3 |[C_{ \nu ddu}^{\tt VL,SL}]_{x321}|^2 
	+ 3.2 |[C_{ \nu udd}^{\tt SR,VL}]^-_{x123}|^2 
	+1.4 |[C_{ \nu ddu}^{\tt SR,VL}]^-_{x231}|^2
	+0.9 \kappa_3^2 |[C_{ \nu udd}^{\tt SR,VL}]^+_{x132}|^2
	\notag\\
	&+ 0.2 \kappa_3^2 \big( |[C_{ \nu ddu}^{\tt SR,VL}]^+_{x231}|^2
	+ |[C_{ \nu udd}^{\tt SR,VL}]^+_{x123}|^2 \big)
	+0.1 \big( |[C_{ \nu ddu}^{\tt VL,SL}]_{x231}|^2 
	+ |[C_{ \nu udd}^{\tt SR,VL}]^-_{x132}|^2 \big)
	\notag\\
	&- 6.5 \Re \big( [C_{ \nu udd}^{\tt SR,VL}]^-_{x123} 
	[C_{ \nu ddu}^{\tt VL,SL}]_{x321}^*  \big)
	+ 4.1 \Re\big( ([C_{ \nu udd}^{\tt SR,VL}]^-_{x123}
	- [C_{ \nu ddu}^{\tt VL,SL}]_{x321} )
	[C_{ \nu ddu}^{\tt SR,VL}]_{x231}^{-*}  \big)
	\notag\\
	&+ 1.6 \kappa_3 \Re\big( ( [C_{ \nu ddu}^{\tt VL,SL}]_{x321}
	- [C_{ \nu udd}^{\tt SR,VL}]^-_{x123} )
	[C_{ \nu udd}^{\tt SR,VL}]_{x132}^{+*}  \big)
	- 1.3 \kappa_3 \Re \big( [C_{ \nu ddu}^{\tt SR,VL}]^-_{x231} 
	[C_{ \nu udd}^{\tt SR,VL}]_{x132}^{+*}  \big)
	\notag\\
	&+ 1.2 \Re\big( ( [C_{ \nu ddu}^{\tt VL,SL}]_{x321}
	- [C_{ \nu udd}^{\tt SR,VL}]^-_{x123} )
	([C_{ \nu ddu}^{\tt VL,SL}]_{x231}^{*}
	- [C_{ \nu udd}^{\tt SR,VL}]_{x132}^{-*} ) \big)
	\notag\\
	&+ 0.9 \kappa_3^2 \Re\big( ( [C_{ \nu udd}^{\tt SR,VL}]^+_{x123}
	-[C_{ \nu ddu}^{\tt SR,VL}]^+_{x231} )
	[C_{ \nu udd}^{\tt  SR,VL}]_{x132}^{+*}  \big)	
	+ 0.7 \kappa_3 \Re \big( [C_{ \nu ddu}^{\tt SR,VL}]_{x231}^- 
	[C_{ \nu ddu}^{\tt SR,VL}]_{x231}^{+*}  \big)\notag\\
	&+ 0.8 \kappa_3 \Re\big( ( [C_{ \nu udd}^{\tt SR,VL}]^-_{x123}
	- [C_{ \nu ddu}^{\tt VL,SL}]_{x321} )
	[C_{ \nu ddu}^{\tt SR,VL}]_{x231}^{+*}  \big)
	- 0.7 \kappa_3 \Re \big( [C_{ \nu ddu}^{\tt SR,VL}]_{x231}^- 
	[C_{ \nu udd}^{\tt SR,VL}]_{x123}^{+*}  \big)
    \notag\\
	&+ 0.8 \kappa_3 \Re\big( ( [C_{ \nu ddu}^{\tt VL,SL}]_{x321}
	- [C_{ \nu udd}^{\tt SR,VL}]^-_{x123}  )
	[C_{ \nu udd}^{\tt SR,VL}]_{x123}^{+*}  \big)
	- 0.5 \kappa_3^2 \Re \big( [C_{ \nu ddu}^{\tt SR,VL}]_{x231}^+ 
	[C_{ \nu udd}^{\tt SR,VL}]_{x123}^{+*}  \big)
    \notag\\
	&+ 0.7 \Re\big( ( [C_{ \nu udd}^{\tt SR,VL}]^-_{x132}
	- [C_{ \nu ddu}^{\tt VL,SL}]_{x231}  )
	[C_{ \nu ddu}^{\tt SR,VL}]_{x231}^{-*}  \big)
	- 0.3 \Re \big( [C_{ \nu udd}^{\tt SR,VL}]_{x132}^- 
	[C_{ \nu ddu}^{\tt VL,SL}]_{x231}^*  \big)\notag\\
	&+ 0.1 \kappa_3 \Re\big( ( [C_{ \nu ddu}^{\tt VL,SL}]_{x231}
	- [C_{ \nu udd}^{\tt SR,VL}]^-_{x132}  )
	[C_{ \nu udd}^{\tt SR,VL}]_{x132}^{+*}  \big)
	\notag\\
	&+ 0.07 \kappa_3 \Re\big( ( [C_{ \nu udd}^{\tt SR,VL}]^+_{x123}
	- [C_{ \nu ddu}^{\tt SR,VL}]^+_{x231} )
	([C_{ \nu ddu}^{\tt VL,SL}]_{x231}^{*}
	- [C_{ \nu udd}^{\tt SR,VL}]_{x132}^{-*} ) \big),
\\%
	{\Gamma_{n \to \nu_x \pi^0 a} \over (0.1\rm GeV)^9} &=
	52 |[C_{ \nu ddu}^{\tt VL,SL}]_{x221}|^2 
	+ 51 |[C_{ \nu udd}^{\tt SR,VL}]^-_{x122}|^2
	+ 14 \kappa_3^2 |[C_{ \nu ddu}^{\tt SR,VL}]_{x221}|^2
	+ 1.1 \kappa_3^2 |[C_{ \nu udd}^{\tt SR,VL}]^+_{x122}|^2
	\notag\\
	&- 103 \Re \big( [C_{ \nu udd}^{\tt SR,VL}]^-_{x122} 
	[C_{ \nu ddu}^{\tt VL,SL}]_{x221}^*  \big)
	  + 15 \kappa_3 \Re\big( ([C_{ \nu udd}^{\tt SR,VL}]^-_{x122}
	- [C_{ \nu ddu}^{\tt VL,SL}]_{x221} )
	[C_{ \nu ddu}^{\tt SR,VL}]_{x221}^*   \big)
	\notag\\
	&+ 9.7 \kappa_3 \Re\big( ([C_{ \nu udd}^{\tt SR,VL}]^-_{x122}
	- [C_{ \nu ddu}^{\tt VL,SL}]_{x221} )
	[C_{ \nu udd}^{\tt SR,VL}]_{x122}^{+*}  \big)
    + 6.4 \kappa_3^2 \Re\big( [C_{ \nu udd}^{\tt SR,VL}]^+_{x122} 
	[C_{ \nu ddu}^{\tt  SR,VL}]_{x221}^*  \big),
\\%
	{\Gamma_{n \to \nu_x \eta a} \over (0.1\rm GeV)^9} &=
	1.7 |[C_{ \nu ddu}^{\tt VL,SL}]_{x221}|^2 
	+ 0.06 |[C_{ \nu udd}^{\tt SR,VL}]^-_{x122}|^2
	+ 0.02 \kappa_3^2 \big( |[C_{ \nu ddu}^{\tt SR,VL}]_{x221}|^2
	+ |[C_{ \nu udd}^{\tt SR,VL}]^+_{x122}|^2  \big)
	\notag\\
	&+ 0.6 \Re \big( [C_{ \nu udd}^{\tt SR,VL}]^-_{x122} 
	[C_{ \nu ddu}^{\tt VL,SL}]_{x221}^*  \big)
	  + 0.2 \kappa_3 \Re\big( ([C_{ \nu udd}^{\tt SR,VL}]^+_{x122}
	- [C_{ \nu ddu}^{\tt SR,VL}]_{x221} )
	[C_{ \nu ddu}^{\tt VL,SL}]_{x221}^*  \big)
	\notag\\
	&+ 0.05 \kappa_3 \Re\big( ([C_{ \nu udd}^{\tt SR,VL}]^+_{x122}
	- [C_{ \nu ddu}^{\tt SR,VL}]_{x221} )
	[C_{ \nu udd}^{\tt SR,VL}]_{x122}^{-*} \big)
    - 0.04 \kappa_3^2 \Re\big( [C_{ \nu udd}^{\tt SR,VL}]^+_{x122} 
	[C_{ \nu ddu}^{\tt  SR,VL}]_{x221}^*  \big),
\\%
	{\Gamma_{n \to \nu_x K^0 a} \over (0.1\rm GeV)^9} &=
	3.2 |[C_{ \nu ddu}^{\tt VL,SL}]_{x321}|^2 
	+ 3.1 |[C_{ \nu udd}^{\tt SR,VL}]^-_{x123}|^2 
	+ 2.1 |[C_{ \nu ddu}^{\tt VL,SL}]_{x231}|^2
	+ 1.3 |[C_{ \nu udd}^{\tt SR,VL}]^-_{x132}|^2
	\notag\\
	&+ 0.9 \kappa_3^2 |[C_{ \nu ddu}^{\tt SR,VL}]^+_{x231}|^2
	+ 0.2 \kappa_3^2 \big( |[C_{ \nu udd}^{\tt SR,VL}]^+_{x132}|^2
	+ |[C_{ \nu udd}^{\tt SR,VL}]^+_{x123}|^2 \big)
	+ 0.1 |[C_{ \nu ddu}^{\tt SR,VL}]^-_{x231}|^2
	\notag\\
	&- 6.3 \Re \big( [C_{ \nu udd}^{\tt SR,VL}]^-_{x123} 
	[C_{ \nu ddu}^{\tt VL,SL}]_{x321}^*  \big)
	+ 5.1 \Re\big( ( [C_{ \nu ddu}^{\tt VL,SL}]_{x321}
	- [C_{ \nu udd}^{\tt SR,VL}]^-_{x123} )
	[C_{ \nu ddu}^{\tt VL,SL}]_{x231}^{*} \big)
	\notag\\
	&+ 3.9 \Re\big( ( [C_{ \nu ddu}^{\tt VL,SL}]_{x321}
	- [C_{ \nu udd}^{\tt SR,VL}]^-_{x123} )
	[C_{ \nu udd}^{\tt SR,VL}]_{x132}^{-*} \big)
	+ 3.3 \Re \big( [C_{ \nu udd}^{\tt SR,VL}]^-_{x132} 
	[C_{ \nu ddu}^{\tt VL,SL}]_{x231}^*  \big)
	\notag\\
	&+ 1.6 \kappa_3 \Re\big( ( [C_{ \nu udd}^{\tt SR,VL}]^-_{x123}
	- [C_{ \nu ddu}^{\tt VL,SL}]_{x321}  )
	[C_{ \nu ddu}^{\tt SR,VL}]_{x231}^{+*} \big)
	- 1.4 \kappa_3 \Re \big( [C_{ \nu ddu}^{\tt SR,VL}]^+_{x231} 
	[C_{ \nu ddu}^{\tt VL,SL}]_{x231}^*  \big)
	\notag\\
	&- 1.3 \kappa_3 \Re \big( [C_{ \nu ddu}^{\tt SR,VL}]^+_{x231} 
	[C_{ \nu udd}^{\tt SR,VL}]_{x132}^{-*}  \big)
	+ 1.2 \Re\big( ( [C_{ \nu ddu}^{\tt VL,SL}]_{x321}
	- [C_{ \nu udd}^{\tt SR,VL}]^-_{x123} )
	[C_{ \nu ddu}^{\tt SR,VL}]_{x231}^{-*} \big)
	\notag\\
	&+ 0.9 \Re \big( [C_{ \nu ddu}^{\tt SR,VL}]^-_{x231} 
	[C_{ \nu ddu}^{\tt VL,SL}]_{x231}^*  \big)
	+ 0.9 \kappa_3^2 \Re\big( ( [C_{ \nu udd}^{\tt SR,VL}]^+_{x123}
	- [C_{ \nu udd}^{\tt SR,VL}]^+_{x132} )
	[C_{ \nu ddu}^{\tt SR,VL}]_{x231}^{+*} \big)
	\notag\\
	&+ 0.8 \kappa_3 \Re\big( ( [C_{ \nu udd}^{\tt SR,VL}]^+_{x132}
	- [C_{ \nu udd}^{\tt SR,VL}]^+_{x123} )
	([C_{ \nu ddu}^{\tt VL,SL}]_{x321}^{*}
	- [C_{ \nu udd}^{\tt SR,VL}]_{x123}^{-*} ) \big)
	\notag\\
	&+ 0.8 \kappa_3 \Re \big( [C_{ \nu udd}^{\tt SR,VL}]^+_{x132} 
	[C_{ \nu ddu}^{\tt VL,SL}]_{x231}^*  \big)
	- 0.7 \kappa_3 \Re \big( [C_{ \nu udd}^{\tt SR,VL}]^+_{x123} 
	[C_{ \nu ddu}^{\tt VL,SL}]_{x231}^*  \big)
	\notag\\
	&+ 0.7 \Re \big( [C_{ \nu ddu}^{\tt SR,VL}]^-_{x231} 
	[C_{ \nu udd}^{\tt SR,VL}]_{x132}^{-*}  \big)
	+ 0.7 \kappa_3 \Re \big( [C_{ \nu udd}^{\tt SR,VL}]^-_{x132} 
	[C_{ \nu udd}^{\tt SR,VL}]_{x132}^{+*}  \big)
	\notag\\
	&- 0.6 \kappa_3 \Re \big( [C_{ \nu udd}^{\tt SR,VL}]^-_{x132} 
	[C_{ \nu udd}^{\tt SR,VL}]_{x123}^{+*}  \big)
	- 0.4 \kappa_3^2 \Re \big( [C_{ \nu udd}^{\tt SR,VL}]^+_{x123} 
	[C_{ \nu udd}^{\tt SR,VL}]_{x132}^{+*}  \big)
	\notag\\
	&- 0.1 \kappa_3 \Re \big( [C_{ \nu ddu}^{\tt SR,VL}]^-_{x231} 
	[C_{ \nu ddu}^{\tt SR,VL}]_{x231}^{+*}  \big)
	  + 0.07 \kappa_3 \Re\big( ( [C_{ \nu udd}^{\tt SR,VL}]^+_{x132}
	- [C_{ \nu udd}^{\tt SR,VL}]^+_{x123} )
	[C_{ \nu ddu}^{\tt SR,VL}]_{x231}^{-*} \big),
\\%
	{\Gamma_{n \to e^- \pi^+ a} \over (0.1\rm GeV)^9} &=
	10 \kappa_3^2 \big( |[C_{ eddd}^{\tt SL,VR}]_{e222}|^2 
	+ |[C_{ eddd}^{\tt SR,VL}]_{e222}|^2 \big) 
	+ 0.05 \kappa_3^2 \Re \big( [C_{ eddd}^{\tt SL,VR}]_{e222}
	[C_{ eddd}^{\tt SR,VL}]_{e222}^* \big),
\\%
	{\Gamma_{n \to \mu^- \pi^+ a} \over (0.1\rm GeV)^9} &=
	9.7 \kappa_3^2 \big( |[C_{ eddd}^{\tt SL,VR}]_{\mu 222}|^2 
	+ |[C_{ eddd}^{\tt SR,VL}]_{\mu 222}|^2 \big) 
	+ 8.5 \kappa_3^2 \Re \big( [C_{ eddd}^{\tt SL,VR}]_{\mu 222}
	[C_{ eddd}^{\tt SR,VL}]_{\mu 222}^* \big),
\\%
	{\Gamma_{n \to e^- K^+ a} \over (0.1\rm GeV)^9} &=
	2.3 |[C_{ eddd}^{\tt SL,VR}]^-_{e232}|^2 
	+ 1.3 |[C_{ eddd}^{\tt VL,SL}]_{e223}|^2
	+ 0.2 \kappa_3^2 \big( |[C_{ eddd}^{\tt SL,VR}]_{e223}|^2 
	+ |[C_{ eddd}^{\tt SL,VR}]^+_{e232}|^2 \big)
	\notag\\
	& - 3.4 \Re\big( [C_{ eddd}^{\tt SR,VL}]^-_{e232} 
	[C_{ eddd}^{\tt  VL,SL}]_{e223}^*  \big)
	- 0.7 \kappa_3 \Re\big( ([C_{ eddd}^{\tt SL,VR}]_{e223}
	+ [C_{ eddd}^{\tt SL,VR}]^+_{e232} )
	[C_{ eddd}^{\tt SL,VR}]_{e232}^{-*}  \big)
	\notag\\
	&+ 0.6 \kappa_3 \Re\big( ([C_{ eddd}^{\tt SL,VR}]_{e223}
	+ [C_{ eddd}^{\tt SL,VR}]^+_{e232} )
	[C_{ eddd}^{\tt VR,SR}]_{e223}^*   \big)
	+ 0.4 \kappa_3^2 \Re\big( [C_{ eddd}^{\tt SL,VR}]^+_{e232} 
	[C_{ eddd}^{\tt  SL,VR}]_{e223}^* \big)
	\notag\\
	&- 0.004 \kappa_3 \Re\big( ([C_{ eddd}^{\tt SL,VR}]_{e223}
	+ [C_{ eddd}^{\tt SL,VR}]^+_{e232} )
	[C_{ eddd}^{\tt SR,VL}]_{e232}^{-*}  \big)
	+ 0.003 \kappa_3 \Re\big( [C_{ eddd}^{\tt SL,VR}]_{e223}
	[C_{ eddd}^{\tt VL,SL}]_{e223}^*  \big)
	\notag\\
	&+ 0.003 \kappa_3 \Re\big( [C_{ eddd}^{\tt SL,VR}]^+_{e232}
	[C_{ eddd}^{\tt VL,SL}]_{e223}^*  \big)
	+ 0.001 \kappa_3^2 \Re\big( [C_{ eddd}^{\tt SL,VR}]^+_{e232}
	[C_{ eddd}^{\tt SR,VL}]_{e223}^*  \big)
	\notag\\
	&+5\cdot10^{-4} \kappa_3^2 \Re\big( [C_{ eddd}^{\tt SL,VR}]_{e223}
	[C_{ eddd}^{\tt SR,VL}]_{e223}^* 
	+ [C_{ eddd}^{\tt SL,VR}]^+_{e232}
	[C_{ eddd}^{\tt SR,VL}]_{e232}^{+*}  \big)    + \tL \leftrightarrow \tR,
\\%
	{\Gamma_{n \to \mu^- K^+ a} \over (0.1\rm GeV)^9} & =
	1.5|[C_{ eddd}^{\tt SL,VR}]^-_{\mu 232}|^2 
	+ 0.9 |[C_{ eddd}^{\tt VL,SL}]_{\mu 223}|^2
	+ 0.2 \kappa_3^2 |[C_{ eddd}^{\tt SL,VR}]_{\mu 223}|^2 
	+ 0.1 \kappa_3^2 |[C_{ eddd}^{\tt SL,VR}]^+_{\mu 232}|^2
	\notag\\
	& - 2.4 \Re\big( [C_{ eddd}^{\tt SR,VL}]^-_{\mu 232} 
	[C_{ eddd}^{\tt  VL,SL}]_{\mu 223}^*  \big)
    - 0.5 \kappa_3 \Re\big( ([C_{ eddd}^{\tt SL,VR}]_{\mu 223}
	+ [C_{ eddd}^{\tt SL,VR}]^+_{\mu 232} )
	[C_{ eddd}^{\tt SL,VR}]_{\mu 232}^{-*}  \big)
	\notag\\
	&- 0.5 \kappa_3 \Re\big( [C_{ eddd}^{\tt SL,VR}]_{\mu 223}
	[C_{ eddd}^{\tt SR,VL}]_{\mu 232}^{-*}  \big)
	+ 0.5 \kappa_3 \Re\big( [C_{ eddd}^{\tt SL,VR}]_{\mu 223}
	[C_{ eddd}^{\tt VR,SR}]_{\mu 223}^*  \big)
	\notag\\
	&+ 0.4 \kappa_3 \Re\big( ([C_{ eddd}^{\tt VL,SL}]_{\mu 223} 
	- [C_{ eddd}^{\tt SL,VR}]^-_{\mu 232} )
	[C_{ eddd}^{\tt SR,VL}]_{\mu 232}^{+*}  \big)
	+ 0.3 \kappa_3^2 \Re\big( [C_{ eddd}^{\tt SL,VR}]^+_{\mu 232} 
	[C_{ eddd}^{\tt  SL,VR}]_{\mu 223}^* \big)
    \notag\\
	&+ 0.3 \kappa_3 \Re\big( ([C_{ eddd}^{\tt SL,VR}]_{\mu 223}
	+ [C_{ eddd}^{\tt SL,VR}]^+_{\mu 232} )
	[C_{ eddd}^{\tt VL,SL}]_{\mu 223}^*  \big)
	+ 0.2 \kappa_3^2 \Re\big( [C_{ eddd}^{\tt SL,VR}]^+_{\mu 232}
	[C_{ eddd}^{\tt SR,VL}]_{\mu 223}^*  \big)
	\notag\\
	&+ 0.1 \kappa_3^2 \Re\big( [C_{ eddd}^{\tt SL,VR}]_{\mu 223}
	[C_{ eddd}^{\tt SR,VL}]_{\mu 223}^* 
	+ [C_{ eddd}^{\tt SL,VR}]^+_{\mu 232}
	[C_{ eddd}^{\tt SR,VL}]_{\mu 232}^{+*}  \big) + \tL \leftrightarrow \tR.
\end{align}	
\end{subequations} }%
Similarly, we explicitly show processes with neutrinos in the final state. The decay widths for antineutrino counterparts can be obtained through $\tL \leftrightarrow \tR$.
When restricted to WCs associated with the $\pmb{8}_{\tL(\tR)} \otimes \pmb{1}_{\tR(\tL)}$ irreps, our above results agree with those given in \cite{Li:2024liy}. 

\bibliography{references_paper.bib}{}

@article{Liao:2025sqt,
    author = "Liao, Yi and Ma, Xiao-Dong and Wang, Hao-Lin",
    title = "{Chiral perturbation theory for baryon-number-violating nucleon decay into a vector meson}",
    eprint = "2506.05052",
    archivePrefix = "arXiv",
    primaryClass = "hep-ph",
    doi = "10.1103/fzq9-tfcp",
    journal = "Phys. Rev. D",
    volume = "112",
    number = "3",
    pages = "L031704",
    year = "2025"
}

@article{Scherer:2002tk,
    author = "Scherer, Stefan",
    editor = "Negele, John W. and Vogt, E. W.",
    title = "{Introduction to chiral perturbation theory}",
    eprint = "hep-ph/0210398",
    archivePrefix = "arXiv",
    reportNumber = "MKPH-T-02-09",
    journal = "Adv. Nucl. Phys.",
    volume = "27",
    pages = "277",
    year = "2003"
}

@article{Bali:2022qja,
    author = {Bali, Gunnar S. and Collins, Sara and S\"oldner, Wolfgang and Weish\"aupl, Simon},
    collaboration = "RQCD",
    title = "{Leading order mesonic and baryonic SU(3) low energy constants from Nf=3 lattice QCD}",
    eprint = "2201.05591",
    archivePrefix = "arXiv",
    primaryClass = "hep-lat",
    doi = "10.1103/PhysRevD.105.054516",
    journal = "Phys. Rev. D",
    volume = "105",
    number = "5",
    pages = "054516",
    year = "2022"
}

@article{Sakharov:1967dj,
    author = "Sakharov, A. D.",
    title = "{Violation of CP Invariance, C asymmetry, and baryon asymmetry of the universe}",
    doi = "10.1070/PU1991v034n05ABEH002497",
    journal = "Pisma Zh. Eksp. Teor. Fiz.",
    volume = "5",
    pages = "32--35",
    year = "1967"
}

@article{Arias-Aragon:2022byr,
    author = "Arias-Arag\'on, Fernando and Smith, Christopher",
    title = "{Leptoquarks, axions and the unification of B, L, and Peccei-Quinn symmetries}",
    eprint = "2206.09810",
    archivePrefix = "arXiv",
    primaryClass = "hep-ph",
    doi = "10.1103/PhysRevD.106.055034",
    journal = "Phys. Rev. D",
    volume = "106",
    number = "5",
    pages = "055034",
    year = "2022"
}

@article{Jenkins:1990jv,
    author = "Jenkins, Elizabeth Ellen and Manohar, Aneesh V.",
    title = "{Baryon chiral perturbation theory using a heavy fermion Lagrangian}",
    reportNumber = "UCSD-PTH-90-23",
    doi = "10.1016/0370-2693(91)90266-S",
    journal = "Phys. Lett. B",
    volume = "255",
    pages = "558--562",
    year = "1991"
}

@article{Fan:2024gzc,
    author = "Fan, Wei-Qi and Liao, Yi and Ma, Xiao-Dong and Wang, Hao-Lin",
    title = "{Baryon number violating hydrogen decay}",
    eprint = "2412.20774",
    archivePrefix = "arXiv",
    primaryClass = "hep-ph",
    doi = "10.1016/j.physletb.2025.139335",
    journal = "Phys. Lett. B",
    volume = "862",
    pages = "139335",
    year = "2025"
}

@article{Liao:2025vlj,
    author = "Liao, Yi and Ma, Xiao-Dong and Wang, Hao-Lin",
    title = "{New chiral structures for baryon number violating nucleon decays}",
    eprint = "2504.14855",
    archivePrefix = "arXiv",
    primaryClass = "hep-ph",
    month = "4",
    year = "2025"
}

@article{Heeck:2019kgr,
    author = "Heeck, Julian and Takhistov, Volodymyr",
    title = "{Inclusive Nucleon Decay Searches as a Frontier of Baryon Number Violation}",
    eprint = "1910.07647",
    archivePrefix = "arXiv",
    primaryClass = "hep-ph",
    reportNumber = "UCI-TR-2019-24",
    doi = "10.1103/PhysRevD.101.015005",
    journal = "Phys. Rev. D",
    volume = "101",
    number = "1",
    pages = "015005",
    year = "2020"
}

@article{Liang:2023yta,
    author = "Liang, Jin-Han and Liao, Yi and Ma, Xiao-Dong and Wang, Hao-Lin",
    title = "{Dark sector effective field theory}",
    eprint = "2309.12166",
    archivePrefix = "arXiv",
    primaryClass = "hep-ph",
    doi = "10.1007/JHEP12(2023)172",
    journal = "JHEP",
    volume = "12",
    pages = "172",
    year = "2023"
}

@article{Li:2024liy,
    author = "Li, Tong and Schmidt, Michael A. and Yao, Chang-Yuan",
    title = "{Baryon-number-violating nucleon decays in ALP effective field theories}",
    eprint = "2406.11382",
    archivePrefix = "arXiv",
    primaryClass = "hep-ph",
    reportNumber = "CPPC-2024-05, DESY-24-082",
    doi = "10.1007/JHEP08(2024)221",
    journal = "JHEP",
    volume = "08",
    pages = "221",
    year = "2024"
}

@article{Irvine-Michigan-Brookhaven:1983iap,
    author = "Bionta, R. M. and others",
    collaboration = "Irvine-Michigan-Brookhaven",
    title = "{A Search for Proton Decay Into e+ pi0}",
    reportNumber = "PRINT-83-0544 (MICHIGAN)",
    doi = "10.1103/PhysRevLett.51.27",
    journal = "Phys. Rev. Lett.",
    volume = "51",
    pages = "27",
    year = "1983",
    note = "[Erratum: Phys.Rev.Lett. 51, 522 (1983)]"
}

@article{Hirata:1988ad,
    author = "Hirata, K. S. and others",
    title = "{Observation in the Kamiokande-II Detector of the Neutrino Burst from Supernova SN 1987a}",
    doi = "10.1103/PhysRevD.38.448",
    journal = "Phys. Rev. D",
    volume = "38",
    pages = "448--458",
    year = "1988"
}

@inproceedings{Takhistov:2016eqm,
    author = "Takhistov, Volodymyr",
    collaboration = "Super-Kamiokande",
    title = "{Review of Nucleon Decay Searches at Super-Kamiokande}",
    booktitle = "{51st Rencontres de Moriond on EW Interactions and Unified Theories}",
    eprint = "1605.03235",
    archivePrefix = "arXiv",
    primaryClass = "hep-ex",
    reportNumber = "UCI-TR-2016-11",
    pages = "437--444",
    year = "2016"
}

@article{DUNE:2020ypp,
    author = "Abi, Babak and others",
    collaboration = "DUNE",
    title = "{Deep Underground Neutrino Experiment (DUNE), Far Detector Technical Design Report, Volume II: DUNE Physics}",
    eprint = "2002.03005",
    archivePrefix = "arXiv",
    primaryClass = "hep-ex",
    reportNumber = "FERMILAB-PUB-20-025-ND, FERMILAB-DESIGN-2020-02",
    month = "2",
    year = "2020"
}

@article{Hyper-Kamiokande:2018ofw,
    author = "Abe, K. and others",
    collaboration = "Hyper-Kamiokande",
    title = "{Hyper-Kamiokande Design Report}",
    eprint = "1805.04163",
    archivePrefix = "arXiv",
    primaryClass = "physics.ins-det",
    month = "5",
    year = "2018"
}

@article{JUNO:2015zny,
    author = "An, Fengpeng and others",
    collaboration = "JUNO",
    title = "{Neutrino Physics with JUNO}",
    eprint = "1507.05613",
    archivePrefix = "arXiv",
    primaryClass = "physics.ins-det",
    doi = "10.1088/0954-3899/43/3/030401",
    journal = "J. Phys. G",
    volume = "43",
    number = "3",
    pages = "030401",
    year = "2016"
}

@article{Theia:2019non,
    author = "Askins, M. and others",
    collaboration = "Theia",
    title = "{THEIA: an advanced optical neutrino detector}",
    eprint = "1911.03501",
    archivePrefix = "arXiv",
    primaryClass = "physics.ins-det",
    doi = "10.1140/epjc/s10052-020-7977-8",
    journal = "Eur. Phys. J. C",
    volume = "80",
    number = "5",
    pages = "416",
    year = "2020"
}

@article{Claudson:1981gh,
    author = "Claudson, Mark and Wise, Mark B. and Hall, Lawrence J.",
    title = "{Chiral Lagrangian for Deep Mine Physics}",
    reportNumber = "HUTP-81/A036",
    doi = "10.1016/0550-3213(82)90401-1",
    journal = "Nucl. Phys. B",
    volume = "195",
    pages = "297--307",
    year = "1982"
}

@article{Yoo:2021gql,
    author = "Yoo, Jun-Sik and Aoki, Yasumichi and Boyle, Peter and Izubuchi, Taku and Soni, Amarjit and Syritsyn, Sergey",
    title = "{Proton decay matrix elements on the lattice at physical pion mass}",
    eprint = "2111.01608",
    archivePrefix = "arXiv",
    primaryClass = "hep-lat",
    reportNumber = "RBRC-1333, KEK-CP-0385",
    doi = "10.1103/PhysRevD.105.074501",
    journal = "Phys. Rev. D",
    volume = "105",
    number = "7",
    pages = "074501",
    year = "2022"
}

@article{Li:2025slp,
    author = "Li, Tong and Schmidt, Michael A. and Yao, Chang-Yuan",
    title = "{Baryon-number-violating nucleon decays in sterile neutrino effective field theories}",
    eprint = "2502.14303",
    archivePrefix = "arXiv",
    primaryClass = "hep-ph",
    reportNumber = "CPPC-2025-02",
    month = "2",
    year = "2025"
}

@article{Fridell:2023tpb,
    author = "Fridell, K\r{a}re and Hati, Chandan and Takhistov, Volodymyr",
    title = "{Noncanonical nucleon decays as window into light new physics}",
    eprint = "2312.13740",
    archivePrefix = "arXiv",
    primaryClass = "hep-ph",
    reportNumber = "KEK-TH-2588, KEK-Cosmo-0336, KEK-QUP-2023-0038, IPMU23-0051,
  IFIC/23-54, IPMU23-0051, ULB-TH/23-18, IFIC/23-54",
    doi = "10.1103/PhysRevD.110.L031701",
    journal = "Phys. Rev. D",
    volume = "110",
    number = "3",
    pages = "L031701",
    year = "2024"
}

@article{SNO:2018ydj,
    author = "Anderson, M. and others",
    collaboration = "SNO+",
    title = "{Search for invisible modes of nucleon decay in water with the SNO+ detector}",
    eprint = "1812.05552",
    archivePrefix = "arXiv",
    primaryClass = "hep-ex",
    doi = "10.1103/PhysRevD.99.032008",
    journal = "Phys. Rev. D",
    volume = "99",
    number = "3",
    pages = "032008",
    year = "2019"
}

@article{KamLAND:2015pvi,
    author = "Asakura, K. and others",
    collaboration = "KamLAND",
    title = "{Search for the proton decay mode $p \rightarrow \overline{\nu} K^{+}$ with KamLAND}",
    eprint = "1505.03612",
    archivePrefix = "arXiv",
    primaryClass = "hep-ex",
    doi = "10.1103/PhysRevD.92.052006",
    journal = "Phys. Rev. D",
    volume = "92",
    number = "5",
    pages = "052006",
    year = "2015"
}

@article{Bijnens:1985kj,
    author = "Bijnens, J. and Sonoda, H. and Wise, Mark B.",
    title = "{On the Validity of Chiral Perturbation Theory for Weak Hyperon Decays}",
    reportNumber = "CALT-68-1221",
    doi = "10.1016/0550-3213(85)90569-3",
    journal = "Nucl. Phys. B",
    volume = "261",
    pages = "185--198",
    year = "1985"
}

@article{SNO:2022trz,
    author = "Allega, A. and others",
    collaboration = "SNO+",
    title = "{Improved search for invisible modes of nucleon decay in water with the SNO+detector}",
    eprint = "2205.06400",
    archivePrefix = "arXiv",
    primaryClass = "hep-ex",
    doi = "10.1103/PhysRevD.105.112012",
    journal = "Phys. Rev. D",
    volume = "105",
    number = "11",
    pages = "112012",
    year = "2022"
}

@article{Super-Kamiokande:2015pys,
    author = "Takhistov, V. and others",
    collaboration = "Super-Kamiokande",
    title = "{Search for Nucleon and Dinucleon Decays with an Invisible Particle and a Charged Lepton in the Final State at the Super-Kamiokande Experiment}",
    eprint = "1508.05530",
    archivePrefix = "arXiv",
    primaryClass = "hep-ex",
    doi = "10.1103/PhysRevLett.115.121803",
    journal = "Phys. Rev. Lett.",
    volume = "115",
    number = "12",
    pages = "121803",
    year = "2015"
}

@article{BESIII:2021slv,
    author = "Ablikim, M. and others",
    collaboration = "BESIII",
    title = "{Search for invisible decays of the $\Lambda$ baryon}",
    eprint = "2110.06759",
    archivePrefix = "arXiv",
    primaryClass = "hep-ex",
    doi = "10.1103/PhysRevD.105.L071101",
    journal = "Phys. Rev. D",
    volume = "105",
    number = "7",
    pages = "L071101",
    year = "2022"
}

@article{KamLAND:2005pen,
    author = "Araki, T. and others",
    collaboration = "KamLAND",
    title = "{Search for the invisible decay of neutrons with KamLAND}",
    eprint = "hep-ex/0512059",
    archivePrefix = "arXiv",
    doi = "10.1103/PhysRevLett.96.101802",
    journal = "Phys. Rev. Lett.",
    volume = "96",
    pages = "101802",
    year = "2006"
}

@article{JUNO:2024pur,
    author = "Abusleme, Angel and others",
    collaboration = "JUNO",
    title = "{JUNO sensitivity to invisible decay modes of neutrons}",
    eprint = "2405.17792",
    archivePrefix = "arXiv",
    primaryClass = "hep-ex",
    doi = "10.1140/epjc/s10052-024-13638-0",
    journal = "Eur. Phys. J. C",
    volume = "85",
    number = "1",
    pages = "5",
    year = "2025"
}

@article{ParticleDataGroup:2024cfk,
    author = "Navas, S. and others",
    collaboration = "Particle Data Group",
    title = "{Review of particle physics}",
    doi = "10.1103/PhysRevD.110.030001",
    journal = "Phys. Rev. D",
    volume = "110",
    number = "3",
    pages = "030001",
    year = "2024"
}

@article{Super-Kamiokande:2005mbp,
    author = "Ashie, Y. and others",
    collaboration = "Super-Kamiokande",
    title = "{A Measurement of atmospheric neutrino oscillation parameters by SUPER-KAMIOKANDE I}",
    eprint = "hep-ex/0501064",
    archivePrefix = "arXiv",
    doi = "10.1103/PhysRevD.71.112005",
    journal = "Phys. Rev. D",
    volume = "71",
    pages = "112005",
    year = "2005"
}

@article{Learned:1979gp,
    author = "Learned, J. and Reines, F. and Soni, A.",
    title = "{Limits on Nonconservation of Baryon Number}",
    reportNumber = "UCI-79-54",
    doi = "10.1103/PhysRevLett.43.907",
    journal = "Phys. Rev. Lett.",
    volume = "43",
    pages = "907",
    year = "1979",
    note = "[Erratum: Phys.Rev.Lett. 43, 1626 (1979)]"
}

@article{Super-Kamiokande:2016exg,
    author = "Abe, K. and others",
    collaboration = "Super-Kamiokande",
    title = "{Search for proton decay via $p \to e^+\pi^0$ and $p \to \mu^+\pi^0$ in 0.31  megaton\textperiodcentered{}years exposure of the Super-Kamiokande water Cherenkov detector}",
    eprint = "1610.03597",
    archivePrefix = "arXiv",
    primaryClass = "hep-ex",
    doi = "10.1103/PhysRevD.95.012004",
    journal = "Phys. Rev. D",
    volume = "95",
    number = "1",
    pages = "012004",
    year = "2017"
}

@article{Super-Kamiokande:2014otb,
    author = "Abe, K. and others",
    collaboration = "Super-Kamiokande",
    title = "{Search for proton decay via $p\to\nu K^+$ using 260  kiloton\textperiodcentered{}year data of Super-Kamiokande}",
    eprint = "1408.1195",
    archivePrefix = "arXiv",
    primaryClass = "hep-ex",
    doi = "10.1103/PhysRevD.90.072005",
    journal = "Phys. Rev. D",
    volume = "90",
    number = "7",
    pages = "072005",
    year = "2014"
}

@article{Super-Kamiokande:2024qbv,
    author = "Taniuchi, N. and others",
    collaboration = "Super-Kamiokande",
    title = "{Search for proton decay via p\textrightarrow{}e+\ensuremath{\eta} and p\textrightarrow{}\ensuremath{\mu}+\ensuremath{\eta} with a 0.37~Mton-year exposure of Super-Kamiokande}",
    eprint = "2409.19633",
    archivePrefix = "arXiv",
    primaryClass = "hep-ex",
    doi = "10.1103/PhysRevD.110.112011",
    journal = "Phys. Rev. D",
    volume = "110",
    number = "11",
    pages = "112011",
    year = "2024"
}

@article{Grojean:2023tsd,
    author = "Grojean, Christophe and Kley, Jonathan and Yao, Chang-Yuan",
    title = "{Hilbert series for ALP EFTs}",
    eprint = "2307.08563",
    archivePrefix = "arXiv",
    primaryClass = "hep-ph",
    reportNumber = "DESY-23-098, HU-EP-23/39",
    doi = "10.1007/JHEP11(2023)196",
    journal = "JHEP",
    volume = "11",
    pages = "196",
    year = "2023"
}

@article{Cherry:1981uq,
    author = "Cherry, M. L. and Deakyne, M. and Lande, K. and Lee, C. K. and Steinberg, R. I. and Cleveland, B. T.",
    title = "{Experimental Test of Baryon Conservation: A New Limit on the Nucleon Lifetime}",
    doi = "10.1103/PhysRevLett.47.1507",
    journal = "Phys. Rev. Lett.",
    volume = "47",
    pages = "1507--1510",
    year = "1981"
}

@article{Quevillon:2020hmx,
    author = "Quevillon, J\'er\'emie and Smith, Christopher",
    title = "{Baryon and lepton number intricacies in axion models}",
    eprint = "2006.06778",
    archivePrefix = "arXiv",
    primaryClass = "hep-ph",
    doi = "10.1103/PhysRevD.102.075031",
    journal = "Phys. Rev. D",
    volume = "102",
    number = "7",
    pages = "075031",
    year = "2020"
}

@article{Brugeat:2024rxe,
    author = "Brugeat, Th\'eo and Smith, Christopher",
    title = "{Dark-matter induced neutron-antineutron oscillations}",
    eprint = "2412.06434",
    archivePrefix = "arXiv",
    primaryClass = "hep-ph",
    doi = "10.1007/JHEP01(2025)132",
    journal = "JHEP",
    volume = "01",
    pages = "132",
    year = "2025"
}

@article{Peccei:1977hh,
    author = "Peccei, R. D. and Quinn, Helen R.",
    title = "{CP Conservation in the Presence of Instantons}",
    reportNumber = "ITP-568-STANFORD",
    doi = "10.1103/PhysRevLett.38.1440",
    journal = "Phys. Rev. Lett.",
    volume = "38",
    pages = "1440--1443",
    year = "1977"
}

@article{Peccei:1977ur,
    author = "Peccei, R. D. and Quinn, Helen R.",
    title = "{Constraints Imposed by CP Conservation in the Presence of Instantons}",
    reportNumber = "ITP-572-STANFORD",
    doi = "10.1103/PhysRevD.16.1791",
    journal = "Phys. Rev. D",
    volume = "16",
    pages = "1791--1797",
    year = "1977"
}

@article{Weinberg:1977ma,
    author = "Weinberg, Steven",
    title = "{A New Light Boson?}",
    reportNumber = "HUTP-77/A074",
    doi = "10.1103/PhysRevLett.40.223",
    journal = "Phys. Rev. Lett.",
    volume = "40",
    pages = "223--226",
    year = "1978"
}

@article{Wilczek:1977pj,
    author = "Wilczek, Frank",
    title = "{Problem of Strong  $P$  and  $T$  Invariance in the Presence of Instantons}",
    reportNumber = "Print-77-0939 (COLUMBIA)",
    doi = "10.1103/PhysRevLett.40.279",
    journal = "Phys. Rev. Lett.",
    volume = "40",
    pages = "279--282",
    year = "1978"
}

@article{Weinberg:1978kz,
    author = "Weinberg, Steven",
    editor = "Deser, S.",
    title = "{Phenomenological Lagrangians}",
    reportNumber = "HUTP-78-A051A",
    doi = "10.1016/0378-4371(79)90223-1",
    journal = "Physica A",
    volume = "96",
    number = "1-2",
    pages = "327--340",
    year = "1979"
}

@article{Gasser:1983yg,
    author = "Gasser, J. and Leutwyler, H.",
    title = "{Chiral Perturbation Theory to One Loop}",
    reportNumber = "CERN-TH-3689",
    doi = "10.1016/0003-4916(84)90242-2",
    journal = "Annals Phys.",
    volume = "158",
    pages = "142",
    year = "1984"
}

@article{Gasser:1984gg,
    author = "Gasser, J. and Leutwyler, H.",
    title = "{Chiral Perturbation Theory: Expansions in the Mass of the Strange Quark}",
    reportNumber = "CERN-TH-3798",
    doi = "10.1016/0550-3213(85)90492-4",
    journal = "Nucl. Phys. B",
    volume = "250",
    pages = "465--516",
    year = "1985"
}

@article{Chikashige:1980ui,
    author = "Chikashige, Y. and Mohapatra, Rabindra N. and Peccei, R. D.",
    title = "{Are There Real Goldstone Bosons Associated with Broken Lepton Number?}",
    reportNumber = "MPI-PAE-PTH-36-80",
    doi = "10.1016/0370-2693(81)90011-3",
    journal = "Phys. Lett. B",
    volume = "98",
    pages = "265--268",
    year = "1981"
}

@article{Hui:2016ltb,
    author = "Hui, Lam and Ostriker, Jeremiah P. and Tremaine, Scott and Witten, Edward",
    title = "{Ultralight scalars as cosmological dark matter}",
    eprint = "1610.08297",
    archivePrefix = "arXiv",
    primaryClass = "astro-ph.CO",
    doi = "10.1103/PhysRevD.95.043541",
    journal = "Phys. Rev. D",
    volume = "95",
    number = "4",
    pages = "043541",
    year = "2017"
}

@article{Niemeyer:2019aqm,
    author = "Niemeyer, Jens C.",
    title = "{Small-scale structure of fuzzy and axion-like dark matter}",
    eprint = "1912.07064",
    archivePrefix = "arXiv",
    primaryClass = "astro-ph.CO",
    doi = "10.1016/j.ppnp.2020.103787",
    month = "12",
    year = "2019"
}

@article{Super-Kamiokande:2005lev,
    author = "Kobayashi, K. and others",
    collaboration = "Super-Kamiokande",
    title = "{Search for nucleon decay via modes favored by supersymmetric grand unification models in Super-Kamiokande-I}",
    eprint = "hep-ex/0502026",
    archivePrefix = "arXiv",
    doi = "10.1103/PhysRevD.72.052007",
    journal = "Phys. Rev. D",
    volume = "72",
    pages = "052007",
    year = "2005"
}

@article{Super-Kamiokande:2013rwg,
    author = "Abe, K. and others",
    collaboration = "Super-Kamiokande",
    title = "{Search for Nucleon Decay via $n \to \bar{\nu} \pi^{0}$ and $p \to \bar{\nu} \pi^{+}$ in Super-Kamiokande}",
    eprint = "1305.4391",
    archivePrefix = "arXiv",
    primaryClass = "hep-ex",
    doi = "10.1103/PhysRevLett.113.121802",
    journal = "Phys. Rev. Lett.",
    volume = "113",
    number = "12",
    pages = "121802",
    year = "2014"
}

@article{Mohapatra:1982tc,
    author = "Mohapatra, Rabindra N. and Senjanovic, Goran",
    title = "{The Superlight Axion and Neutrino Masses}",
    reportNumber = "BNL-30711",
    doi = "10.1007/BF01577819",
    journal = "Z. Phys. C",
    volume = "17",
    pages = "53--56",
    year = "1983"
}

@article{Preskill:1982cy,
    author = "Preskill, John and Wise, Mark B. and Wilczek, Frank",
    editor = "Srednicki, M. A.",
    title = "{Cosmology of the Invisible Axion}",
    reportNumber = "HUTP-82-A048, NSF-ITP-82-103",
    doi = "10.1016/0370-2693(83)90637-8",
    journal = "Phys. Lett. B",
    volume = "120",
    pages = "127--132",
    year = "1983"
}

@article{Dine:1982ah,
    author = "Dine, Michael and Fischler, Willy",
    editor = "Srednicki, M. A.",
    title = "{The Not So Harmless Axion}",
    reportNumber = "UPR-0201T",
    doi = "10.1016/0370-2693(83)90639-1",
    journal = "Phys. Lett. B",
    volume = "120",
    pages = "137--141",
    year = "1983"
}

@article{Abbott:1982af,
    author = "Abbott, L. F. and Sikivie, P.",
    editor = "Srednicki, M. A.",
    title = "{A Cosmological Bound on the Invisible Axion}",
    reportNumber = "PRINT-82-0695 (BRANDEIS)",
    doi = "10.1016/0370-2693(83)90638-X",
    journal = "Phys. Lett. B",
    volume = "120",
    pages = "133--136",
    year = "1983"
}

@article{Super-Kamiokande:2022egr,
    author = "Matsumoto, R. and others",
    collaboration = "Super-Kamiokande",
    title = "{Search for proton decay via $p\rightarrow \mu^+K^0$ in 0.37 megaton-years exposure of Super-Kamiokande}",
    eprint = "2208.13188",
    archivePrefix = "arXiv",
    primaryClass = "hep-ex",
    doi = "10.1103/PhysRevD.106.072003",
    journal = "Phys. Rev. D",
    volume = "106",
    number = "7",
    pages = "072003",
    year = "2022"
}

@article{Sjostrand:2006za,
    author = "Sjostrand, Torbjorn and Mrenna, Stephen and Skands, Peter Z.",
    title = "{PYTHIA 6.4 Physics and Manual}",
    eprint = "hep-ph/0603175",
    archivePrefix = "arXiv",
    reportNumber = "FERMILAB-PUB-06-052-CD-T, LU-TP-06-13",
    doi = "10.1088/1126-6708/2006/05/026",
    journal = "JHEP",
    volume = "05",
    pages = "026",
    year = "2006"
}

@article{Bierlich:2022pfr,
    author = "Bierlich, Christian and others",
    title = "{A comprehensive guide to the physics and usage of PYTHIA 8.3}",
    eprint = "2203.11601",
    archivePrefix = "arXiv",
    primaryClass = "hep-ph",
    reportNumber = "LU-TP 22-16, MCNET-22-04, FERMILAB-PUB-22-227-SCD",
    doi = "10.21468/SciPostPhysCodeb.8",
    journal = "SciPost Phys. Codeb.",
    volume = "2022",
    pages = "8",
    year = "2022"
}

@article{Helo:2018bgb,
    author = "Helo, Juan C. and Hirsch, Martin and Ota, Toshihiko",
    title = "{Proton decay and light sterile neutrinos}",
    eprint = "1803.00035",
    archivePrefix = "arXiv",
    primaryClass = "hep-ph",
    doi = "10.1007/JHEP06(2018)047",
    journal = "JHEP",
    volume = "06",
    pages = "047",
    year = "2018"
}

@article{Heeck:2020nbq,
    author = "Heeck, Julian",
    title = "{Light particles with baryon and lepton numbers}",
    eprint = "2009.01256",
    archivePrefix = "arXiv",
    primaryClass = "hep-ph",
    doi = "10.1016/j.physletb.2020.136043",
    journal = "Phys. Lett. B",
    volume = "813",
    pages = "136043",
    year = "2021"
}

@article{Fajfer:2020tqf,
    author = "Fajfer, Svjetlana and Susi{\v{c}}, David",
    title = "{Colored scalar mediated nucleon decays to an invisible fermion}",
    eprint = "2010.08367",
    archivePrefix = "arXiv",
    primaryClass = "hep-ph",
    doi = "10.1103/PhysRevD.103.055012",
    journal = "Phys. Rev. D",
    volume = "103",
    number = "5",
    pages = "055012",
    year = "2021"
}

@article{Domingo:2024qoj,
    author = {Domingo, Florian and Dreiner, Herbi K. and K{\"o}hler, Dominik and Nangia, Saurabh and Shah, Apoorva},
    title = "{A novel proton decay signature at DUNE, JUNO, and Hyper-K}",
    eprint = "2403.18502",
    archivePrefix = "arXiv",
    primaryClass = "hep-ph",
    reportNumber = "BONN-TH-2024-08",
    doi = "10.1007/JHEP05(2024)258",
    journal = "JHEP",
    volume = "05",
    pages = "258",
    year = "2024"
}

@article{Heeck:2025uwh,
    author = "Heeck, Julian and Shoemaker, Ian M.",
    title = "{Nucleon Decays into Light New Particles in Neutrino Detectors}",
    eprint = "2506.08090",
    archivePrefix = "arXiv",
    primaryClass = "hep-ph",
    month = "6",
    year = "2025"
}

@article{Davoudiasl:2014gfa,
    author = "Davoudiasl, Hooman",
    title = "{Nucleon Decay into a Dark Sector}",
    eprint = "1409.4823",
    archivePrefix = "arXiv",
    primaryClass = "hep-ph",
    doi = "10.1103/PhysRevLett.114.051802",
    journal = "Phys. Rev. Lett.",
    volume = "114",
    number = "5",
    pages = "051802",
    year = "2015"
}

@article{Fornal:2018eol,
    author = "Fornal, Bartosz and Grinstein, Benjamin",
    title = "{Dark Matter Interpretation of the Neutron Decay Anomaly}",
    eprint = "1801.01124",
    archivePrefix = "arXiv",
    primaryClass = "hep-ph",
    doi = "10.1103/PhysRevLett.120.191801",
    journal = "Phys. Rev. Lett.",
    volume = "120",
    number = "19",
    pages = "191801",
    year = "2018",
    note = "[Erratum: Phys.Rev.Lett. 124, 219901 (2020)]"
}

@article{Bastero-Gil:2024kjo,
    author = "Bastero-Gil, Mar and Huertas-Roldan, Teresa and Santos, Daniel",
    title = "{Neutron decay anomaly, neutron stars, and dark matter}",
    eprint = "2403.08666",
    archivePrefix = "arXiv",
    primaryClass = "astro-ph.CO",
    doi = "10.1103/PhysRevD.110.083003",
    journal = "Phys. Rev. D",
    volume = "110",
    number = "8",
    pages = "083003",
    year = "2024"
}

@article{Bagger:1994hh,
    author = "Bagger, Jonathan and Poppitz, Erich and Randall, Lisa",
    title = "{The R axion from dynamical supersymmetry breaking}",
    eprint = "hep-ph/9405345",
    archivePrefix = "arXiv",
    reportNumber = "JHU-TIPAC-940005, JHU-TIPAC-94005, MIT-CTP-2309, NSF-ITP-94-48",
    doi = "10.1016/0550-3213(94)90123-6",
    journal = "Nucl. Phys. B",
    volume = "426",
    pages = "3--18",
    year = "1994"
}

@article{Branco:2011iw,
    author = "Branco, G. C. and Ferreira, P. M. and Lavoura, L. and Rebelo, M. N. and Sher, Marc and Silva, Joao P.",
    title = "{Theory and phenomenology of two-Higgs-doublet models}",
    eprint = "1106.0034",
    archivePrefix = "arXiv",
    primaryClass = "hep-ph",
    doi = "10.1016/j.physrep.2012.02.002",
    journal = "Phys. Rept.",
    volume = "516",
    pages = "1--102",
    year = "2012"
}

@article{Witten:1984dg,
    author = "Witten, Edward",
    title = "{Some Properties of O(32) Superstrings}",
    reportNumber = "Print-84-0838 (PRINCETON)",
    doi = "10.1016/0370-2693(84)90422-2",
    journal = "Phys. Lett. B",
    volume = "149",
    pages = "351--356",
    year = "1984"
}

@article{Ringwald:2012cu,
    author = "Ringwald, Andreas",
    editor = "Mondrag{\'o}n, Myriam and Bashir, Adnan and Delepine, David and Larios, Francisco and Loaiza, Oscar and de la Macorra, Axel and Nellen, Lukas and Sahu, Sarira and Salazar, Humberto and Velasco-Sevilla, Liliana",
    title = "{Searching for axions and ALPs from string theory}",
    eprint = "1209.2299",
    archivePrefix = "arXiv",
    primaryClass = "hep-ph",
    reportNumber = "DESY-12-151",
    doi = "10.1088/1742-6596/485/1/012013",
    journal = "J. Phys. Conf. Ser.",
    volume = "485",
    pages = "012013",
    year = "2014"
}

@article{Bonnefoy:2022rik,
    author = "Bonnefoy, Quentin and Grojean, Christophe and Kley, Jonathan",
    title = "{Shift-Invariant Orders of an Axionlike Particle}",
    eprint = "2206.04182",
    archivePrefix = "arXiv",
    primaryClass = "hep-ph",
    reportNumber = "DESY-22-096, HU-EP-22/22",
    doi = "10.1103/PhysRevLett.130.111803",
    journal = "Phys. Rev. Lett.",
    volume = "130",
    number = "11",
    pages = "111803",
    year = "2023"
}
\bibliographystyle{JHEP}

\end{document}